\documentstyle[12pt]{article}

\catcode`\@=11
\@addtoreset{equation}{section}
\catcode`@=12

\newcommand{\be}{\begin{equation}}
\newcommand{\ee}{\end{equation}}
\newcommand{\bef}{\begin{figure}}
\newcommand{\enf}{\end{figure}}

\def\spose#1{\hbox to 0pt{#1\hss}}
\def\ltapprox{\mathrel{\spose{\lower 3pt\hbox{$\mathchar"218$}}
 \raise 2.0pt\hbox{$\mathchar"13C$}}}
\def\gtapprox{\mathrel{\spose{\lower 3pt\hbox{$\mathchar"218$}}
 \raise 2.0pt\hbox{$\mathchar"13E$}}}
\def\inapprox{\mathrel{\spose{\lower 3pt\hbox{$\mathchar"218$}}
 \raise 2.0pt\hbox{$\mathchar"232$}}}

\begin{document}

\centerline{\LARGE FACTS AND IDEAS IN MODERN COSMOLOGY}
\centerline{}
\centerline{}
\centerline{}
\centerline{\Large Yu. V. Baryshev$^{(1,2)}$
F. Sylos Labini$^{(3,4)}$ M. Montuori$^{(4,5)}$  L. Pietronero$^{(4)}$
\footnote{Send offprint request: to F. Sylos Labini}}
\centerline{}
\centerline{}
\centerline{($^1$) Astronomical Institute of St.
Petersburg University,
St. Petersburg 198904, Russia}
\centerline{($^2$) Scientific-Educational Union
"Earth \& Universe", St.
Petersburg, Russia}
\centerline{($^3$) Dipartimento di Fisica, Universit\`a di
Bologna, Italy}
\centerline{($^4$) Dipartimento di Fisica, Universit\`a di Roma
" La Sapienza''}
\centerline{P.le A. Moro 2, I-00185 Roma, Italy.}
\centerline{($^5$) Dipartimento di Fisica, Universit\`a di
Cosenza, Italy}
\centerline{}
\centerline{}
\centerline{}
\centerline{\it Accepted for publication in}
\centerline{\Large {\it Vistas In astronomy} {\bf Vol.38} {\it Part.4}, 1994}

\centerline{}
\centerline{}
\centerline{}
\begin{abstract}
A review of the principles of observational testing
of cosmological theories
is given with a special
emphasis on the distinction between observational facts
and theoretical hypotheses. A classification of
modern cosmological theories
and possible observational
tests for these theories is presented.
The main rival cosmological
models are analyzed from the point of view of observational testing
of their initial hypothesis.
A comparison of modern observational data
with theoretical predictions is presented.
In particular we discuss in detail the validity
of the two basic assumptions of modern cosmology
that are the Cosmological Principle and the Expanding
Space Paradigm.
It is found that classical
paradigms need to be reanalyzed
and  that
it is necessary to develop crucial cosmological tests to
discriminate alternative theories.
\end{abstract}

\section{INTRODUCTION}

Cosmology as a part of physics is an experimental science.
For this reason  all reasonable
relations in cosmology must have an experimental confirmation.
The fast growth of observational data in the last two decades
now has made possible the comparison between observable
quantities and theoretical predictions.
The generally accepted basic assumptions
for interpretation of observational data in
cosmology remain the {\it Cosmological Principle}
and the {\it Expanding Space Paradigm}.
But recently acute discussions about
the validity of alternative theories
have been  resumed in literature
(see e.g. Arp et al., 1990; Peebles et al., 1991; Hoyle et al., 1993;
1994a; 1994b).
This is not accidental and indicates a new situation in
observational cosmology.

The main cosmological theories are based on the assumption
of the homogeneity of matter distribution.
The reason is the following.
The basic hypothesis of a post-Copernican Cosmological theory is that
{\em all the points} of the Universe have to be essentially equivalent:
this hypothesis is
required in order to avoid any privileged {\em observer}.
This assumption has been implemented
by Einstein in the so-called
 Cosmological Principle (CP): {\em all the positions}
in the Universe have to be essentially equivalent,
so that the Universe is homogeneous.
This situation implies also the condition of spherical
symmetry about every point, so that the Universe is also Isotropic.
There is a hidden assumption in the formulation of the CP with regard to
the hypothesis
that all the points are equivalent.
The condition that all the occupied points
are statistically equivalent with
respect to their environment corresponds in fact
to the property of
Local Isotropy. It is generally believed that the
Universe cannot be isotropic about every
point without being also homogeneous (Weinberg, 1972).
Actually Local Isotropy does not necessarily imply homogeneity
(Sylos Labini, 1994);
in fact a topology theorem
states that homogeneity
is implied by the condition of local isotropy
together with {\em the assumption
of the analyticity or regularity} for the distribution of matter.
Up to the seventies analyticity was an obvious implicit assumption
in any physical  problem. Recently however we have learned
about intrinsically
irregular structures for which analyticity should be considered as a
property to be tested with appropriate analysis of experiment
(Mandelbrot, 1982; Pietronero \& Tosatti, 1986).

The current idea is that in the observable
Large-Scale Structure distribution,
isotropy and homogeneity do not apply to the Universe
in detail but only to a "smeared-out " Universe,
averaged over regions of order
$\:\lambda_{0}$.
One of the main problems
of observational cosmology is therefore
the  identification of $\:\lambda_{0}$,
but the observational situation appears highly
problematic. In fact in the available redshift surveys
a clear cut-off towards homogenization has not been identified.
On the other hand the Cosmic Microwave Background Radiation (CMBR),
that is one of the most important experimental facts in
modern cosmology, has a perfect black-body spectrum
(Mather et al., 1994) and it
is exceptionally isotropic (Strukov et al., 1992;
Smoot et al., 1992),
so that many theories of galaxy formation in the
framework of the Big Bang model have difficulties
in considering the small temperature fluctuations of the CMBR
as the seeds that give rise to such complex large-scale structures of matter.

Moreover, long term hopes on  classical tests such as
$\:\Theta(z), m(z), N(m)$ are destroyed by recent
observations. A lot of
observational data are now available for
classical tests up to $\: z \approx 3$
and $\: m  \approx 28$
(while Hubble began with
$\: z \approx 0.001$ and $\:m \approx 19$; Hubble, 1929) but
there is no clear empirical answer to the question about the
geometry of the Universe. What is more, such
fundamental questions as
the nature of the cosmological redshift are still without
empirical answer. Indeed the first point in  Sandage's list of
unsolved problems
for the next two decades is the "proof or not that the redshift is
a true expansion" (Sandage, 1987)
and this concerns the {\it Expanding Space Paradigm}.
 This is why a number of
cosmological models are discussed
now in the framework of the Big Bang as well as
of alternative theories  (see e.g. Narlikar 1987, 1989, 1993;
Harrison, 1993;
Hoyle et al., 1993; 1994a; 1994b;
Narlikar et al., 1993; Peebles et al., 1991, 1994;
Coles \& Ellis, 1994).

The present paper reviews cosmological
theories and observational tests which could be performed
to decide between world models.
In {\em section 2} we describe the empirical basis of cosmology
and the theoretical framework of modeling the Universe.
{\em Section 3} is devoted to the analysis of the basic
hypothesis of the cosmological models
and their predictions for observations.
In {\em section 4} we compare the observational data now
available with the predictions of different models.
Finally in {\em section 5}
we summarize the situation and formulate the main conclusions.

\section{GENERAL BASIS OF COSMOLOGY}

The observational data and contemporary theoretical physics
are the basis of cosmology. Any given cosmological model
must be checked by comparison with observations.
We divide the cosmological tests into crucial and parametric ones. The
first deal with the fundamental basis of any cosmological theory.
The tests of the second kind give an estimate of the parameters of
different models. It is important to stress that in real
astronomical data several biases and selection effects occur, and
hence, the comparison of theoretical
prediction with observational evidence is a very difficult task.

\subsection{Empirical basis for cosmological theories}
Well established observations are the basis of any
theoretical cosmological model.
The experimental results
which one can consider as
fundamental empirical facts are:

\begin{itemize}
\item the local laboratory physics;
\item the global cosmological redshift
 and its linear dependence from distance;
\item the space distribution and motion of galaxies;
\item the cosmic background radiation,
and in particular the microwave radiation;
\item the mean chemical composition of matter;
\item ages of different kinds of celestial objects.
\end{itemize}

It is not a trivial statement that the
local laboratory physics is a part of
the cosmologically
distributed matter. Such deep hypotheses, as Mach's
principle and Feynman-Wheeler electrodynamics, are
based on this fact.
The connection between local and global
Universe is the most difficult problem in cosmology
and it is still without a definite answer.

The  cosmological  redshift is
the second very important fact. The definition of redshift
$\:z$ is
\be
\label{1}
z= \frac {\lambda_{obs}- \lambda_{emit}}
 {\lambda_{emit}}=\frac{\nu_{emit}-\nu_{obs}}{\nu_{obs}}
\ee
where $\:\lambda_{emit}$
(or $\nu_{emit}$) is the
wavelength (or frequency) of
the light emitted from a light
source, and  $\:\lambda_{obs}$
(or $\nu_{obs}$) is the
wavelength (or frequency) of
the light observed.
This observed redshift $\:z$ has many
interpretations, but, in any case, any
physical interpretation should be
kept well distinct from the
observational measurement, as stressed by many
authors
 (see e.g. North, 1965; Harrison, 1981, 1993).
The observed redshift-distance relation
 was established by Hubble (1929)
for small $\:z$ and small distance $\:r$ in the form
\be
\label{2}
z =\frac{H_{0}}{c}r=\frac{r}{R_{H_{0}}}
\ee
where $\:H_{0}$ is the Hubble constant ($\:50-100 km sec^{-1}Mpc^{-1}$),
$\: c $ is the velocity of light and
$\:R_{H_{0}}= c/H_{0}$ is the Hubble radius.
Modern estimates of the value $\:H_{0}$
\footnote{ We shall use in this paper the normalized
Hubble constant $\:h=H_{0}/100 km sec^{-1} Mpc^{-1}$.}
will be discussed in {\em section 4.1.1.}.

The distribution and motion of galaxies
in space has been investigated very intensively
in the last few years.
Several recent galaxy redshift
surveys such as {\it CfA1} (Huchra et al., 1983),
{\it CfA2} (De Lapparent et al., 1988;
Da Costa et al., 1994; Park et al., 1994),
 {\it SSRS1}
(Da Costa et al., 1988),
{\it SSRS2} (Da Costa et al., 1994),
{\it Perseus-Pisces} ( Giovanelli \& Haynes, 1986)
and also pencil beams surveys
(Broadhurst et al., 1990) and {\it ESP}
 (Vettolani et al., 1994),
have uncovered remarkable structures such as
filaments, sheets, superclusters
and voids. These galaxy catalogues
now probe scales from $\: \sim 100-200 h^{-1} Mpc$,
for the wide angle galaxy surveys, up to $\:\sim  1000 h^{-1} Mpc$
for the deeper pencil  beam surveys,
and show that
the large-scale structures are the common features of
the visible Universe.
 One of the most important issues
raised by these catalogues is that the scale of {\em the
largest inhomogeneities} is comparable with {\it the
extent of the surveys}, so that the largest known
structures are limited by {\em the boundaries of
 the survey} in which they are detected.
It is remarkable to note, for example,  that
Tully et al. (1992), analyzing the combined Abell and ACO cluster
catalogues, provide evidence of
structures on a scale of $\:\sim 450 h^{-1} Mpc$,
lying in the plane of the Local Supercluster (see {\it section 4.2.3.}).
Hence from these data  emerges a new picture in which the
scale of homogeneity seems to shift to a very large value, not
yet identified.

The best known background radiation is the
Cosmic Microwave Background Radiation
(CMBR). After Relict and COBE experiments,
the CMBR anisotropies and spectrum are well
known
(Mather et al., 1990; 1994;
Melchiorri \& Melchiorri, 1994; Gush et al., 1990;
 Strukov et al., 1992; Smoot et al., 1992).
The perfect thermal Planck spectrum
 of the CMBR and its very small
anisotropies
are among the most important
observational cosmological facts ({\it section 4.2.4.}).

The relative
abundance of  chemical elements in
the observable Universe represents the most stringent
proof on the processes of generating
and destroying atomic nuclei. We discuss the experimental data
in {\it section 4.1.5.4.}
Another important experimental evidence
concerns the age of stellar systems:
the age estimates of celestial objects are based on
stellar evolution, stellar dynamics and radioactive decay
theories. Hence the ages are not purely observational, but
rather an indirect evidence.
In addition one should not forget
that the age of the
system can be much more than the age of an element, if the
system is open and there is the possibility of
replacing dead elements with new ones. A
discussion of the modern data will be
done in {\em section 4.1.5.1.}.

Now we will consider possible theoretical schemes
which could be proposed to explain these data.

\subsection {A classification of cosmological theories}
We shall classify cosmological models  according to
 their answers to the following questions (see Fig.\ref{fg1}):

\bef
\vspace{12cm}
\caption{\footnotesize
A classification of cosmological models in accordance
with basic initial assumptions.
We have discussed four different cosmological
models: Standard Friedmann model, Steady State model,
Fractal model and Tired Light model.
The basic hypothesis of these models are:
the interpretation of gravity,
the matter distribution, the nature of
cosmic microwave background radiation,
the nature of cosmological redshift,
the interpretation of evolution and
the arrow of time.}
\label{fg1}
\enf

\begin{itemize}
\item What is gravity?
\item How is  matter distributed in space?
\item What is the nature of  the redshift?
\item What is the nature of  the Cosmic Microwave
Background Radiation (CMBR)?
\item What is  evolution and the arrow of time ?
\end {itemize}

\subsubsection{Gravitation theories}

The heart of any cosmological  theory  is the  gravitation theory.
Gravity is the only important force beyond galaxy scale.
  The gravitation theory can be constructed in different ways.
Presently there are at least three main approaches to
relativistic gravity theories. The first one considers
gravity as a property of space-time, i.e. it is the
geometry of curved space-time. The second one treats
gravity as a kind of matter within the space-time:
 it is the relativistic field theory in
flat space-time. The third approach is
based on the direct interaction between gravitating particles.
It is important to note that
up to now relativistic gravity has been tested experimentally
 only in  weak field approximation. Such well known
 relativistic effects as the bending of light, the
 gravitational frequency shift, the pericenter advance,
 the delay of light signals, rotational effects and gravitational
 radiation from binary systems may be derived  in
many  reasonable gravitational theories.

According to the geometrical description,
 gravity is curvature of space-time and test
particles move along geodesic lines in the
curved space.
 The metric $\:g^{ik}$ of space-time
 (hereafter simply "space") is determined
 by the distribution of matter via Einstein-Hilbert
equations  (see e.g. Weinberg,
 1972; Misner, Thorn \& Wheeler, 1973).
In the framework of geometric theories there are
 several generalizations  of Einstein's
equations, such as the famous   $\:\Lambda-term$,
(see Weinberg, 1989),
the additional scalar field in Brans
 - Dicke's theory (Brans \& Dicke, 1961),
 the $\:C $-field in Hoyle's theory (Hoyle,
1948, 1991, 1993b),
the variable mass Narlikar-Hoyle theory, the non
symmetrical affine connection, etc.
 An interesting class of quasi-geometrical theories is
represented by the bimetric-theories, in which two metrics
coexist; one is the Minkowski metric $\:\eta^{ik}$ and
 the other is the metric of a Riemannian (effective)
 space $\:g^{ik}$  (see e.g. Logunov \& Mestvirishvili, 1989).
Cosmological solutions of these theories could
differ significantly from classic  Friedmann  models
 despite the fact that there is an agreement in the
weak field approximation.
It should be stressed that
 although the geometry has great success in
gravity physics, there are several conceptual
difficulties within the geometrical description
of gravity, such as:
what is the curved measuring rod in curved space ?
How do conservation laws work without
flat space-time symmetry ?
What is the graviton, i.e. the quantum
of the gravitational field?
Really, we know that the photon is the massless particle of
electromagnetic field which carries
energy-momentum within the space, but if gravity
is  the geometry of space, which particle
carries energy-momentum through the space?
Another problem is the pseudotensorial character of
the gravitation energy-momentum.

The tensor field description of gravity
 by means of symmetric  tensor field  $\:\psi^{ik}$
in flat Minkowski space-time,
could resolve these problems, because gravity
 may be thought as a material field in the space-time.
 In such an approach, due to the  symmetry of Minkowski space,
 all conservation
 laws exist; graviton is a massless material particle
 in space, energy density of gravitational field is
 positive for free and static fields.
The main difficulty, in this approach,
is that Lagrangian formalism does not
 determine uniquely the energy momentum tensor
(EMT) of
 any field. But in the case of
 gravitation, the EMT is the source of the field
(nonlinear character of gravity)
 and must determine the right side of
 the field equations.
The tensor
field theory of gravitation (TFT)
 has been discussed by many authors
 (see e.g. Thirring, 1961;  Kalman, 1961;
Weinberg,
1965; Ogievetsky \&
 Polubarinov, 1965; Whitrow \&
Morduch, 1965; Deser, 1970;
Bowler, 1976; Cavalleri \& Spinelli, 1980;
Sokolov \& Baryshev, 1980; Baryshev, 1988; Sokolov, 1992;
 Baryshev, 1994b),
but it is poorly
known among astrophysicists.
Within the context of the tensor field theory
 there are many possibilities for choosing the
field and interaction Lagrangians
and one can get  Einstein-like
field equations (Ogievetskij \& Polubarinov, 1965)
as another scalar-tensor possibility.
A detailed  review of the TFT and its
astrophysical applications will be given by Baryshev (1995a).

The third possibility is the relativistic direct-interaction
theory of gravitation, which may be considered by
analogy with classical electrodynamics
in terms of direct inter-particle action
 (Wheeler \& Feynman, 1949).
Within the scope of this theory,
 there is no such concept as the field and no
problem of an infinity in the energy
of the electromagnetic field of the point charge.
This approach has been considered in
 details in the textbook by Hoyle \&  Narlikar
 (1974) and in the thesis of Pantyushin (1972).
The direct interaction theory is
 particularly attractive for cosmology. In fact,
one can interpret the independent degrees
 of freedom of the field, in the field theory, as
 the interaction of local systems with the
whole Universe.

\subsubsection{Matter distribution}

In this section we discuss the distribution of
visible matter (galaxies) as an experimental fact
and only later we consider its theoretical implication.
It is important to specify this perspective because this field
is often strongly influenced by theoretical
expectations or "principles".
The Cosmological Principle was introduced to avoid schemes in
which we would be in a special point of the universe.
This is indeed a reasonable requirement that, however, was
interpreted in practice as corresponding to the too strict
requirements of isotropy and homogeneity. The present
discussion will show that it is risky to adopt
principles that have not been tested experimentally.
However we are going to see in the end
that the essence of the principle is reasonable
and correct. What is not correct is the usual
mathematical interpretation that was due to the lack of
other more subtle concepts that have been
developed only recently.

With the Cosmological Principle in mind people began to
look at the observational data about galaxy positions
holding the
idea that, obviously, the distribution of galaxies
must become homogeneous above some relatively short length scale.
The first observations, like the Lick catalogue, referred only
to the angular positions of galaxies in the
sky and indeed showed a well defined tendency towards
a smooth distribution at relatively large angular scales.
The situation appeared to be in agreement with expectations
and these data were analyzed via Limber equation to go from
angular to space coordinates. Limber equations are
based on the assumption that the scale of eventual correlations
is much smaller than the size of the catalogue and allow us
to reconstruct the real three dimensional
properties from the angular ones. In this
way a correlation length of {\it 5 Mpc}
was identified as the characteristic
length for homogenization.

This well behaved situation began to shake with the
extensive
redshift measurements. The direct test of the actual
positions of galaxies in three dimensional
space showed in fact the presence of structures like
clusters and superclusters, as well as voids, whose sizes were
much larger than the previously derived characteristic
length. At first the data were questioned as incomplete or referring
to rare fluctuations. Contrary to these expectations, however,
the improvements of the data made these structures sharper and
showed that they are actually everywhere. The first
three dimensional catalogues,
like {\it CfA1 }, are much more irregular than the
angular catalogues. This appeared to be a critical situation,
but then, a statistical analysis, based on the Limber type hypothesis,
was performed on these three dimensional catalogues and
confirmed the existence of a correlation length of
$\: \sim 5 Mpc$ above which the distribution is expected to approach
to the homogenization. This was, in our opinion, the crucial
point when the field
entered into a situation of ambiguity and confusion
 that persists still today. As we are going to
see, the question is simple but subtle and
 requires new concepts that were
not available at that time (see {\it section 3.4.}).

It may be interesting to
note that the confrontation with intrinsic
irregularity and non-analyticity has been
 quite dramatic also in other fields of
physics like, for example, critical
 phenomena that required conceptual revolution
 of the renormalization group to be understood.
 The additional problem with
cosmology, that explains in part the resistance to
these new ideas, is that intrinsic irregularities
and non-analyticity appear to be in contrast
with the Cosmological Principle: we are
 going to see that this is actually not the
 case and that, in many ways, the new
picture, that will arise from our discussion, is
much simpler than the standard one.

\subsubsection {The nature of the redshift}

As we have emphasized above, the cosmological redshift
is an observational fact which can be interpreted
in different ways. Harrison (1981, 1993)
 has shown a clear distinction between different possibilities
 to interpret the observational redshift-distance relation.
There are at least four possible mechanisms for cosmological redshift:
space expansion,
Doppler effect, gravitational effect, and tired light effect.

The space expansion redshift is due to increasing
of space volume ("space creation") in an
expanding universe.
The expansion redshift  is
determined by the Lemaitre expansion-redshift law:
\be
\label{9}
\nu_{obs} = L \nu_{emit}
\ee
\be
\label{10}
L=\frac{R(t)}{R_{0}(t_{0})}
\ee
\be
\label{11}
(1+z)_{expansion}=\frac{R_{0}(t_{0})}{R(t)}
\ee
where $\nu_{emit}$ is the emitted frequency
 and $\nu_{obs}$ is the observed one, $\:L$ is the
expansion Lemaitre factor, $\:R$ is the value of the
scaling factor at the time of emission and $\:R_{0}$ is the value at
the time of reception. The ratio $\:R_{0}/R$ means how much the
universe has expanded during the time of photon traveling. The
redshift (Eq.\ref{11}) could be understood
as a result of the effective recession velocity, but not
as a result of the ordinary
physical velocity in ordinary physical space.
In particular the effective recession
velocity can be much greater than velocity of light (see e.g.
Murdoch, 1977;  Harrison, 1981, 1993).

The
expanding space paradigm, even if wildly accepted, has been criticized
 by several authors (see e.g.  Milne, 1934),
who emphasized that space itself has no existence
and probably it is
more physically correct to use static
space, as in ordinary physics, and consider the expansion
of matter as  motion in this space. In this perspective
the observed cosmological redshift could be due to the
Doppler effect, gravitation and tired light effects.

It was shown by Bondi (1947), that in the case
of spherically symmetrical distribution
 of dust-like matter, the space expansion redshift
can be expressed as a sum of two parts.
The first one is the Doppler effect, due
to the relative velocity of the
source and the observer.
The second part is due to gravitation effect
of the total mass inside the spherical
 ball, with the light source at the center
 and the observer at the ball surface.
 So for small masses ($\:G M / r\ll c^{2}$)
and small velocities ($\:v \ll c$), the cosmological redshift
is given by:
\be
\label{r1}
z_{cos}\approx \frac {v}{c} + \frac
 {1}{2}\left( \frac {v}{c}\right)^{2} + \frac {\delta \Phi(l)}
{c^{2}}
\ee
where
\be
\label{r2}
\delta \Phi(l) = \Phi(l) - \Phi(0) =
\int^{l}_{0} \frac{G M(l')}{l'^{2}}dl'
\ee
where $\delta \Phi(l)$ is the Newtonian potential
difference between the observer at the
surface and the source at the center of a
finite ball with proper radius $\l$, and
 $\Phi(l)$ is the Newtonian potential
($\Phi(l) = -G M(l)/l$).
In the special case of the critical
homogeneous matter distribution,
 $\rho = const.$, $v^{2} = 2G m/l$
from Eq.\ref{r1} and \ref{r2}, we get:
\be
\label{r3}
z_{cos} = z_{Dopl} + z_{grav}
\ee
 where
\be
\label{r4}
z_{Dopl} = \frac{v}{c} +
\frac{1}{2} \left( \frac{v}{c} \right) ^{2}
\ee
and
\be
\label{r5}
z_{grav} = \left(\frac{2 \pi G \rho}
{3 c^{2}}\right) l^{2}
\ee
It means that the cosmological redshift
does not depend only on the conditions
 at the source and at the observer,
 but also on the distribution of matter
in the intervening space around the source.
Note that the gravitational part of the cosmological shift
is redshift and not blueshift, as
it was supposed by Zel'dovich \&
Novikov (1977).
It is important that for the calculation
of $z_{grav}$ in Eq.\ref{r5},
we must consider the mass distribution
around the source, but not around the
observer, and
use proper distance to calculate the mass effect.
It also strictly follows from the
general Mattig's relation between
proper distance and redshift
 (see {\it section 3.2.4.1}).

 To generalize Eq.\ref{r5}
in the case of the fractal matter distribution,
we can use the equivalence of the
 points of the structure and the
the statistically
average spherical symmetry around
each point.
As in the homogeneous case,
 for calculation of the gravitational
 redshift, we must consider mass around
the source.
 Hence, for simple fractal structure with
 fractal dimension $D$, the
 mass in a ball of radius $l$ scales as:
\be
\label{r6}
M(l)\sim l^{D}
\ee
An important consequence of the fractal
distribution of matter is the possibility
of linear gravitational redshift-distance relation
for $D=2$ (Baryshev, 1981, 1994a).
Indeed from Eq. \ref{r5}, for $D=2$,
we get:
\be
\label{r7}
z_{grav} = \left(\frac{2 \pi G \rho_{0}R_{0}l}
{c^{2}} \right) = \frac{H_{grav}\cdot l}{c}
\ee
where
\be
\label{r8}
H_{grav} = \frac{2 \pi G \rho_{0}R_{0}}{c}
\ee
 and $\rho_{0}$, $R_{0}$ is the lower
cut-off of the fractal structure.
To give an order of magnitude for
$H_{grav}$, one can choose values for
$\rho_{0}$ and  $R_{0}$ corresponding to the
typical galaxy parameters, then:
\be
\label{r9}
H_{grav} = 68.6 \left(\frac{\rho_{0}}{5.2\cdot 10^{-24}
gr cm^{-3}}\right)
\left(\frac{R_{0}}{10 kpc}\right)
km\cdot sec^{-1}\cdot Mpc^{-1}
\ee
\smallskip
In the case of a
static spherically symmetric
mass distribution of radius
$r$ and mass $M$, for
the gravitational redshift of
the light by the source at the
surface and observed at infinity is given by:
\be
\label{6}
\nu_{obs}=E\nu_{emit}
\ee
\be
\label{7}
E=\left(1-\frac{2GM}{r c^{2}}\right)^{1/2}
\ee
\be
\label{8}
(1+z)_{Gravity}=\left(1-\frac{2GM}{r c^{2}}\right)^{-1/2}
\ee
 $\:E$ is the
Einstein gravitation factor,
and $\:z_{g}\rightarrow\infty$ when $\:r\rightarrow R_{g}$.

The exact formula for the Doppler effect
caused by the relative motion of bodies in
space is:
\be
\label{3}
\nu_{obs} = D \nu_{emit}
\ee
\be
\label{4}
D=\gamma^{-1}\:(1-\vec{\beta}\cdot \vec{n})^ {-1}
\ee
where $\:D$ is the Doppler factor,
$\:\gamma= (1-\beta^{2}) ^{-\frac{1}{2}}$
is the Lorentz
factor, $\:\vec{\beta}= \vec{v}/{c}$ is
the velocity vector of the moving body
and $\:\vec{n}$ is the unit vector to observer direction.
For purely radial receding motion the angle
$\:\Theta $ between $\:\vec{n}$ and
$\:\vec{\beta}$ equals $\:180^{\circ}$,
 then according to  Eq.\ref{3}
we have redshift
\be
\label{5}
(1+z)_{Doppler}=\left( \frac{c+v}{c-v}\right) ^{1/2}
\ee
Note that Doppler redshift $\:z\rightarrow \infty$
for  $\:v\rightarrow c$.

The tired light redshift is a result of the photon energy loss due to
some unknown physical process. This idea was first suggested by Zwicky
(1929) and later discussed by many authors (see e.g. Hubble \&
Tolman 1935, Geller \& Peebles, 1972; Jaakkola et al., 1979;
LaViolette, 1986; Vigier, 1988).
According to the tired light mechanism, photon
energy is depleted in a linear fashion.
Hence for a finite distance $\:r$ we get:
\be
\label{12}
h\nu_{obs} = Zh\nu_{emit}
\ee
\be
\label{13}
Z = e^{-\alpha r}
\ee
\be
\label{14}
(1+z)_{Tired} = e^{\alpha r}=Z^{-1}
\ee
Where $\:Z$ is the Zwicky factor, $\:\alpha=H_{0}/c$ is the energy
attenuation coefficient, $\:H_{0}$ is the Hubble constant.
 Of course, at the present time,
 there is not any experimental evidence for
this kind of energy dissipation, but in principle it could
be checked in future experiments.

\subsubsection {CMBR nature}

The three main characteristics of the CMBR are the nearly
perfect blackbody spectrum, its exceptional isotropy and that it has
an energy content comparable with that of local radiations, such as
galactic dust, zodiacal light emission and starlight background,
despite the fact
that it is an extragalactic radiation. There are basically two different
approaches to explain the nature of the CMBR.
The first refers to the hypothesis that CMBR is originated
in the hot early Universe. In fact
in the standard picture of Hot Big Bang theory (HBB),
the Universe had expanded from a very dense
and hot state. Consequently the space was filled with blackbody
radiation. As the Universe
expands, the radiation preserves a blackbody spectrum with a
temperature that decreases with redshift:
\be
\label{15}
T_{obs}= T_{s}/(1+z)
\ee
where $\:T_{obs}$ is the observed temperature, $\:T_{s}$ is the
temperature of the epoch with redshift $\:z$.
The present temperature of the CMBR cannot be deduced
without free parameters:
for $\:z \approx 10^3$
and $\:T_{s} \approx 3 \cdot 10^{3} K$
we get the CMBR with the temperature $\:T_{obs} \approx 3K$.
Thus it is pre-stellar radiation which has been cooled by
the expansion of space.

According to the second approach  the CMBR is the
result of an integration of the contributions
from an appropriate set of celestial objects.
This opposite point of view on CMBR was discussed in
alternative cosmological models. The most interesting one is the
post-stellar thermalized radiation (see e.g. Hoyle, 1991). If the observed
abundance of $\:^4He$, that is $\: \approx 7.5 \cdot
10^{-32} gram \cdot  cm^{-3}$, has been synthesized from
hydrogen in stellar interiors, with an energy release of
$\: 6 \cdot 10^{18} ergs$
per gram of  $\:^4He$ produced, then it gives a cosmic
electromagnetic energy density of
$\:4.5 \cdot 10^{-13} erg \cdot  cm^{-3}$. If
this electromagnetic energy has been thermalized,
the resulting
temperature $\:T$ would be about $\:3 K$.
It is obvious that the alternative redshift mechanisms could lead to
cooling process of any radiation in the Universe. Eq.\ref{15}
 is valid for Doppler, gravitational and expansion redshifts. In
addition, there must be some
thermalising processes in the Universe to
convert radiation to its thermodynamic state.
The mechanism of thermalization is the harder
problem in this scenario. In fact dust emission differs
substantially from
that of a pure blackbody and one is forced to introduce
an ad hoc mechanism of thermalization. The simplest way, that is the
presence of a huge dust density to make the Universe opaque,
is forbidden by the observed transparency up to $\:z \sim 4 - 5$.
Hence the blackbody spectrum can be recovered only
if an appropriate mechanism
of thermalization is assumed. Very long grains  must be
hypothesized in order to get the needed emissivity
but their existence in the extragalactic environment is doubtful.
For a discussion of the experimental
data on the CMBR see {\it section 4.2.4}

\subsubsection {Evolution and the arrow of time}

The best known basis for an evolutionary approach to cosmology
is the second law of thermodynamics. Irreversible events determine
the notion of the thermodynamic arrow of time and the
thermodynamic evolution of  matter in the Universe. Another
type of irreversible event has to be found in the generation of
electromagnetic and gravitational radiation. These radiations
 carry
energy from accelerated sources outwards. The time-reversed
version of these events, i.e. the spherically symmetric convergence to a
common point, has never been observed. The absence of incoming radiation
determines the electromagnetic and gravitational arrow of time in
the sense of the absorber theory of Wheeler and Feynman.
A discussion of the "correct response" of the Universe has
been done by Narlikar (1977) in connection with the definition of the
three arrows of time: thermodynamic, electromagnetic and space
cosmological expansion.
He argued the expansion of the Universe as a whole to prevent
thermodynamic equilibrium being reached.

The problem of evolution is also connected with the origin of
the chemical
elements and with possible variation with time of fundamental
physical constants. The origin of the different types of
astrophysical
objects, such as stars, standard galaxies, extremely active galaxies
and quasars, clusters of galaxies, large-scale structures of the
Universe, is a matter of discussion for any cosmological model, and one
needs to construct evolution scenarios for the life of these objects.

Note also that thermodynamic evolution
of the self-gravitating open systems
could be a cause of the origin of an order
from chaos. This subject has been very
poorly developed up to now, while in the real
Universe this is the main type of astrophysical
object.

\subsubsection {Main rival cosmological theories}

In Fig.\ref{fg1} a
classification of cosmological theories is shown according
to their  answers to the above mentioned questions.

The Big Bang model, based on the geometrical gravitation theory
(General Relativity), has thermodynamic and cosmological evolutions
due to the expansion of space.
The matter distribution is homogenous,
the redshift is due to space
expansion and the CMBR is the relict of a hot beginning.

The Steady-State model uses geometrical
scale invariant gravitation theory
 or, in some variant, direct interaction
theory in curved space-time. The model has three self-consistent
arrows of time: thermodynamic, electromagnetic and cosmological.
The Universe is globally in a steady
state, due to the creation of space with matter.
The matter distribution is
homogenous and
the redshift is caused by the expansion of space.
The CMBR can be a relic of newly
created particles or
a post-stars thermalized radiation.
\smallskip

Fractal cosmology
is now at the beginning of its developments.
There are some attempts
based
on the General Relativity framework and
many possibilities are still under consideration:
it is possible to include
this cosmology  in the Big Bang theory even if one
has to consider that the homogeneity of the
luminous matter
distribution is not reached up to some hundreds of
Mpc. For example, if  the dark matter
turns out to be predominant
and distributed
homogeneously  at smaller scales, there
may be basically no problems with the Friedmann metric,
the Big-Bang model, etc., but in this case it should be
difficult to explain the forming of very large-scale
structures with power law correlations within the Hubble time.
In the opposite
case, the problem becomes very hard and it has to
be reconsidered from the very beginning.
Some preliminary ideas will be presented
in {\it section 3.4}. Even if no definite
theory is now available, it is interesting to
discuss a new approach based on tensor field
relativistic gravitation in Minkowski space-time.
In any case some fundamental aspects
of the current theories of galaxy formation,
such as the biased galaxy formation (Kaiser, 1984)
and related theories, have fundamental problems
with the new picture that now emerges from the data.
The main point that we stress in {\it section 3.4}
is that one cannot discuss the
properties of a
self-similar distribution in terms of amplitudes of correlation.
The only meaningful physical quantity is the
exponent that characterizes the power law behavior,
 while the amplitude is an essential part not only of the
data analysis, but also of the theoretical models.

The tired light cosmology is based on an alternative redshift
phenomenon, due to the still unknown physical process of photon energy
depletion in space. The matter distribution is
homogenous, the CMBR is post-stellar. Space expansion is excluded, but
local evolution is possible. Unfortunately in the tired light
cosmology, the gravitation theory is not, generally, discussed.

\subsection {Classification of cosmological tests}

As we have emphasized above,
the main aim of observational cosmology is to
compare predictions of
cosmological models with available  observational
data. There are two kinds of cosmological tests according to the two
parts of cosmological theories, namely, the initial hypotheses and
predicted relations between observable quantities (e.g. redshift $\:z$,
flux density $\:F_{\nu}$ or integral flux $\:F$, magnitudes $\:m_{i}$ in
filter band {\it i}, angular size $\:\Theta$, surface brightness $\:J_{i}$,
galaxy number counts $\:N_{i}$).
In this connection we introduce the nomenclature of crucial and parametric
cosmological tests. Crucial tests allow us to judge the validity
of the basic assumptions of the theories, and parametric ones give
experimental estimations of the model parameters .
Well known classical cosmological tests, such as angular size-redshift
$\:\Theta (z)$, visual magnitudes-redshift $\:m(z)$ and
number counts of galaxies $\:N(m)$, are parametric ones.

According to our classification of cosmological models (Fig.\ref{fg1}),
we could
consider as crucial the following tests: experimental testing of the
relativistic gravity,
observations of matter distribution, testing of the nature of the
redshift,
testing of the nature of the CMBR,
direct measurements of an evolution effect.
 Note that even parametric tests,
considered as a unified system, may have properties of crucial tests if
there are no parameters of the models which satisfy this system of tests.
Of course crucial tests are the most interesting experiments in
observational cosmology, but obviously, they are limited by
observational technical limits and are not easy to perform.
Among these tests the distribution of luminous matter
in the three dimensional space has had a strong
impulse in this last decade and it will be
rapidly developed in the following years. From these
data one should have the opportunity to test,
at very large scale, the basic assumption
of  homogeneity of the matter distribution.
As we discuss in {\it section 4.2.2.} these
observations can have important consequences also
for the nature of the redshift.
 In modern cosmology there are a lot of parametric tests which deal
with different kinds of astrophysical objects in a wide range of
the electromagnetic spectrum from radio to gamma rays. In this paper
we shall consider only classical tests, because they have furnished the
most reliable data (see {\it section 4}).

\subsection {Biases and selection effects in astronomical data}

For testing a cosmological model, we need to choose astrophysical
objects and fix the values of their parameters, which will be used as
standards in the test. The main difficulties in this approach are the
possible evolution effects of the objects, i.e. the time variation of
parameters, and different kinds of biases and selection effects
which are present in all astronomical data.
Astrophysical biases or selection effects can be divided into two
groups: physical effects and technical  effects. In their turn, physical
effects can be intrinsic and intervening.

The technical selection effects
are caused by the technical limits of astronomical devices: telescopes,
light receivers, processing electronics. The main ones are: finite
angular, spectral and time resolution, finite flux sensitivity, fixed
aperture and spectral band, and receiver's noise.
A very important selection effect
is the so-called Malmquist bias, that  arises in flux limited
samples and states that the average absolute luminosity of the
nearby members of the sample is fainter than that of
more distant members (see e.g. Teerikorpi, 1987; 1993; Sandage, 1988).

The next selection effect is the K-effect, which results from
the cosmological redshift of the spectrum of distant objects.
The K-correction is the magnitude difference between
a redshifted and non-redshifted spectral energy
distribution, when observed through a fixed spectral
interval.
The selection is due to the combination of
the wavelength shift and the fixed detector
effective wavelength.
The K-correction for galaxies of different morphological
types is necessary to interpret
magnitude-redshift relation
and the luminosity function of galaxies
(see e.g. Pence, 1976; Sandage, 1988; Yoshii \& Takahara 1988;
 Guideroni \& Rocca-Volmerange, 1990).

The background radiation, including extragalactic, galactic, solar
system space and earth atmospheric radiation (for ground
observations) is a physical intervening selection effect too.  Really physical
and technical effects work together and can produce very "strange"
behavior of directly observed quantities.

An important bias in the detection of galaxies, is
related to their surface brightness. Very distant galaxies,
that have compact images, will appear just like a star on  the
photographic plate, while galaxies of low surface brightness
(LSB) look small because one cannot distinguish them from the
noise due to the non zero sky brightness. This effect
limits strongly the ability to recognize galaxies of
different surface brightness (Disney \& Phillips, 1987).
Surveys of LSB galaxies
 (Binggeli et al, 1990;
Eder et al., 1989;  Thuan, 1987) have shown that dwarf and LSB
 galaxies fall into the structures delineated by
the luminous ones and that there
are not evidences that these galaxies fill voids.
However this effect can be relevant in the determination
of  number of galaxies for unit magnitude (number-count relation);
 in fact McGaugh (1994)
stressed that, due to this selection bias, one underestimates the
number of LSB
 in local surveys (see {\it section 4.1.4.})
while it does not happen in deeper samples.

Different evolution effects could
be treated as an intrinsic physical selection effects. In this case we
must add supplement terms in the theoretical  predictions to take into
account evolution (see e.g., Yoshii \& Takahara, 1988; Guideroni
\& Rocca-Volmerange, 1990; Yoshii, 1993).
The relativistic beaming effect is another example of intrinsic
physical selection effect. It must be taken into account for active
galaxies which contain plasma jets with relativistic bulk motion. The
effect has a strong influence on measured fluxes, angular sizes and
number counting of objects (see e.g. Padovani \& Urry, 1992; Baryshev
\& Teerikorpi, 1995).
The gravitational lensing is an intervening physical selection effect
which easily shifts observable relations for fluxes and angular sizes.
The detailed description of the gravitational lensing and its modern
applications may be found in the review  articles of Blanford \&
Narayan (1992), Refsdal \& Surdej (1994),
and in the monograph of Schneider et al.(1992)
(see {\it section 4.1.5.3.}).

These examples demonstrate that the check of cosmological theories
with observations, and the rejection of some of them, is not an easy task. It
needs great caution, because one must take into account all the essential
selection effects.
The evolution, for example, has been sometimes invoked
to explain the difference between theoretical
predictions and observational data (e.g. in the case of number
counts), but there is not a self consistent model of evolution that takes
into account  different experimental facts (Yoshii, 1993).
Moreover, if one defines evolution as the difference between
observable and theoretical quantities, one must be sure
that the theoretical model has been widely verified.
That is why different cosmological
alternatives should be closely studied to derive more definite
predictions.

\section{THEORETICAL PREDICTIONS OF COSMOLOGICAL MODELS}

The development of cosmology is basically determined by
developments of the  observational tests of world models.
Unfortunately, in spite of a great growth of observational
information, there is no agreement between
the different cosmological tests and one has
to add to the models new free parameters.
Furthermore, the last years have
generated more critical attitudes towards
 the standard cosmological model.
{}From many recent redshift surveys
({\em section 3.4}) it follows that the length
$\:\lambda_{0}$, above which the distribution
should be smooth and essentially structureless,
is not identified and that it
is much greater  than the previously considered size
of about $\:10 Mpc$.
In particular the most important
assumption that has to be reconsidered
is the homogeneity of matter distribution
at least up to a scale 10 times larger
than the previously believed scale. In many cases
this situation gives rise to
huge problems for the standard cosmology.

An important problem of the Standard Model is the
 perfect linearity of the redshift-distance
 relation, that is, in the traditional
interpretation, a consequence of
homogeneity,
deeply inside the inhomogeneity cell,
i.e. for distances smaller  than $\:\lambda_{0}$
(Baryshev, 1994a).
The fact that there is no generally
accepted clear observational determination
of the Universe geometry within the framework
of the standard Big Bang theory, while there
 are tremendous collections of observational
data, suggests that we have to consider more
carefully  alternative cosmological theories.
We shall consider
in the following possible cosmological
theories and their predictions for observations.

\subsection{Paradoxes of Newtonian cosmology
and origin of modern cosmological ideas}

The Newtonian cosmology deals with the Euclidean
space and the Newtonian theory of gravitation (for historical
references see the excellent book by North, 1965).
At the end of the last century, three
paradoxes of the Newtonian cosmology had been
formulated: Neumann-Seeliger gravitational paradox,
 Cheseaux-Olbers photometric
paradox and Boltzmann thermodynamic paradox.

The {\it gravitational paradox} is
the following: as the volume of  a matter
distribution of finite density tends to infinity,
the Newtonian gravitational potential at any point has
no definite value and the gravitational force also becomes indefinite.
To put it in another way, the solution
of the Poisson equation inside the homogenous ball is:
\be
\label{16}
\Phi(r) = -2 \pi G \rho_{0} (R^{2}-\frac{r^{2}}{3})
\ee
where $\:\rho_{0}$ is the mass density, $\:R$
is the radius of the ball, $\:r$ is the radial distance from
the center of the ball.  Hence the gravitational potential
in the center of the ball is:
\be
\label{17}
\Phi(0)  = -2 \pi G \rho_{0} R^{2}_{R \rightarrow \infty}
\rightarrow - \infty
\ee
i.e. it has an infinite value for infinite radius of the ball.
The gradient of the gravitational potential
at the surface of the ball is also infinite, i.e. it leads
to an infinite gravity force. In terms
of modern field theory, it means infinite energy density
of the gravitational field, because
the latter is $(d\Phi / dr)^{2} / 8\pi G$.

The {\it background radiation paradox} is the following:
if an
uniform distribution of  stars covers
the whole sky up to some finite radius $r$,
the night sky should be as bright as the  mean star surface.
If $\:n_{0}$ is the mean number density of stars in space, then
the sky fraction
$\:f(r)$ (dilution factor) covered by stars  is:
\be
\label{18}
f(r) =\frac{1}{4\pi} \int^{r}_{0} \frac{A}{r^{2}}n_{0} 4\pi r^{2} dr
\ee
where $\:A=\pi R_{*}^{2}$ is the surface of the star cross section
and $\:\Omega_{*} =A/r^{2}$ is the solid angle of the star
at distance $\:r$. Therefore the sky will be completely covered ($\:f(r)=1$)
if the stars are distributed up to the radius:
\be
\label{19}
r_{ph} = \frac{1}{An_{0}}
\ee
The total energy density of the background radiation observed
from the Earth is:
\be
\label{20}
\rho_{BR}(r) =\frac{1}{c} \int_{0}^{r}
\frac{L_{*}}{4\pi r^{2}} n_{0} d\vec{r} = \frac{L_{*}n_{0}r}{c}
\ee
where $\:L_{*}$ is the typical luminosity of a
 star. For $\:r=r_{ph}$ (Eq.\ref{19})
$\:\rho_{BR}(r)$ equals the value at the surface of a star, and from
Eq.\ref{20}:
\be
\label{21}
\rho_{BR}(r_{ph}) =\frac{L_{*}} {cA}
\ee

The {\it thermodynamic paradox}
 is the thermal death
of  the Boltzmanian Universe (elastic interacting molecules),  i.e.
the existence of an asymptotic
homogeneous state with small thermodynamics fluctuations.
On the contrary we observe a highly inhomogeneous Universe with
evolution of rather complex systems.

These paradoxes have been of great importance in cosmological researches
because they stimulated very active
search for self consistent models of the Universe
(for historical background see North, 1965; and for
a modern point of view see Harrison, 1981).
In order to resolve a paradox, one has to take into account
the initial postulates of the model used. For example
for the gravitational paradox resolution one may consider
some modification of the Newtonian theory of gravitation.
The first solution of the gravitational paradox
 was found by Neumann and
Seeliger (see North, 1965). They proposed that gravity
falls off at large distances faster than the
inverse-square law. For the gravitational
potential they took the expression of the
usual Newtonian form multiplied by an
additional factor $\: e^{-\alpha r}$, where
$\:\alpha$ is sufficiently small
to be consistent with the usual
Newtonian theory for small distances.
This makes possible an infinite quasi-Newtonian
Universe without
gravitational paradox. In fact, the same
idea was used by Einstein (1917), who
introduced the $\:\Lambda-$term in
 General Relativity.
Instead of the Poisson equation he suggested:
\be
\label{22}
\Delta\phi - \Lambda\phi = 4 \pi G \rho_{0}
\ee
where $\:\lambda$ is the universal constant. The solution of Eq.\ref{22} is:
\be
\label{23}
\phi = - \frac {4\pi G}{\Lambda} \rho_{0}=const
\ee
It corresponds to an infinite distribution
of homogeneous matter which is in
 equilibrium with internal forces.
 Local inhomogeneities in matter
distribution will add a local
potential $\:\phi$ which will
 be like the Newtonian one for
 sufficiently large $\:4\pi G \rho_{0}$
relative to $\:\Lambda\phi$. In {\em section 3.4.7.}
it will be shown that this
seemingly   "ad hoc" hypothesis has a very
clear foundation within the framework of
the tensor field gravitation theory.

The second idea which has been recently used
in cosmological papers again, is the
hierarchical distribution of matter
 (known now as fractal). It was shown by
Fournier D'Albe (1907) and Charlier
 (1908, 1922) that the Newtonian Universe
may be build up without gravitational and
photometric paradoxes if the matter is distributed
according to the
 unlimited clustering hierarchy. In the model the
 typical values of the mass within the distance $\:r$
 of hierarchy scale as:
\be
\label{24}
M(r) \approx r^{1}
\ee
The size and the mass of the Universe are arbitrarily
 large, but the mean density $\:M(r)/r^{3} \approx r^{-2}$
 converges to zero and Olbers's paradox is avoided.
The virial velocity in a cluster of size $\:r$
is $\:v^{2} \approx M/r = const $, independent of $\:r$.
 The modern development of this idea will be discussed
 in {\em section 3.4} and the modern observational data
 which confirm the fractal structure of matter distribution at
 least up to $\:100 Mpc$ will be considered in {\em section 3.4.3.,
3.4.4., 4.2.3}.

The third basic idea in modern cosmology was the
 expanding Universe model. The idea was a consequence of
 the geometrical revolution in gravitation theory,
 which had been made by Einstein's General Relativity
(Einstein, 1916). Non static solutions
of Einstein's equations (Friedmann, 1922) had
shown that the Universe can expand or contract
in the sense of continuous space creation or annihilation.
This idea has been developed by many authors and it
is one of the main initial hypotheses in the most modern
cosmological models.

The next important idea was Milne's discovery
of a purely Newtonian derivation of the Friedmann
equations (Milne, 1934; McCrea \& Milne, 1934).
Milne and Mc Crea showed that the expanding Universe
 equations, which had been previously derived from
 General Relativity, can be obtained directly from
simple Newtonian theory (using flat space, static
 Euclidean space, Newtonian time and Newtonian
dynamics). All one has to do is to consider a ball of
matter with finite radius $\:R$ ("cosmic ball") and
forget about the matter outside the ball. Then the
 radius $\:R$ of the expanding spherical symmetric
 cosmic ball will be governed by the equation:
\be
\label{25}
\frac{d^{2}R}{d t^{2}}= - \frac{4 \pi G}{3} R \rho
\ee
or after integrating
\be
\label{26}
\left(\frac{d R}{d t}\right )^{2}=
 - \frac{8\pi G}{3} R ^{2}\rho - kv_{0}^{2}
\ee
where constant $\:k$ has the value $\: 0, 1$ or $\: -1$,
 which corresponds to parabolic, elliptic or hyperbolic
 motion of the ball's particles, and $\:v_{0}^{2}$ is the
 module of the integration constant. This discovery
has created a problem in cosmology: How can it be that the
non-relativistic Newtonian theory yields
exact general relativistic results ?
Where are such restrictions as the limit on
the velocity of light and general retarded response effects ?

A very important idea is the suggestion made by
Gamow, Ivakenko and Landau (1928) and by Bronstein (1934) about
the relativistic-quantum-gravity,
({\it Ghc}), character of the future cosmological theory.
In this approach, gravitation theory must be a quantum
relativistic field theory, because it is based on the existence
of the quanta of the field, i.e. gravitons, which carry
energy-momentum in space and may transform into other
 elementary particles (Ivanenko \& Sokolov, 1947).
Modern attempts to join quantum field
particle physics and General Relativity
(hereafter GR) (such
as inflation and quantum cosmology) continue these ideas.

Another cosmological idea that needs to be mentioned is the
absorber theory of Wheeler and Feynman (1949).
In this theory  local and  global Universe
could be described as a single system.
The theory is free from an action of an elementary charge
upon itself and provides an  experimentally satisfactory
account of the behaviour of a system of  point
charges in electromagnetic interaction with one another.
Cosmological implications of the theory have been
considered by Hoyle \& Narlikar (1974).

Now we shall consider how the ideas
mentioned above work in
different cosmological models.

\subsection{Big Bang models}

 The Hot Big Bang (HBB) scenario is the currently
most accepted model so that one refers to it as the
"Standard Model". A complete introduction to this
model can be found in the books of Weinberg (1972) and
of  Peebles (1980; 1993), while discussions on
its observational tests and its state of art
are found in the reviews of Sandage (1988),
Peebles et al. (1991; 1994),
Coles \& Ellis (1994).

\subsubsection{Initial Hypotheses of the standard Friedmann model}

The {\it  first hypothesis}
of the Standard Friedmann model (hereafter  SM )
is that GR is a correct
relativistic gravitation theory,
and that it can be applied to the Universe as a whole.
According to this hypothesis
gravity is described by a metric tensor $\:g^{ik}$
of a Riemannian space. The "field" equations of GR
(Einstein-Hilbert equations) have the form:
\be
\label{27}
R^{ik}-\frac{1}{2} R g^{ik} = \frac{8 \pi G}{c^{4}} T^{ik}_{(m)}
\ee
where $\:R^{ik}$ is the Ricci tensor,
 $\:T^{ik}_{(m)}$ is the energy-momentum tensor
(hereafter EMT) for matter only.
Solutions of the Eq.\ref{27} for
unbounded matter distribution are cosmological
ones and are the basis of all cosmological interpretations.

The {\it  second hypothesis} is the homogeneity
of the matter distribution in space, i.e.:
\be
\label{28}
\rho(\vec{r},t) = \rho(t)
\ee
\be
\label{29}
p(\vec{r},t)=p(t)
\ee
Peebles (1980) discussed in detail if the
homogeneity
of the Universe has to be expected from general physical arguments.
He emphasized that
within the SM we cannot account for the
homogeneity and we must accept it as an assumption
(see also Harrison, 1981).
This means that homogeneity must be accepted as a
phenomenon to be explained by some future deeper theory.
On the other hand, modern observations
of the large-scale structure of the Universe
point to the existence of very
large fractal like inhomogeneities. Hence
the homogeneity assumption can be tested experimentally
and it
can be postulated only for
sufficiently large scales (see {\it section 4.2.3.} and
{\it section 3.4}).

{}From the assumption of homogeneity in
Friedmann cosmology it follows also the isotropy
of the matter distribution.
The requirement for Isotropy and Homogeneity is
implemented by the Cosmological Principle in order to avoid
a privileged observer. We show in {\it section 3.4.1.}
that this last condition is ensured only by local isotropy
and the requirement of homogeneity is a very strong one
and it is not necessary in  order to avoid any
privileged observer in the Universe.

As a consequence of homogeneity and isotropy one can gets the
Robertson-Walker line element in the form:
\be
\label{30}
ds^{2} = c^{2}dt^{2} - R(t)^{2} [dr^{2}+ I(r)^{2}
(d\theta^{2}+sin^{2}\theta d\phi^{2})]
\ee
where $\:r,\theta,\phi$ are the
comoving space coordinates, $\:I(r) = sin(r),r,sinh(r)$
corresponding to curvature constant
values $\:k=+1,0,-1$ respectively and $\:R(t)$ is the scale factor.
The {\em  expanding space paradigm} is that the
{\em metric} (proper) {\it distance} of a comoving body of fixed
coordinate distance $\:r$ from a comoving observer is:
\be
\label{31}
l = R(t) \cdot r
\ee
and increases with time $\:t$ as the scale factor $\:R(t)$.
Note that physical dimension of proper distance
$\:[l]=cm$, hence if $\:[R]=cm$ then $\:r$ is
a dimensionless coordinate distance.

The recession "velocity", or the space expansion velocity,
of a comoving
(i.e., with $\:r$ constant) body is:
\be
\label{32}
V = \frac{dl}{dt} =  \frac{dR}{dt} r = \frac{dR}{dt} \frac{l}{R} =  H(t) l =
c \frac{l}{R_{H}}
\ee
where $\:H(t)= (1/R) dR/dt$ is the rate of the
expansion (Hubble constant)
and $\:R_{H} = c/H(t)$ is the Hubble distance at the time $\:t$.
The "velocity" $\:V$ - distance $\:l$ relation (Eq.\ref{32}) is
linear for all distances $\:l$: therefore for $\:l>R_{H}$ we get $\:v>c$
(Harrison, 1993). In connection with this the quantity $\:V$
would be better called the {\em "space expansion velocity"},
because it is not the usual physical velocity of a body in space.
The expansion of space induces
the wave stretching of the travelling photons via the Lemaitre's
equation (Eq.\ref{11}), i.e.:
\be
\label{33}
 (1+z) = \frac{\lambda_{0}}{\lambda_{1}} = \frac{R_{0}}{R_{1}}
\ee
where $\:\lambda_{1}$
and  $\:\lambda_{0}$ are the wavelengths at the
emission and reception respectively,
and $\:R_{1}$ and $\:R_{0}$ the corresponding values of the scale factor.
Equation \ref{33} may be obtained from
the radial null geodesic ($\:ds=0$, $\:d\theta=0$,
$\:d\phi=0$) of the Robertson-Walker line element Eq.\ref{30}.

The behavior of the scale factor
 with time $\:R(t)$ is governed by the
Einstein-Hilbert equations
(Eq.\ref{27}) which, in the case of homogeneity,
gives
the Friedmann equation:
\be
\label{35}
\frac{d^{2}R}{dt^{2}}=
- \frac{4 \pi G}{3} R \left( \rho + \frac{3p}{c^{2}} \right)
\ee
or after integrating
\be
\label{36}
\left(\frac{d R}{d t}\right )^{2}=
 - \frac{8\pi G}{3} R ^{2}\rho - kc^{2}
\ee
where $\:k=0,+1,-1$ for flat, closed and open space geometry.
Solving this equation we find the dependence of the scale factor
from time, i.e. $\:R(t)$.
The model has two parameters that are  the Hubble
parameter:
\be
\label{37}
H = \frac{dR}{d t} \frac{1}{R}
\ee
and the deceleration parameter:
\be
\label{38}
q = - R \frac{d^{2}R}{dt^{2}}
 \left( \frac{dR}{dt} \right)^{-2}
\ee
which, for the present time $\:t_{0}$, are $\:H(t_{0})=H_{0}$ and
$\:q(t_{0})=q_{0}$
respectively. In this theory we have also the density parameter:
\be
\label{39}
\Omega = \frac{\rho}{\rho_{c}}
\ee
\be
\label{40}
\rho_{c} = \frac{3 H^{2}}{8\pi G}
\ee
and the space curvature parameter:
\be
\label{41}
K = \frac{kc^{2}}{(HR)^{2}}
\ee
These parameters satisfy the equations:
\be
\label{42}
\Omega =  K+1
\ee
and
\be
\label{43}
q = \frac{1}{2} \Omega \left(1+\frac{3p}{\rho c^{2}}\right)
\ee
Thus for
$\:p \ll \rho c^{2}$ the SM is fixed by two parameters
$\:\Omega$ and $\:H$ or $\:q$ and $\:H$. Eq.\ref{43}
means that dynamics and geometry of
the model are uniquely defined.
{}From Eq.\ref{42} it follows that
\be
\label{44}
H(z) = H_{0} (1+z)(1+z\Omega_{0})^{1/2}
\ee
and it may be shown that Eq.\ref{32} can be written in the
form (Harrison, 1993):
\be
\label{45}
V = c(\Omega_{0}-1)^{-1/2} sin^{-1}
\left( \frac{2(\Omega_{0}-1)^{1/2}}{\Omega_{0}^{2}(1+z)}
\left(z\Omega_{0} + (\Omega_{0}-2)(1+z\Omega_{0})^{1/2}-1\right)\right)
\ee
In the limit of $\:z \ll 1$ this yields:
\be
\label{46}
V \approx cz
\ee

The {\it third hypothesis} of the SM is that
the observed redshift in Eq.\ref{46} is due to the space
expansion, i.e. given by Eq.\ref{33}, and that the real
distance equals the proper distance in Eq.\ref{31}.
Hence for $\:z \ll 1$ and $\:l \ll R_{H}$
from Eq.\ref{46} and Eq.\ref{32} we get
the relation:
\be
\label{47}
z \approx \frac{V}{c} = \frac{H_{0}}{c} l = \frac{l}{R_{H}}
\ee
which may be interpreted as the observable Hubble relation.
We stress again that the quantity $\:V$ is not the velocity of
a receding galaxy in the usual sense, but $\:V$ is the velocity of the
space expansion ("creation of space")
 and such a phenomenon has never been tested in laboratory physics.

In the SM it is suggested
({\it the fourth hypothesis})
that the Universe has expanded
from a very hot state, and due
 to the creation of space the matter cools as in
usual thermodynamics.
Note that as in the case of
the expansion velocity, in this case we have no "usual"
thermodynamics, because the equation $\:dU = - pdV$ is
laboratory tested only for changes of the gas volume $\:dV$
in the static Euclidean space without space creation.

\subsubsection{Successes of the standard model}

There are several great successes  of the application
of the SM to the real
observed Universe. Review of the
evidences in favor of the relativistic
HBB  models can be found,
for example in Weinberg, 1972; Sandage, 1987, 1988;
 Peebles,
 1980, 1993; Peebles et al. 1991, 1994. There are no
gravitational, photometric and thermodynamic
 paradoxes in the  SM, because the age
 of the Universe is finite and rather small
 in comparison  with any reasonable time
 scale.
In the  SM, space has been filled
 with blackbody radiation, the cosmic microwave
background
radiation
(CMBR).
As the Universe expands, the
 number of CMBR photons per unit volume drops as
\be
\label{48}
n_{0}=\frac {n(z)}{{(1+z)}^3}
\ee
when $\:n_{0}$ is the present value of the number density.
The photon wavelength is stretched by
 the expansion as in Eq.\ref{33}. Thus the
CMBR preserves a black body spectrum with
 a temperature that decreases as the Universe expands
\be
\label{49}
T_{0}=\frac {T(z)}{(1+z)}
\ee
where $\:T_{0}$ is the present temperature
of the CMBR and its observed value is about $\:3 K$.
 Eq.\ref{48} and Eq.\ref{49}  can be used
 back to $\:z\sim10^{10}$. The observed
thermal spectrum of the CMBR is the greatest
success of the SM predictions, even if its fundamental
quantity, $\:T_{0}$, cannot be deduced
from any calculations of the early Universe.
We stress that the SM determines only one of the
three characteristics of the CMBR: the blackbody spectrum. In fact,
as we shall see in {\it section.3.2.5}, the extreme isotropy
of CMBR is a paradox for the  SM .

In the  SM, the Universe was hot and dense enough
to drive thermonuclear reactions that changed the
 chemical composition of the matter. The values
of the abundances left over from this hot epoch
depend on the cosmological parameters.
Knowing the present temperature and
 assuming a value for the present matter
 density, we have fixed the thermal history
of the Universe.
If the matter is uniformly
 distributed and lepton members are comparable
to the baryon number, this is sufficient to fix the
 final abundances of the light elements.
The observed light element abundances
of $\:^4He, ^2H, ^3He$ and $\:^7Li$ are
 in good agreement with SM predictions,
provided that the baryonic density is
in a defined and narrow range.
In particular the abundance of $\:He$ and $\:D$
are explained with high accuracy, while there are
more uncertainties in the abundances of $\:Li, Be$ and $\:B$
(Boesgaard \& Steigman, 1985).

\subsubsection {Crucial Tests}

According to our definition
(see {\it section 2.3}), crucial tests
concern
the validity of the different cosmological
 theories. Therefore crucial tests for
 SM will be the following:
\begin{itemize}
\item experimental testing of General Relativity
\item determination of the matter distribution in space
\item testing of the reality of space expansion
\item measuring the temperature of the CMBR at high redshift
\item determination of the ages of the oldest objects
\item measuring the evolution
of the chemical composition of matter at high redshift
\end{itemize}

We shall consider the tests in {\it section 4.2}.

\subsubsection {Parametric tests}

The most reliable observational data have
been collected for classical cosmological tests.
Here we discuss the predicted relations among
 observable quantities in the framework of the
standard Friedmann model (for more details see
e.g. Weinberg, 1972; Sandage, 1987, 1988;
Yoshii \& Takahara, 1988).
In {\it section 4.1.} we consider the comparison
of these predictions with observational data.
 \bigskip

{\em  3.2.4.1 The proper distance-redshift relation}
 \smallskip

The basic relation for the calculation of different
 observable quantities is the connection between
metric (proper) $\:l$, angular $\:l_{a}$ and bolometric
$\:l_{bol}$ distances in the expanding Universe:
\be
\label{50}
l=l_{a}(1+z) =\frac{l_{bol}}{(1+z)}
\ee
For metric distance (Eq.\ref{31}) we have

\be
\label{51}
{\cal L}(z,q_{0})
\equiv \frac{l}{R_{H_{0}}}
=\frac{zq_{0}+(q_{0}-1)
((2q_{0}z+1)^{1/2}-1}{q_{0}^{2}(1+z)}
\ee

where $\:R_{H_{0}}=c/H_{0}$.
Note that for the case $q_{0}=1/2$ and $z<<1$
from Eq.\ref{32} we get:
\be
\label{a1}
z_{cos} = \frac{v}{c} + \frac{1}{2}\left(\frac{v}{c}\right)
^{2} + \frac{1}{2}\frac{GM(l)}{c^{2}l}=z_{D} +z_{grav}
\ee
where we used the relation $x\sim z\sim v/c$ and the
energy equation $v^{2} = 2G M(l)/l$.
So the cosmological redshift can be considered
as the sum of the Doppler and the gravitational parts,
al least, in the small $\:z$ approximation.
 \bigskip

{\em  3.2.4.2 The angular size-redshift relation}
\smallskip

{}From Eq.\ref{50} and Eq.\ref{51} we get
theoretical angular size-redshift relation in the form:
\be
\label{56}
\Theta (z, q_{0}) = \left( \frac{d}{R_{H_{0}}}
\right) \left( \frac{{\cal L}(z, q_{0})}{(1+z)}\right)^{-1}
\ee
where $\:\Theta$ is the angular size  of the body
with linear size $\:d$, and the $\:{\cal L}$
 function is given by Eq.\ref{51}.
 \bigskip

{\em  3.2.4.3. The magnitude-redshift relation}
\smallskip

Let $\:L(\nu_{emit})d\nu_{emit}$ be
the spectral luminosity between $\:\nu_{emit}$ and $\:\nu_{emit}+d\nu_{emit}$
emitted from a source at redshift $\:z$. Then the observed
 luminosity will be:
\be
\label{59}
L(\nu_{obs})d\nu_{obs} =
 \frac{L(\nu_{emit})d\nu_{emit}}
{(1+z)^{4}}= \frac{L(\nu_{obs}(1+z))}{(1+z)^3}d\nu_{obs}
\ee
The apparent bolometric flux received
at Earth from a source with redshift $\:z$,
with its absolute bolometric luminosity
\be
\label{60}
L_{bol} = \int L(\nu)d\nu
\ee
is given by
\be
\label{61}
S_{bol} = \frac{L_{bol}}{4\pi l_{bol}^{2}} = \frac {L_{bol}}
{4 \pi R_{H_{0}}^{2} {\cal L}^{2} (z, q_{0}) (1+z)^{2}}
\ee
Converting Eq.\ref{61} into magnitude one obtains the
theoretical magnitude-redshift $\:m(z)$ relation
\be
\label{62}
m_{bol}=5 \log(R_{H_{0}}{\cal L}(z,q_{0})(1+z))+25+M_{bol}
\ee
where $\:{\cal L}$ is given by Eq.\ref{51}, $\:R_{H_{0}}$ is
 measured in Mpc, $\:M_{bol}$ is the
 bolometric absolute magnitude of the source.
 For $\:z \ll 1$ we get
\be
\label{66}
m_{bol}\approx 5 \log z + 1.086 (1-q_{0}) z + const.
\ee
The first term in the right side of
Eq.\ref{66} is the usual Hubble  $\:m(z)$-diagram
and the second is the cosmological correction.
In the experimental astrophysics, a fixed
detector effective wavelength {\em "i"} and a finite
bandwidth $\:\zeta_{i}$ are used. Then the flux in the filter {\it i}
 is given by
\be
\label{67}
S_{i}=\int S_{\nu}d\nu=\frac{L_{i}}{4\pi l_{bol}^{2}}
\left( \frac{\tilde{ L_{i}}}{L_{i}} \right)
\ee
where
\be
\label{68}
L_{i}=\int L(\nu)\zeta_{i}(\nu) d\nu
\ee
\be
\label{69}
\tilde{L_{i}}=(1+z)\int L(\nu (1+z)) \zeta_{i}(\nu) d\nu
\ee
or in magnitudes
\be
\label{70}
m_{i}=5 \log l_{bol} + 25 + M _{i} + K_{i}(z)
\ee
with
\be
\label{71}
K_{i}(z)=-2.5 \log \left( \frac{\tilde{L_{i}}}{L_{i}} \right)
\ee
$\:K_{i}(z)$ is the $\:K$-correction.
In evaluating the correction at
fairly large redshifts the spectral energy distribution
$\:L( \nu )$ and the sensitivity
 function $\:\zeta_{i}$ are necessary.
In the case of spectral observations the observed
flux density of an object
with spectral luminosity $\:L(\nu)$ is given by:
\be
\label{72}
S_{\nu_{obs}}=\frac{L (\nu_{obs}(1+z))}{(1+z)^{3}4\pi l_{a}^{2}}
\ee
For power law luminosity $\:L_{\nu_{emit}}
\sim \nu_{emit}^{-\alpha}$  we have
\be
\label{73}
S_{\nu_{obs}}=\frac{L  (\nu_{obs})} {(1+z)^{1 + \alpha} 4 \pi  l^{2}}=
\frac{L ( \nu_{obs})} {4\pi l_{sp}^{2}}
\ee
Hence in that case the spectral distance $\:l_{sp}$ is
\be
\label{74}
l_{sp}= R_{H_{0}}
{\cal L}(z,q_{0}) (1+z)^{\frac{1 +\alpha}{2}}
\ee
The measured surface brightness of the source is the received
flux divided by the apparent area obtained by combining
Eq.\ref{56} and Eq.\ref{61}, which give the bolometric one:
\be
\label{75}
J_{bol}(z)=\frac{S_{bol}}{\Theta^{2}}=\frac{const}{(1+z)^{4}}
\ee
{}From Eq.\ref{61} and Eq.\ref{67} the filtered surface brightness is
\be
\label{76}
J_{i}(z)=\frac{S_{i}}{\Theta^{2}}=\frac{const}
{(1+z)^{4}}\cdot 10^{-0.4K_{i}(z)}
\ee
where $\:K_{i}(z)$ is given by  Eq.\ref{71}.
\bigskip

{\em  3.2.4.4. Time-redshift relation}
\smallskip

In the framework of the SM the age
of the Universe at redshift $\:z$  is given by
\be
\label{77}
t(z)= \frac{1}{H_{0}}\int_{0}^{\frac{1}{1+z}}
\frac {dy} {(1+2q_{0}+2q_{0}/y)^{1/2}}
\ee
or in differential form
\be
\label{78}
dt= - \frac {dz} {H_{0} (1+z)^{2} (1 + 2 q_{0} z)^{1/2} }
\ee
The present age of the Universe $\:t_{0}$ for $\:q_{0}=0.5$
is $\:t_{0}=\frac{2}{3}H_{0}^{-1}$,  and for $\:q_{0}=0$
$\:t_{0}=H_{0}^{-1}$.
The age of an object of redshift $\:z$ which was
formed at $\:z_{f}$ is given by
\be
\label{81}
t_{obj}(z)=t(z)-t(z_{f})
\ee
\bigskip

{\em  3.2.4.5. Count-redshift and count-magnitude relations}
\smallskip

Let $\:n_{0}$ be the number density of objects in space,
then the differential number counts of the objects
with the redshift between $\:z$ and $\:z+dz$
will be
\be
\label{82}
dN(z,q_{0})= 4 \pi  n_{0} \frac{dV}{dz} dz
\ee
where $\: dV/dz $ is the differential comoving
volume element of the standard model, which is
given by:
\be
\label{83}
\frac{dV(z,q_{0})}{dz}=\frac{4\pi c^{3}
{\cal L} ^{2}(z, q_{0})}{H_{0}^{3}(1+z)(1+2q_{0}z)^{1/2}}
\ee
Eq.\ref{83} takes into account expansion and non-
Euclidean geometry of the space.
For example, for the case of $\:q_{0}=0.5$
we can get integral count-redshift relation $\:N(z,q_{0})$
integrating Eq.\ref{82} over $\:z \in [0,z]$ :
\be
\label{84}
N(z,0.5)=\frac{32 \pi c^{3}}{3H_{0}^{3}}n_{0}
\left( \frac {(1+z)^{1/2}-1} {(1+z)^{1/2}} \right)^{3}
\ee
The $\:N(z,q_{0})$ relation
can be transformed to  $\:N(m,q_{0})$
using  Eq.\ref{70}. Let $\:d^{2}A(m,z,q_{0})$
be the
number of objects with redshift $\:[z,z+dz]$
contributing to the counts per steradians
and magnitude bin around apparent magnitude $\:m_{i}$, i.e.:
\be
\label{85}
d^{2}A(m_{i},z,q_{0}) = \Phi(M_{i}) \frac{dV}{dz} dm_{i} dz
\ee
where $\:\Phi(M_{i})$ is the luminosity function per volume unit,
$\:M_{i}$ is the absolute magnitude through the {\em "i"}
filter and is given by Eq.\ref{70}:
\be
\label{86}
M_{i}=m_{i} - 5 \log \left( {\cal L}(z,q_{0})(1+z)
R_{H_{0}}(Mpc) \right)-25-K_{i}(z)
\ee
The differential number count per steradian of objects
in the {\em "i"} band and magnitude
bin $\:dm_{i}$ is obtained by integrating
Eq.\ref{85} on $\:z$:
\be
\label{87}
N_{d}(m_{i},q_{0}) = \int^{z_{max}}_{0} d^{2}A(m_{i},z,q_{0})
\ee
where $\:z_{max}$ is the maximum redshift for the objects.
The number of points per steradian and redshift
bin $\:dz$ can be computed by integrating Eq.\ref{85}
on $\:m_{i} \in [m_{1},m_{2}]$:
\be
\label{88}
N_{d}(z,q_{0}) = \int^{m_{2}}_{m_{1}} d^{2}A(m_{i},z,q_{0})
\ee
\bigskip

{\em 3.2.4.6. Cosmic Background Radiation}
\smallskip

The effective brightness of the diffuse background
radiation expected from a population of sources
in a volume extending from $\:z=0$ to $\:z=z_{max}$ depends on
the luminosity function $\:\Phi(L_{\nu})$ and the spectral energy
distribution of the source $\:L_{\nu}$:
\be
\label{89}
J_{\nu} = c^{3}H_{0}^{-3} \int^{\infty}_{0}   dL_{\nu}\int_{0}^{z_{max}}
S_{\nu}(z') L^{2}(z',q_{0}) (1+z')^{-4} (1+ 2q_{0}z')^{-1/2} \Phi(z',L_{\nu})
dz'
\ee
where $\:S_{\nu}(z)$ is the flux density of an object at frequency $\:\nu$
and is given by Eq.\ref{72} and
\be
\label{90}
\Phi(z,L_{\nu}) = \Phi(0,L_{\nu}) ( 1 + z)^{3}
 \ee
is the volume density of objects.
The dimension of $\:I_{\nu}$ is $\:Jy\cdot sr^{-1} Hz^{-1}$.
The energy density of the
background radiation can be expressed in the form:
\be
\label{91}
\rho_{\nu} =\frac{4 \pi}{c} I_{\nu}
= \frac{L_{0\nu}n_{0}}{H_{0}}b(z_{max},q_{0})
\ee
where $\: L_{0\nu} n_{0}/H_{0}=\rho_{\nu}$
is the characteristic radiation energy density
of the fixed type of objects
with mean number density $\:n_{0}$ and luminosity density $\:L_{0\nu}$
and $\:b(z,q_{0})$ is the model dependent factor.
\smallskip

\subsubsection{Paradoxes of the standard model }
The Standard Hot Big Bang model
has been accepted by most physicists.
We shall confront the SM with
modern observational data
in {\em section 4}. Here we discuss
 several paradoxes of the  SM .
For a more detailed description of these paradoxes see
e.g. von Horner, 1974; Guth, 1992; Linde et al., 1994;
Blau \& Guth, 1987;
 Harrison, 1993; 1995;
Baryshev, 1994a).
\smallskip

i.) The first of these paradoxes is called the {\it flatness paradox}.
The behavior of $\:(\Omega -1)/\Omega$,
with time, can be written as
\be
\label{92}
\frac{\Omega - 1}{\Omega} = \frac{3k}{8 \pi G \rho R^{2}}
\ee
Then one can calculate the allowed range of $\:\Omega$ at the Planck time
$\:t_{pl} = (G h/c)^{1/2} = 5.4 \cdot 10^{-44} sec$
\be
\label{93}
|\Omega -1 | < 10^{-59}
\ee
The flatness problem is then the difficulty in understanding
why $\:\Omega$ was so close to one, while $\:\Omega=1$
is a set of measure zero on the real line of all possible values
of $\:\Omega$. The question is: why is the Universe close to flat, i.e.
why is the geometry of the real space nearly Euclidean ($\:k=0$) ?
\smallskip

ii.) The second paradox is the {\it isotropy paradox}.
The problem is related to the large-scale isotropy of the observable
Universe, seen most strikingly in the isotropy of the
CMBR. In the context of the standard model
the horizon distance, i.e. the distance that a light pulse could
have traveled since the singularity
at the time $\:t=0$, is $\:r=ct$,
while the scale factor is $\:R(t) \sim t^{1/2}$
(for early stages). Hence the ratio
$\:R/r \sim t^{-1/2} \rightarrow \infty$ for
$\:t \rightarrow 0$. At the time of recombination $\:t_{r}$
two emitters $\:A$ and $\:B$ of CMBR
 photons, arriving at the earth today from two opposite
directions in the sky, are separated from each other by
more than $\:70$ times the distance that light could have traveled up until
that time. Thus, there is no way that a point $\:A$ and $\:B$
could have communicated with each other, and no physical process
that would bring them to the same temperature.
The paradox is that the large-scale isotropy of the Universe,
proved by the isotropy of the CMBR,  cannot be
explained by the SM and it must be assumed as an initial condition.

iii.) The third paradox is the {\it superluminal velocity paradox}.
The velocity of space expansion
$\: V \sim dR(t) / dt \sim t^{-1/2} $
 and $\: V \rightarrow \infty$ for
$\:t \rightarrow 0$. Hence as
noted by Von Horner (1974) and Murdoch (1977),
the SM  permits superluminal motions
of any distinct point at early epoch. More recently Harrison (1993)
has analyzed  the redshift-distance and the velocity-distance relations
and concluded that the linear $\:V(t) = c\cdot l/R_{H}$ law applies quite
generally in expanding homogeneous and isotropic cosmological models,
and the recession velocity $\:V$ can exceed the velocity of light
if $\:l>R_{H}$ (see Eq.\ref{32}). This violates the premises of
Special Relativity, but it is permitted by GR. In GR
 space is not rigid and can bend, twist and stretch; there is nothing
in General Relativity that places any limit on the speed with which
such stretching can take place (Guth, 1992). However the paradox
is that the permanent space creation demands
the violation of the  maximum
motion velocity principle, because in this case
we get information about increasing volume of space
via local  physical  parameters, such as the matter density.

iv.) The fourth paradox is the {\it inhomogeneity paradox}.
In the
framework of the SM
 the observed linear Hubble law ($\:z \sim l^{1}$)
is a consequence of  the homogeneity of  matter distribution.
It is not clear from observation
at which scale the cut-off towards homogeneity is reached
(see {\em section 4.2.3., 3.4.3.}), but it is
evident that at small scales, i.e.
for distances lower than
$\: \sim 100 Mpc$, the distribution of matter is fractal.
It was shown by Haggerty \& Wertz (1972),
Fang et al. (1991) and Ribeiro (1992a, 1992b, 1993)
that density fluctuations inside the fractal
inhomogeneity cell will lead to strong
disturbance of pure Friedmann behavior.
 However, observations
suggest the opposite conclusion: according to Sandage (1986; 1994)
a striking linearity of the $\:z-l$ relation is observed in the distance
range $\:(2-25) Mpc$ (see {\it section 4.1.1.}).
 This is the inhomogeneity paradox:
a highly inhomogeneous galaxy distribution at scales
where the $\:z-l$ relation
is linear means that the Hubble law should not
be  a consequence of homogeneity
(see Baryshev, 1994a).
\smallskip

v.) The fifth paradox is the {\it global energy paradox}.
If one divides
the Universe in "cosmic boxes" or "cells" which are representative
samples of the Universe, then
what is inside is always in the same
state as what is outside. Hence the Universe is not
like a steam engine
and pressure is not the cause of expansion. The Universe has no
edge and the pressure everywhere is therefore impotent and unable
to produce mechanical energy.
As was emphasized by Harrison (1981):
"the conservation of energy principle
serves us well in all science
except in cosmology" and " the total
energy decreases in an expanding Universe and increases in a collapsing
Universe. To the question where the energy goes
in an expanding Universe and where it comes
from in a collapsing Universe the answer is - nowhere,
because in this case energy
is not conserved."
Hence due to
the creation of space the total energy in the Universe is not
conserved (see Harrison, 1995).

\subsubsection {Non-standard models}

There are several modifications of
the  SM . Among these,
we can mention the cosmological constant theory,
 inflationary cosmology, rotation of the
Universe and others. Now we consider
only the first two which
are the more developed models.
\vspace{2cm}

{\it 3.2.6.1.  $\:\Lambda$ - term}
\smallskip

In General Relativity, the
gravitational field equations may be
written  as :
\be
\label{94}
R^{ik} - \frac{1}{2} g^{ik} R - g^{ik}
\Lambda = \frac{8 \pi G}{c^{4}}
T^{ik}_{(m)}
\ee
where the $\:\Lambda$ is the
 famous cosmological constant.
During the last twenty years
the $\:\Lambda$ -term has appeared
 and  disappeared repeatedly
in cosmological papers. Note
that Einstein called it {\it "the
biggest blunder of my life"}. Now the
$\:\Lambda$-term is popular again,
because it gives additional
possibilities
 to fit observational data (see e.g.
Carroll et al., 1992; Croswell, 1993).
Instead of Eq.\ref{42} and Eq.\ref{43} we get
(for $\:p \ll \rho c^{2}$):
\be
\label{96}
\Omega - \lambda = K + 1
\ee
\be
\label{95}
q =  \frac{1}{2} \Omega - \lambda
\ee
where $\:\lambda =\Lambda
c ^{2}/(3H^{2})$.
The bolometric distance
$\:l_{bol}$ is given by:
\be
\label{97}
l_{bol}=\frac{c(1+z)}{H_{0}} \left[ \begin{array}{l}
\frac{1}{(-K_{0})^{1/2}}
sinh \left( \int^{1}_{\frac{1}{1+z}}\frac{(-
K_{0})^{1/2}dy}{y( \Omega_{0}/y
- K_{0} + \lambda_{0} y^{2})^{1/2}}\right) \\
\int^{1}_{\frac{1}{1+z}}\frac{dy}
{y(\Omega_{0}/y+ \lambda_{0} y^{2})^{1/2}} \\
\frac{1}{(K_{0})^{1/2}}
sin\left(\int^{1}_{\frac{1}{1+z}}
\frac{(K_{0})^{1/2}dy}
{ y (\Omega_{0} / y - K_{0} +
\lambda_{0} y^{2})^{1/2}}\right)\end{array} \right.
\ee
for $k=$ -1, $k=$ 0, $k=$ +1 respectively.
In terms of $\:l_{bol}$,
 the angular distance $\:l_{a}$
and the comoving volume
element $\:dV/dz$ are expressed
respectively as :
\be
\label{98}
l_{a}=l_{bol}/(1+z)^{2}
\ee
\be
\label{99}
\frac{dV}{dz}=
\frac{4\pi c l_{bol}^{2}}{H_{0}
(1+z)^{3}(\Omega_{0}(1+z)
- K_{0} + \lambda_{0}/
(1+z)^{2})^{1/2}}
\ee
The age of the Universe at
redshift $\:z$ is given by:
\be
\label{100}
t(z) = \int^{\frac{1}
{1+z}}_{0}\frac{dy}
{ ( \Omega_{0} / y - K_{0} +
\lambda_{0} y^{2} )^{1/2} }
\ee
A Universe with  a cosmological
constant can behave in different
ways than SM allows.
There is a loiter phase which
occurs when the cosmological
constant $\:\lambda$ and the mass
density $\:\Omega$ nearly balance
each other. The loiter phase can last
so long that the
Universe can be far older than
the ages that comes from standard
calculation based on its rate of
expansion and mass density.
Moreover the present age of the
Universe depends upon the $\:\Lambda$-constant
and can be compatible with the current
estimates ({\it section 4.1.1.}) for an appropriate
choice of $\:\Lambda$.

Theoretical expectations for the cosmological
constant exceed observational limits by some $\:120$ orders of magnitude.
This is because anything that contributes to the energy density
of the vacuum acts just like a cosmological constant. In fact in the vacuum
the EMT can be written  as:
\be
\label{lambda1}
T_{\mu\nu} = <\rho> g_{\mu\nu}
\ee
so that the effective cosmological constant
can be written as (Eq.\ref{94})
\be
\label{lambda2}
\Lambda_{eff} = \Lambda + 8 \pi G <\rho>
\ee
Then the cosmological constant contributes to the total vacuum energy with
 a term
\be
\label{lambda3}
\rho_{v} = <\rho> + \frac{\Lambda}{8 \pi G} = \frac{\Lambda_{eff}}{8 \pi G}
\ee
A crude experimental upper bound on $\:\Lambda_{eff}$ is provided by
the value of  the Hubble constant, the flatness of space and the mass
density of the Universe (Weinberg, 1989; Carroll et al., 1992)
\be
\label{lambda4}
|\rho_{v}| < 10^{-47} Gev^{4}
\ee
The energy density of the empty space is the sum of the zero point energies
of the normal modes of some field up to a wave number cutoff $\:k_{max}$
 ($h=c=$ 1):
\be
\label{lambda5}
<\rho> \sim \frac{k_{max}^{4}}{16 \pi^{2}}
\ee
If we take the Planck mass $\:k_{max} \sim (8 \pi G)^{-1/2}$
from Eq.\ref{lambda5} it follows that
$\:<\rho> \sim   2 \cdot 10^{71} GeV^{4}$
so that there is a difference of  118 times with
the observational upper limit (Eq.\ref{lambda4}).
If we take the zero point energy of quantum chronodynamics
$\:<\rho> \sim 10^{-6} GeV^{4}$ the term $\:\Lambda/(8 \pi G)$
in Eq.\ref{lambda3} cancels this term to about 41 decimal places !
Weinberg in his review (1989) considers various possible
solutions of this problem based on different approaches:
all these approaches show that the cosmological constant problem
has great impact on other areas of physics
or astronomy.
Weinberg notes: "More discouraging than any theorem is the fact
that many theorists have tried to invent adjustment mechanisms
to cancel the cosmological constant,
but without any success so far" .
\bigskip

{\it 3.2.6.2. Inflationary Universe}
\smallskip

The inflationary model was proposed by Guth (1981)
to avoid the flatness and the
isotropy  paradoxes.
There are several modifications of
the inflation scenario, which improve
Guth's original model (see e.g. Linde, 1990; Guth, 1992).
The main features
of the various inflationary universe scenarios is the
existence of some stage of evolution at which the universe
expands exponentially, while it is in a vacuum-like state
containing some classical homogenous fields but
(almost) no particles. After the
inflation the vacuum-like state decays into particles and
heats up the Universe; from then the evolution can be
described by the standard Hot-Big-Bang theory.
The homogenous classical field responsible for
inflation is present
in all the Grand Unified Theories (GUT)
of elementary particles, and can play the role of an unstable
vacuum state.

If the Universe has ever been
dominated by a false (unstable) vacuum
characterized by a pressure $\:p = - \rho \cdot c^{2}$, then the
 pressure term in the Friedmann
Eq.\ref{35} has negative sign,
and overcomes the gravitational
attraction caused by  the usual
energy density term.
The force of gravity actually
becomes repulsive and the
scale factor has an exponential
behavior:
\be
\label{101}
R(t) = R_{0} e^{(\chi t)}
\ee
where:
\be
\label{102}
 \chi = \left(\frac{8\pi G \rho_{vac}}
{3}\right)^{1/2}
\ee
and $\:\rho_{vac}$ is the mass density of
the vacuum at time $\:t$.
The precise value of the $\:\rho_{vac}$
 is not well constrained, and  for a
typical GUT $\:\rho_{vac}\sim(10^{14}Gev)^{4}/
(\hbar c)^{3} \approx 10^{74}gr  \cdot cm^{-3}$
(Blau \& Guth, 1987),
while it also has been used the value of
the
Plank
density $\:\rho_{vac}\sim(10^{19}Gev)^{4}
/(\hbar c)^{3} \approx 10^{94}gr \cdot cm^{-3}$
(Novikov, 1988).

Thus the Universe expands exponentially
due to the gravitational repulsion of this
false vacuum. During the inflationary
period, the density of any particles
that may have been present before
inflation, is diluted so much that
it becomes completely negligible.
The isotropy paradox is solved by
an enormous expansion: the process
of inflation magnified very small regions
to become large enough to encompass
the entire observed Universe. Thus, the
emitters of the background radiation
arriving today from opposite directions
in the sky, had time to reach a common
temperature during the inflationary era.
The flatness paradox is solved because the inflationary
process determines a density $\:\rho=\rho_{crit}$, so that
 after inflation and vacuum decay one
 gets $\:\Omega=1$, or in the case of
non-zero
$\:\Lambda$-term $\:\Omega +\Lambda =1$.
Thus the inflation model predicts
 large-scale isotropy, homogeneity and flatness
of the Universe.
Moreover during inflation, from the quantum fluctuations
of the scalar fields,
 could be generated the adiabatic perturbations with a flat
spectrum that
are the origin of the inhomogeneities of matter in the present
universe.  However,
the simplest  GUT
predicts the value of density
 perturbations with a magnitude
 $\:10^{5}$ times larger than
 what is observed (Guth, 1992): indeed some theories lead to
too large density fluctuations after
inflation and therefore
have been  rejected. On the contrary,
in some other theories
the amplitude of density perturbations has the magnitude
which is just necessary for galaxy formation and compatible
with the limits of COBE on the anisotropies of the CMBR.

Unfortunately, the inflation paradigm
does not solve the superluminal
expansion and the inhomogeneity
paradoxes. Moreover, it rather
magnifies superluminal expansion.
 For example, if a distance between
two particles before inflation was $\:l_{0}=10^{-35}
 cm$, after the inflationary era it will be $\:10^{4 \cdot 10^{8}}
cm$, hence the velocity of expansion
 will be (Novikov, 1988):
\be
\label{103}
v \sim l_{0}/ \Delta t \sim 10^{4 \cdot 10^{8}} cm/sec \gg c
\ee
There are two explanations of the relation in Eq.\ref{103}.
The first is that there is  nothing in General
 Relativity that constraints the speed of space
expansion (see Guth, 1992).
The second is that there is no possibility
 to measure the relative velocity of distant
objects in a strong gravitational field, i.e. there
is no sense in the notion of the relative
velocity for the two particles which
was considered above (Novikov, 1988).
However, if there is no possibility
to measure the relative velocity
of distant particles, then one can not measure the distance
between the particles too, because
one has no rigid measuring rod.
 Hence there is no such concept
 as distance in inflating Universe.
So the inflation
paradigm uses the  highly superluminal
expansion of space to carry information
about local physical conditions for
almost the whole Universe.
Hence superluminal expansion
paradox is still unsolved within
inflation models.

Recently a modified version
of the Inflationary scenario has been introduced
 by Linde (1990, 1994).
In this model, due to quantum fluctuations of the scalar fields,
the universe exists eternally as a
huge self-reproducing entity and it is extremely
inhomogeneous, with a fractal like structure, on
scales larger than the casually connected horizon.
This fractal structure has nothing to do with the fractal
distribution of matter in the visible Universe, but it concerns
the distribution of many universes that undergo, at different times,
an inflationary process.

According to Linde (1994) there are some testable predictions
of the inflationary model. First of all the inflation theories predict
that the Universe is extremely flat with an average density equal
to the critical one. A crucial point of the inflationary scenario
is that the density has to be very close to the critical one:
this is connected with
the famous dark matter problem that is still unsolved.
Another testable prediction is related
to density perturbations produced during
inflation that, in some models, are in agreement with the
results of COBE. The scalar fields
themselves are not directly observable quantities,
if they fill homogeneously the Universe but their presence affects
the properties of elementary particles.
In the GUT
of weak and strong interactions,
there are two scalar fields and it would
be necessary to verify experimentally the GUT theories.
Unfortunately, this appears to be a very difficult task
as the energy scale is about that of the Planck mass rest
energy $\:\sim 10^{19} Gev$, while
the currently available particle accelerators reach
$\:\sim 10^{4} Gev$. There are
some indirect tests like the proton decay and the
search for supersymmetric
particles, but there are no definite results.
In any case there is not a well defined
candidate for the inflation scalar field.

\subsection{Steady-State Models}

The Steady State Model (hereafter SSM) was proposed
by Bondi and Gold (1948) and Hoyle (1948) as a
competing alternative to the Standard
Hot Big Bang Model. More recent
reviews of the  SSM  can be found in Hoyle (1982, 1991),
Narlikar (1977, 1987, 1993) and Hoyle et al. (1993, 1994a, 1994b).

\subsubsection{Basic Hypotheses of the SSM}

 The {\it first hypothesis}
of SSM is the Perfect Cosmological
Principle (PCP), i.e. globally unchanging Universe in space and
time. According to PCP there is no global evolution  of the
Universe so that the mean matter density does not change with
time. Bondi and Gold emphasize that the main reason for PCP is
the conservation of the cosmological external conditions
which allow us to use terrestrial physics unambiguously in
cosmology.
   The {\it second  hypothesis} is that gravitation theory
is a modified version of General Relativity, i.e. it is
a geometrical
description of gravity. There are several versions of the
gravitational field equation within the SSM. In
Hoyle-Narlikar's  approach a scale invariant gravitation theory
is used so that the creation of matter
by means of the {\it C-field}
is possible. It does not violate the law of energy conservation
because the theory is derived from an action principle and
therefore, because of
Noether's theorem, it obeys the conservation laws. The
{\it C-field} has negative energy and negative stresses and it is
conserved together with usual matter.
   The {\it third hypothesis } is that the observed cosmological
redshift is caused by the space expansion, i.e. increasing space
volume with time ("creation of space"). To compensate the decreasing
matter density it is assumed the creation of matter ("creation of
space with matter") so the PCP is fulfilled.

\subsubsection{Prediction for testing}

In the SSM the Universe had no
beginning and so it is infinitely old, the
 density of the Universe remains constant,
the matter and space are being
continuously created, and  the Universe is
not evolving.
   The Robertson-Walker line element in the SSM is
characterized by the following parameters: 1) the space
curvature $\:k=0$; 2)the Hubble constant $\:H$ is a fundamental
constant; 3) the deceleration parameter $\:q=-1$. Hence
the scale factor is :
\be
\label{105}
R(t)=Ae^{Ht}
\ee
where $\:A$ is constant, and the
density of the Universe has the critical value:
\be
\label{106}
\rho = \frac{3H^{2}}{8\pi G}
\ee
For metric distance we get
\be
\label{107}
l = R_{H} z
\ee
where $\:R_{H} = c/H$.
Hence the angular size-redshift
relation has the form
\be
\label{108}
\Theta(z) = \frac{d}{R_{H}}\frac{(1+z)}{z}
\ee
the bolometric magnitude-redshift relation is:
\be
\label{109}
m_{bol}(z) = 5 \log(z(1+z)) + const.
\ee
and the integral count-redshift is
\be
\label{110}
N(z) = const \cdot \left( \log(1+z) -
\frac{z(2+3z)}{2(1+z)^{2}}\right)
\ee
It is clear that there are no flatness
and isotropy paradoxes in the
SSM. However
the superluminal expansion,
inhomogeneity and global energy
paradoxes still exist in the SSM.
The main observational difficulty
that the SSM faces,is, however, the CMBR: an adequate
theory that should produce the correct spectrum
and isotropy has not been developed yet.

But, as it was mentioned by Hoyle (1991), a
small fractional conversion of
baryonic energy into electromagnetic
energy (e.g. by stellar nuclear reactions)
inevitably yields the correct order of
magnitude for the energy of microwave
background.
The problem consists of the
thermalization of  this energy up to its
thermodynamic value $\:2.73 K$.

\subsubsection{Modern development of the SSM}

An interesting argument in favor of the SSM
 has been found by
Hoyle and Narlikar (1963)
(see also Hoyle \& Narlikar, 1974; Narlikar, 1977).
They considered the Wheeler-Feynman
absorber theory in the context of various
cosmological models to see which
of them give the correct response of the
Universe.
We can divide the rest of the
Universe with respect to the present
position of the electric charge $\:A$ into two
light cones in the future and the past.
The cone and its interior in the future
is the "future absorber", while the cone
in the past is the "past absorber". The rule for the correct
response from a Universe model is that the past
absorber is imperfect and
the future absorber is perfect
(it must totally absorb all disturbances
generated by $\:A$) .

In the static Universe without redshift both absorber are
perfect and the result is ambiguous, i.e. one can get retarded
or advanced waves. In the open Friedmann models
the density of the
matter in the future absorber decreases
 to zero and there is not enough matter
to absorb radiation. Thus the future
absorber is imperfect and we get advanced
radiation. In the closed
Friedmann models the result is
 ambiguous as in the Static Universe.
 SSM only has the correct
response of the Universe and hence
predicts the retarded radiation.

Recently Hoyle, Burbidge and Narlikar
(1993; 1994a; 1994b) have modified the SSM.
In the new "Quasi Steady
State " model there are scale invariant
gravitational equations, which reduce
to those of General Relativity for a
particular choice of the scale. The authors
have shown that the model can explain
the $\:2.73 K$ CMBR and the abundances
 of the light elements $\:D, ^3He,
^4He, ^6Li, ^7 Li, ^9Be, ^{11}B$.
The model is based on the idea of
discrete creation events. These
 creation events occur throughout the
whole Universe, in creation centers. The most
 important creation centers have masses
of the order $\:10^{16} M_{\odot}$.
The newly created particles have
 Planck mass $\:(hc/2\pi G)^{1/2}$
and are unstable over Planck time scale
$\:(hG/2\pi c^{5})^{1/2}$.
The  created masses are much less than
$\:10^{16} M_{\odot}$, related to such
well- known events as young
galaxies, active galactic nuclei, galactic
 groups and clusters with positive
total energy. Within the model there was
a major creation episode when the
mean universal density was $\:10^{-27}
gr  \cdot cm^{-3}$. Since then the universal
expansion has been slowing down
with the parameter $\:q_{0}$
rising
from -1 to its present day value $\:q_{0}=+1 $.

\subsection {Fractal models}

At present a fractal model  has not yet  been
developed, but there are some attempts and many
possibilities are still under consideration.
For this reason we firstly illustrate the main
 properties of self-similar structures, showing why
a change of perspective is required  if one deals
with fractals.
In fact, self-similar structures are
intrinsically irregular (non-analytic) at all scales and fractal geometry
allows us to characterize them mathematically.
There are some deep implications of these
concepts not only on the theoretical side
but also on the data analysis.
The main theoretical consequence is that
one should not discuss a self-similar structure in terms
of amplitudes of correlation, but the only
meaningful quantity is the exponent that
characterizes the power law behaviour.
We discuss in detail the observational properties
of  large-scale structure distribution that one can
obtain from the available redshift surveys
with an appropriate statistical analysis.

{}From a theoretical point of view
the dark matter plays  a crucial  role, because if  it
is homogeneously distributed the
Friedmann metric can still be the right one, even if
 a deep revision of  the
models of the large-scale structure
formation is required.
On the contrary, if the dark matter turns
 out to be
distributed as the visible one,
the problem becomes
very hard and has to be reconsidered from
the beginning. The test on the whole matter
distribution, provided by the analysis of bulk flows and
gravitational lensing effects, is therefore
a crucial one.

We briefly illustrate the old idea of  hierarchical cosmology,
based on the Newtonian theory of gravitation,  that
has behind it the same concept of scale-invariance of fractal structures.
Hence we consider some attempts to reconcile the fractal
distribution in  the framework of General Relativity, in spherically
symmetric models described by the Tolman-Bondi metric.
Finally we discuss a new idea
based on tensor field relativistic gravitation theory
in Minkowski space-time.

\subsubsection {Fractal geometry}

Fractals are simple but subtle.
In this section we provide a brief
description of their essential properties.
 This description is not  intended to
 represent a theory, but only to illustrate
 the consequences of the properties of
self-similarity so that, if the property is
actually present in the experimental data, we will
 be able to detect it correctly; on the contrary,
if the data would be in contrast with the fractal
properties, we have to know well the properties
of fractals in order to eventually conclude that
observations are actually in contrast with them.
Fractal geometry has a long history and some
elements of it can be found already in the work
 of  Poincar\'e  and Hausdorff  of about a
century ago. The introduction of the name "{\it Fractals}"
and the realization that fractal geometry is a
powerful tool to characterize intrinsically
irregular systems is due to Mandelbrot and
it refers mainly to the past twenty years (Mandelbrot, 1982).
Nature is full of strongly irregular structures;
trees, clouds, mountains and lightning are quite
familiar objects but are very different from the
 structures of euclidean geometry.
A common element of these and many other
 structures is that if one magnifies a small portion
of them, this reveals a complexity comparable
to that of the entire structure. This is  geometric
 self-similarity and it has deep implications for
 the non-analyticity of these structures. In fact
analyticity or regularities would imply that at
some small scale the profile becomes smooth
 and one can define a unique tangent. Clearly
this is impossible in a self-similar structure
 because at any small scale a new structure
appears and the structure is never smooth.
 Self-similar structures are therefore intrinsically
 irregular at all scales and this is why many familiar
phenomena have remained at the margins of scientific
investigation.

 The usual mathematical concepts in
 physics are mostly based on analytical functions
and, in this perspective, irregularities are seen as
imperfections. Fractal geometry changes completely
this perspective by focusing exactly on these intrinsic
irregularities and it allows us to characterize them in a
 quantitative mathematical way. In this way it has
extended the frontiers of scientific investigation to
 all the intrinsically irregular structures. It should
be made clear however that fractal geometry is
not a physical theory because it does not explain
 why nature generates fractal structures. It permits
 us however to look at irregularities in a new way and
 to pose the correct questions for a theory.
Fractal geometry has had a large impact in all
sciences and in particular in physics. In fact in
physics the concept of scale invariance was
already familiar from the study of critical
phenomena in phase transitions. This is an example
 of self-similarity that can be understood with the
Renormalization Group theory. This method
 however does not seem to be very effective
for the more familiar fractal structure that
arises from irreversible dynamics and this is
the main theoretical challenge at the moments.
Fractal concepts have become very popular
and even fashionable in many areas for various
 reasons. The characterization of intrinsic
irregularities represents a new powerful tool
to understand the properties of nature. Also
the aesthetic appeal of these structures and the
return of visual intuition into science have played an important role.

In astrophysics however these ideas have been
considered with skepticism and sometimes even
 with opposition. The main reason is that they
may appear to contrast with concepts like the
 Cosmological Principle and the whole standard
theoretical framework based on homogeneity
and the Friedmann metric. On the other hand
the large-scale structures observed experimentally
provide a strong evidence for self-similar
and fractal properties. So we have to hope
that after the first period of emotional debate
 the situation can evolve into constructive confrontation.

\subsubsection {Fractal Structures}

A fractal consists of a system in which more
and more structures appear at smaller and
smaller scales and the structures at small
scales are similar to the one at large scales.
In Fig.\ref {f1}(a), we show an elementary (deterministic)
fractal distribution of points in space whose
construction is self-evident. Starting from a
 point occupied by an object we count how
many objects are present within a volume
characterized by a certain length scale in
 order to establish a generalized "mass-length"
relation from which one can define the fractal
dimension.

\begin {figure}
\vspace {8 cm}
\caption{\footnotesize (a) A simple example of a deterministic fractal
with dimension D=1.2 . The same
structures repeats at different scales in a self-similar way. (b) Example
of a stochastic
fractal with dimension D= 1.2 generated by the "random $\:\beta-$model".}
\label {f1}
\end {figure}

Suppose that in the structure of
Fig.\ref {f1}(a)  we can find no objects in a volume of
 size $\:r_{0}$. If we consider a larger volume
of size $\:r_{1} = k \cdot r_{0}$ we will find
$\:N_{1} ={\tilde  k} \cdot N_{0}$ objects.
In a self-similar structure the parameters $\:k$
 and $\:{\tilde k}$ will be the same also for
other changes of scale. So, in general in a
structure of size $\:r_{n} = k^{n}\cdot r_{0}$
we will have $\:N_{n} ={\tilde  k^{n}} \cdot N_{0}$
objects. We can then write a relation
 between $\:N$ ("mass") and $\:r$ ("length") of type:
\be
\label{l2}
N(r) = B\cdot r^{D}
\ee
where the fractal dimension:
\be
\label{l3}
D = \frac{log{\tilde k}}{log k}
\ee
depends on the rescaling factors $\:k$
and $\:{\tilde k}$. The prefactor $\:B$
is instead related to the lower cut-offs $\:
N_{0}$ and $\:r_{0}$,
\be
\label{l4}
B =\frac{ N_{0}}{r_{0}^{D}}
\ee
It should be noted that Eq.\ref{l2} corresponds
to a smooth convolution of a strongly
fluctuating function as evident
from Fig.\ref {f1}. Therefore a fractal
structure is always connected with large
 fluctuations and clustering at all scales.
{}From Eq.\ref{l2} we can readily compute the
average density $\:<n>$ for a sample of
 radius $\:R_{s}$ which contains a portion
of the fractal structure. The sample volume
is assumed to be a sphere ($\:V(R_{s}) = (4/3)\pi R_{s}^{3}$) and therefore
\be
\label{l5}
<n> =\frac{ N(R_{s})}{V(R_{s})} = \frac{3}{4\pi } B R_{s}^{-(3-D)}
\ee
{}From Eq.\ref{l5} we see that the average density
is not a meaningful concept in a fractal
because it depends explicitly on the sample
size $\:R_{s}$. We can also see that for
$\:R_{s} \rightarrow \infty$ the average density
$\:<n> \rightarrow  0 $,
therefore a fractal structure is asymptotically
dominated by voids. We see therefore that the
average density $\:<n>$ is not a well defined
quantity but the {\it conditional average} density
as given by Eq.\ref{l6} is well defined in terms of
its exponent, the fractal dimension. The
amplitudes of these functions essentially
refer to the unit of measures given by the
lower cut-offs but they have no particular
 physical meaning because they are not intrinsic quantities.
We can also define the conditional density from any
occupied point as:
\be
\label{l6}
\Gamma (r)= S^{-1}\frac{ dN(r)}{dr} = \frac{D}{4\pi } B r ^{-(3-D)}
\ee
where $\:S(r)$ is the area of  a spherical shell of radius $\:r$.
Usually the exponent that defines the decay
 of the conditional density $\:(3-D)$ is called
the codimension and it corresponds to the
exponent $\:\gamma$ of the galaxy distribution.
In Fig.\ref {f1}(b) we show a stochastic fractal
(generated with the random-$\:\beta$-model algorithm
in the  two dimensional euclidean space)
constructed with a probabilistic algorithm
that nevertheless has a well defined fractal dimension $\:D = 1.2$.
\bigskip

{\em 3.4.2.1.  Mathematical Self-similarity}
\smallskip

{}From Fig.\ref {f1}(a) the geometrical self-similarity
is evident in the construction. From a
mathematical point of view self-similarity
implies that a rescaling of the length
by a factor $\:b$
\be
\label {l7}
r \rightarrow r ' = br
\ee
leaves the correlation function unchanged
apart from a rescaling that depends on
$\:b$ but not on the variable $\:r$.
 This leads to the functional relation
\be
\label {l8}
\Gamma ( r') = \Gamma (b\cdot r) =
A(b)\cdot \Gamma (r)
\ee
which is clearly satisfied by a power law
with any exponent.
In fact for
\be
\label {l9}
\Gamma (r) = \Gamma_{0} r ^{\alpha}
\ee
we have
\be
\label {l10}
\Gamma (r') = \Gamma_{0}(br)^{\alpha}
= (b)^{\alpha}\Gamma (r)
\ee
The same does not hold, for example,
for an exponential behavior
\be
\label {l11}
\Gamma ( r ) = \Gamma _{0} e ^{-r/r_{0}}
\ee
This reflects the fact that power laws do
not possess a characteristic  length
while for the exponential decay
$\:r_{0}$ is a characteristic
length. Note that the characteristic
length has nothing to do with the
prefactor of the exponential
and it is not defined by the condition
$\:\Gamma ( r_{0} ) = 1$, but from the
intrinsic behavior of the function.
This brings us to a common
misconception that sometimes
occurs in the discussion of
galaxy correlations. Even for a
perfect power law as Eq.\ref{l9}
one might use the condition
$\:\Gamma(r_{0})=1$ to derive
a "characteristic length":
\be
\label {l12}
r_{0}=\Gamma_{0}^{-1/\alpha}
\ee
This however is completely meaningless
because the power law refers to a fractal
structure constructed as self-similar
and therefore without a characteristic
length.
In Eq.\ref{l12} the value of $\:r_{0}$
is just related to the amplitude of the
power law that, as we have discussed,
has no physical meaning.
The point is that the value
1 used in the relation $\:\Gamma(r_{0})=1$
is not particular in any way so one
could have used as well the
condition $\:\Gamma(r_{0})=10^{10}$
or $\:\Gamma(r_{0})=10^{-10}$
to obtain other lengths. This
is the subtle point of self-similarity;
there is no reference value (like
the average density) with respect
to which one can define what is big or
small.
\bigskip

{\it 3.4.2.2. "Linear and Non-linear Dynamics"}
\smallskip

Such a discussion naturally extends
to the much discussed value of $\:\delta N / N$
the ratio of density fluctuations
with respect to the average density.
Another argument often mentioned
in the discussion of large-scale
structures is that it is true that
larger samples show larger structures
but their amplitudes are smaller
and the value of $\:\delta N / N$
tends to zero at the limits of the
sample; therefore one expects
that just going a bit further,
homogeneity would finally be
observed. Apart from the fact
that this expectation has been
systematically disproved,
the argument is conceptually
wrong for the same reasons
of the previous discussion.
In fact we can consider a
portion of a fractal structure
of size $\:R_{s}$ and study
the behavior of  $\:\delta N / N$.
The average density $\:N$ is
just given by Eq.\ref{l5} while
the overdensity
$\:\delta N$, as a function
of the size $\:r$ of a given
in structure is ($\:r\leq R_{s}$):
\be
\label {l13}
\delta N = \frac{N(r)}{V(r)} - <n>
= \frac {3}{4\pi} B (r^{-(3-D)}-R_{s}^{-(3 - D)})
\ee
We have therefore
\be
\label {l14}
\frac {\delta N}{N} =
\left(\frac{r}{R_{s}}\right)
^{-(3 - D)} - 1
\ee
Clearly for structures that
approach the size of the small
sample, the value of  $\:\delta N / N$
becomes very small and eventually
becomes zero at $\:r = R_{s}$
as shown in Fig.\ref {f2}.
\begin {figure}
\vspace {8 cm}
\caption{\footnotesize  Behavior of
$\:\delta N / N$ as a function of the size  r
in a portion of a fractal structure
for various depths of the sample: $\:R_{s}=100,200,300 Mpc$.
The average density is computed over the whole sample
of radius $\:R_{s}$.
The fact that $\:\delta N/N$ tends
to zero does not mean that the
fluctuations are small and a homogenous distribution
has been reached. The scale at which $\:\delta N/N =1$ scales
with sample depth and has no physical meaning.
In the case of a fractal distribution the normalization
factor, i.e. the average density, is not an intrinsic quantity.}
\label {f2}
\end {figure}
This behavior however could be
interpreted as a tendency towards
homogeneity because again
the exercise refers to a self-similar
fractal by construction. Also
in this case the problems
come from the fact that
one defines an "amplitude"
arbitrary by normalizing
with the average density
that is not an intrinsic
quantity.
A clarification of this
point is very important
because the argument that
since  $\:\delta N / N$ becomes
smaller at large scale, there
is a clear evidence of
homogenization is still
quite popular (Peebles et al., 1991; Peebles, 1993;
 Peebles et al., 1994)
and it adds confusion to the
discussion.

The correct interpretation of
 $\:\delta N / N$ is also
fundamental for the development
of the appropriate theoretical
concepts. For example a popular
point of view is to say that
 $\:\delta N / N$ is large $\:( \gg 1)$
for small structure and this
implies that a non linear
theory will be necessary to
explain this. On the other hand
 $\:\delta N / N$ becomes small
$\:(<1)$ for large structures,
which require therefore a linear
theory.
The value of  $\:\delta N / N$
has therefore generated a conceptual
distinction between small structures
that would entail non linear dynamics
and large structures with small
amplitudes that correspond instead to
 linear dynamics.
If one would apply the same
reasoning to our example of a
fractal structure we would conclude
that for a structure up to a size
(from Eq.\ref {l14}):
\be
\label {l15}
r^{*} = 2^{-\left( \frac{1}{3-D} \right)} R_{s}
\ee
we have  $\:\delta N / N$ so that a non linear
theory is needed. On the other hand for
large structures $\:(r>r^{*})$ we have
 $\:\delta N / N$ corresponding to a
linear dynamics. Since the fractal
structure that we have used
to make this conceptual exercise
has scale invariant structures
by construction, we can see
the distinction between
linear and non linear dynamics
is completely artificial
and wrong.
The point is again that the value
of $\:N$ we use to normalize
the fluctuations is not
intrinsic but it just
reflects the size of the sample
that we consider ($\:R_{s}$).

If we have a sample with  depth $\:\tilde{R_{s}}$
greater than the eventual scale of homogeneity $\:\lambda_{0}$,
then the average density will be constant in the range
$\:\lambda_{0} < r < \tilde{R_{s}}$ apart from small amplitude
fluctuations. The distance at which $\:\delta N/N =1$ will be
given by:
\be
\label {l15a}
r^* = 2^{-\left( \frac{1}{3-D} \right)} \lambda_{0}
\ee
If, for example, $\:D=2$ and $\:\lambda_{0}=200 Mpc$ then
$\:r^{*} = 100 Mpc$:
therefore a homogeneity scale of this order of magnitude is
incompatible with the standard normalization of
$\:\delta N/N =1$ at $\:8 h^{-1} Mpc$.

We can see therefore that the whole
discussion about large and small
amplitudes and the corresponding
non linear and linear dynamics
has no meaning until an unambiguous
value of the average density has been defined, so
that the concepts like large and small amplitudes
can take a physical meaning and be
independent of the size of the
catalogue.
It is also an instructive
exercise to consider the
behavior of the function
\be
\label {l16}
\xi (r) = \frac{<n(r_{0})n(r_{0} + r)>}
{<n>^{2}} - 1
\ee
for a fractal structure. The
result is:
\be
\label {l17}
\xi (r) = \left(\frac{3 - \gamma}
{3}\right)\left(\frac{r}{R_{s}}\right)
^{- \gamma} - 1
\ee
with $\:\gamma = 3 - D$. In this
case also if one defines a length by the
relation $\:\xi (r_{0}) =1$, one
obtains:
\be
\label {l18}
r_{0} = \left( \frac{3 - \gamma}{6} \right)^{\frac{1}{\gamma}} R_{s}
\ee
that has no physical meaning
because it is dependent on the
sample size $\:R_{s}$ and not on the real
correlations of the system.
The basic point of all this
discussion is that in a
self-similar structure one cannot
say that correlations are
"large" or "small", because
these words have no physical meaning due
to the lack of a characteristic
quantity with respect to which
one can normalize these properties.
The deep implication of this
fact is that one cannot
discuss a self-similar structure in terms
of amplitudes of correlation. The only
meaningful physical quantity is the
exponent that characterizes the
power law behaviour. Note that this
problem of the "amplitude" is not
only present in the data analysis but
also in the theoretical models.
Meaningful amplitudes can only be
defined once one has unambiguous
evidence for homogeneity but this
is clearly not the case for galaxy and cluster distributions.

In this section we have seen with
a few examples what type of problems and
misleading conclusions one may derive
from statistical analysis performed in a naive way.
Of course the example of a fractal structure
that we have used, is just for conceptual
clarification and it should not be confused
with a theory. We hope however that this discussion
will help to clarify misconceptions like
that if  $\:\delta N / N$ becomes small
at large scales this is an evidence
for homogenization with all the related
consequences and theoretical implications.

It is therefore clear that a crucial element to
consider in the analysis of galaxy and cluster
distributions is whether a real homogenization
is achieved and average density can be derived.
Only in this case in fact do the amplitudes of correlation
acquire a physical meaning and one can consider it
as a theoretical problem.
In the opposite case and anyhow for the range of scales
in which the structure is self-similar
(even if homogeneity is eventually achieved at large
scale) it is necessary to change the theoretical
language and perspective and adopt the one that
is appropriate for self-similar and non-analytical
structures.
\bigskip

{\it 3.4.2.3. Fair Sample}
\smallskip

In this perspective it is important
to clarify the concept of "Fair Sample".
Often this concept is used as synonymous
of a homogeneous sample (see for example
da Costa et al., 1994). So the analysis of
catalogues along the traditional lines
often leads to the conclusion that we
still do not have a fair sample
and deeper surveys are needed to
derive the correct correlation
properties. A corollary of
this point of view is that since
we do not have a fair sample
its statistical analysis cannot be taken
too seriously.

This point of view is highly misleading
because we have seen that self-similar
structures never become homogeneous, so
any sample containing a self-similar
(fractal) structure would automatically
be declared "not fair" and therefore
impossible to analyze.
The situation is actually much more
interesting otherwise the statistical
mechanics of complex systems would not
exist. Homogeneity is a property and not
a condition of statistical validity of the
sample. A non homogeneous system can
have well defined statistical
properties in terms of scale invariant
correlations, that may be perfectly
well defined.
The whole studies of fractal structures
are about this (Pietronero \& Tosatti, 1986). Therefore one
should distinguish between a "statistical
fair sample" which is a sample in which
there are enough points to derive
some statistical properties unambiguously
and a homogeneous sample that is a property
that can be present or not but that has nothing
to do with the statistical validity of the sample.
In the following we are going to see that even
the small sample like {\it CfA 1} is
statistically fair up to a distance that can be
defined unambiguously.
Also the combined sample
{\it CfA2} and {\it SSRS2} has
been declared to be {\it "not fair"}
(da Costa et al., 1994).
\bigskip

{\it 3.4.2.4. Isotropy, Homogeneity and Cosmological
Principle}
\smallskip

The Cosmological Principle (CP) implies
that all the mass points (galaxies) should
be statistically equivalent with respect
to their environment. This condition corresponds
to the property of local isotropy and it is
generally believed that the universe cannot be isotropic
about every point without being also homogeneous.
Actually local isotropy does not necessarily imply
homogeneity.
In fact a topology theorem states that homogeneity
is implied by the condition of local isotropy
together with the assumption of analyticity or
regularity for the distribution of matter
(Weinberg 1972).
It is easy to see (Fig.\ref {f1}) and to prove
that in fractal structure the condition
of local isotropy, or statistical equivalence
of all occupied points, is actually satisfied but,
since the structure is non-analytical, the property
of homogeneity is not implied. This means
that a fractal structure is in perfect agreement
with the CP.
In addition this also implies that the tests
of dipole moment saturation are only tests of local
isotropy, but not homogeneity. We have shown
in fact that the dipole moment saturates
quickly also in a fractal structure while the monopole
moment grows as the power law of the sample size
(Sylos Labini, 1994).
\bigskip

{\it 3.4.2.5. Power Spectrum}
\smallskip

The power spectrum (PS)
 measures the fluctuation amplitude of the density field
as a function of scale. The amplitudes of the Fourier
modes with different
wavenumber $\:\vec{k}$ are uncorrelated: the phases carry all
information about higher moments of the spatial distribution.
The main problem of the PS
of the normalized density fluctuations is that it
is not suitable to describe a
highly irregular systems
and does not measure
the cut-off towards homogeneity, but it is useful to characterize
a system with a well defined
average density and small fluctuation around it.

For fractal distribution Sylos Labini et al.(1995) found that
the amplitude of the PS
is a power law function of the sample depth
so that it has no physical meaning but it
is only the reflex of the sample size.
Moreover the shape of the PS
 is characterized by two  scaling regimes:
the first one, at high wavenumbers, is
related to the fractal dimension of the
distribution in real space.
The second one, that can be seen as a flattening of the PS
for $\: k \rightarrow 0$, arises because of
a finite size effect and it is spurious. See for example
Park et al. (1994) for an analysis of the PS of
real data.
In order to perform an analysis that does
not imply any a priori assumption one should study
the PS  of the density not normalized
to the average, that is characterized,
in the case of fractal distributions,  by a  single
power law behaviour; moreover
 the amplitude does not depend on the sample size.

\subsubsection {Fractal Properties
of visible matter}

At this point we have all the elements
to perform a correlation analysis for the galaxy
and cluster distribution. Of course this has
been performed extensively in the past using
the function $\:\xi (r)$ (Eq.\ref{l16}) and
related concepts. As we have seen, this approach
can be misleading if the distribution is
not really homogeneous within the sample
limits. Since this is clearly not the case,
the usual analysis of correlations is
considered problematic. This situation
has induced various authors to consider
alternative methods of analysis like the power
spectrum, topological methods, void distribution,
percolation analysis and even more.

However, given the distribution of the
previous section, we are now in the position to understand
what is the real reason for the ambiguity of the analysis
with $\:\xi (r)$ as well as the problem of "fair" samples
versus "homogeneous " samples. In our opinion therefore, the most
important statistical information that we should
derive from the catalogues is still the fair correlation
properties. In particular, there is general agreement
that, at relatively small scales, correlations
show a power law behaviour consistent with clustering
and fractal properties.
The crucial question is, therefore, about the
eventual tendency or evidence for homogenization.
Since this is the property we want
to test, it is not wise to assume it
"a priori", as it is usually done
in the $\:\xi (r)$ analysis, but
also in the various "weighting schemes".
In practice the analysis is quite
simple:
one should consider the conditional
density or the average conditional
density and check whether the
power law behaviour ($\:fractal$)
at small scales is followed by
a well defined constant behavior
over appreciable distances, within
the limits of statistical validity
of the sample.
\bigskip

{\it 3.4.3.1. Galaxies and Clusters}
\smallskip

For the {\it CfA 1} redshift survey
(Huchra et al., 1983) we performed this
analysis in 1988 (Coleman et al., 1988) and the
result was quite surprising:
\\
i.)  the limit of statistical validity
of this sample was found to be
\be
\label {l19}
R_{G}\sim 20 Mpc
\ee
that is about the radius of the
maximum sphere that can be contained
in the sample. This means that, contrary
to the common opinion, we are in the
presence of a statistically "fair
sample" up to this distance.

ii.) The analysis of the appropriate
correlation function shows a well
defined power law behavior up to
$\:R_{s}$, without any tendency
towards homogenization (Fig.\ref{f3}).
\begin {figure}
\vspace {8 cm}
\caption{\footnotesize $\:\Gamma(r), \Gamma^*(r)$ and $\:\xi(r)$
plotted as a function of the length scale for the $\:CfA1$ catalog
 limited to $\:v \le 8000 km sec^{-1}$. The dashed line indicates a
reference slope of  1.7. The CfA does not show any real
correlation length within the limit of its statistical
validity but it shows well defined power law correlations
(from Coleman \& Pietronero, 1992). }
\label {f3}
\end {figure}
This means that {\it CfA 1}
essentially contains a portion
of a fractal structure with
dimension
\be
\label {l20}
D \sim 1.4
\ee
This means that
all the warnings about possible
inconsistency
of the $\:\xi (r)$ analysis are actually
justified and that, the well known
"correlation length" $\:r_{0}\sim 5 Mpc$
(Davis \& Peebles, 1983),
is a spurious result that reflects just the size
of the sample and not the real correlation
properties of the system.

The same type of analysis
was later (Coleman \& Pietronero, 1992)
performed also for the {\it Abell}
catalogue of clusters (Abell, 1958; Bahcall, 1988). Also in this
case we have observed fractal correlation up to the effective
size of the sample ($\:R_{s}\sim 80 Mpc$),
with  a value of the fractal dimension similar to that of galaxies
(Fig.\ref {f4}). This leads us now to an important
point that shows, in a very concrete way, how,
looking at the same experimental data with a broader
conceptual scheme, leads to a physical problematic
of a completely different nature.
We have seen in the previous section that for
fractal structures the amplitude of correlations is physically
meaningless, because it reflects just the unit of measures
and the artificial cut-off, corresponding to the size
of the considered sample.
Then we have seen the {\it CfA 1} catalogue
of galaxies just contains a portion of fractal
distributions.
This means that galaxies are clustered
in a self-similar way. By taking as
units some of these clusters, one
obtains the {\it Abell} catalogue
of clusters for which we also have a fractal distribution
with about the same fractal dimension.
Since clusters are made by
galaxies, it is then quite natural to consider that
cluster correlations are just the continuation
of galaxy correlations at larger scales. This
would imply that the correlations
lengths (derived from correlation amplitudes)
$\:r_{0}^{G}\sim 5 Mpc$ for galaxies and
$\:r_{0}^{C}\sim 25 Mpc$ for
clusters are just a finite
fraction of the depth of the galaxy
$\:R_{G}$ and cluster $\:R_{C}$
catalogues.
We find (Coleman \& Pietronero, 1992) in fact that
\be
\label {l21}
\frac{r_{0}^{C}}{r_{0}^{G}}\sim \frac{R_{C}}{R_{G}}
\sim 5
\ee
that provides, together with
the coincidences of the exponents, a strong evidence for the
fact that galaxy and cluster catalogues
correspond to a single fractal distribution
of galaxies, that is
looked at from different points of view in the two
catalogues (Fig.\ref{fg1a}).
\begin {figure}
\vspace {8 cm}
\caption{\footnotesize Ideal catalogue of galaxies and clusters.
The correlation of clusters appears to be the continuation of the
galaxy correlation to larger scales: the points are galaxies and the groups
of galaxies are clusters. The depth of clusters catalogue is deeper
than that of the galaxy catalogues because clusters are
brighter than galaxies. }
\label {fg1a}
\end {figure}
Therefore, the famous problem of the
galaxy cluster mismatch (Bahcall, 1986)
was just due
to the inappropriate statistical
analysis and it is automatically
eliminated by a correct statistical description.
\begin {figure}
\vspace {8 cm}
\caption{\footnotesize  $\:\Gamma(r), \Gamma^*(r)$ and $\:\xi(r)$
plotted as a function of the length scale for the Abell clusters catalog.
 As the $\:CfA1$ galaxies catalog  the sample shows well
defined power law correlations and no correlation length
(from Coleman \& Pietronero, 1992). }
\label {f4}
\end {figure}

This work has generated a debate and it
may be worth  to report some of its elements.
For example, some authors claim (Lemson \& Sanders, 1992;
Provenzale
et al., 1994)
that the use of
weighting schemes allows us to extend the limits
of statistical validity of the samples and, by doing so,
they claim to observe trends towards homogenization
outside our limits for both galaxy and cluster
catalogues.
Our opinion, about this point, is that any weighting
schemes must be based on a model and the basic idea is to
replicate what is missing in the sample, using
the knowledge of what is available.
Unavoidably, this introduces artificial
homogenization effects so, it is
not a surprise that the claimed scales of
homogenization are about $\:35 Mpc$ for
galaxies and $\:150 Mpc$ for clusters.
Both values are just about twice the
limits of statistical validity
of these catalogues and, therefore,
they refer to length scales that are
strongly affected by the weighting
schemes. In addition to these
comments, we are going to see that
deeper catalogues clearly
contradict these claims for
homogeneity.

{}From a theoretical point of view
the mismatch between galaxy and cluster
correlations was interpreted in many ways.
One is the so called {\it biased galaxy
formation} model (Kaiser, 1984).
This model starts from a reasonable
idea, namely a galaxy grows by taking
matter around it. The point is
that its present mathematical description,
in terms of an uncorrelated random process
(Poisson), does not reproduce even
the qualitative features of the power
law behaviour of galaxy and
cluster correlations.
However an argument is made about the amplitude of
correlation between small and large fluctuations
of a Poisson process and it is then
related
to the fictitious amplitude mismatch of power
law correlations.
A related concept, that was also generated
by these fictitious amplitudes, is the
called "luminous segregation effect".
This was actually not even defined
theoretically and it is in fact
impossible to generate this effect as a
continuous distribution of masses.
However it appeared as the continuum
generalization of the
galaxy cluster mismatch. This confused
situation can be now clarified naturally on the
basis of the previous discussion
and the objective of a theory can finally
be identified in a clear and consistent
way.
\bigskip

{\it 3.4.3.2.  The CfA 2 redshift survey}
\smallskip

Recently an extension of the
{\it CfA 1} catalogues up to a magnitude
limit of $\:15.5$, corresponding
to a depth of about twice, has became
available for an appreciable
angular volume.
This survey confirms and extends our previous
results for {\it CfA 1}. The power law (fractal)
correlations observed up to the sample limit
of the {\it CfA 1} are present also up to the
depth of {\it CfA 2} (Park et al., 1994;
da Costa et al., 1994).
This implies that, also in the extended
{\it CfA 2} catalogue, there is no
tendency towards homogenization (Pietronero \& Sylos Labini, 1995).
Unfortunately, this clear trend
in the correlation properties
is misinterpreted with the
statement that, even {\it CfA 2},
is not a "fair sample" (Da Costa et al. 1994),
simply because it is not homogeneous.
The exponent of the correlation
function is lower than in {\it CfA 1}
and it is $\:\gamma \sim 1.1$, so that the fractal
dimension is $\: D = 3-\gamma \sim 1.9$,
somewhat higher than in {\it CfA 1}.
The amplitude of the correlation
function $\:\xi (r)$ is a linear
function of the sample depth up
to $\:130 Mpc$ (Park et al., 1994).
{}From the comparison of the volume
limited subsamples of {\it CfA 2}
with those of {\it CfA 1} with the same
limiting absolute magnitude, it follows that,
the linear growth of the correlation
function amplitude (or the power
spectrum amplitude) in deeper subsamples
cannot be due to a luminosity bias
otherwise it should be constant is those subsamples.
On the contrary, the observed linear dependence
of $\:r_{0}$ with the sample depth can be naturally explained
by the fractal nature of the galaxy distribution in the {\it CfA2}
survey (Pietronero \& Sylos Labini, 1995).
The availability of the {\it CfA2} catalogue
 brings us to another
point that has been until now used in favor of homogeneity.
Some authors (Peebles, 1993)  in fact argued that,
a certain rescaling of the angular
correlations would actually imply homogeneity.
This argument is based on functions that are
the analogues of $\:\xi(r)$ for angular correlations,
so it suffers from the problems we have previously
discussed with the additional one that the
angular correlation corresponds
to complex projections,
that lead to further complications.
In any case, one of the three angular catalogues used in the discussion
of the angular correlations is the Zwicky catalogue,
that corresponds to the angular properties of the {\it CfA2} catalogue.
Now we have the full three dimensional distribution
for the catalogue and we can clearly see
that nothing happens in the correlations at $\:5 Mpc$
and that no tendency towards homogeneity is present in the whole sample.

Concerning the fact that the angular catalogues
are really smoother than the three dimensional ones, this is due to the
complex properties of angular projections that, contrary to
orthogonal projections, mix different length scales and produce an
artificial
effect of homogenization. For example, we have shown
(Dogterom \& Pietronero, 1991; Coleman \& Pietronero, 1992)
that the angular projection
of a fractal structure becomes really homogenous at large angular scales.
This shows that a smooth angular projection does not
imply the same property in the real distribution.
\bigskip

{\it 3.4.3.3. Pencil beams and very deep surveys}
\smallskip

In the past few years there has been a large
interest in tiny but very deep samples (pencil beams).
These samples extend over length scales of the order of
$\:1000 Mpc$, covering a small solid angle in the sky:
they show a strong irregularity of the distribution
of galaxies over their entire length
(Broadhurst et al., 1990; Kirshner et al., 1978).
\begin {figure}
\vspace {8 cm}
\caption{\footnotesize Wedge diagram of the {\it ESP} survey
(from Vettolani et al., 1994).}
\label {f5}
\end {figure}
This is, again, a
confirmation of the absence of homogeneity up to very large distances.
Curiously however, the attention has been focused
mainly on the apparent periodicity of the peaks
in the galaxy density. Apart from the fact that
the peak intensities are extremely different, the
period is very large ($\: \sim 130 Mpc$)
so that the eventual homogeneity, corresponding
to this periodicity, could be achieved above $\:1000 Mpc$.
In addition, new surveys of this type
show less and less evidence for the periodicity,
while they confirm the irregularities (for a
more detailed discussion of pencil beams
data see also Coleman \& Pietronero, 1992).

The accumulation of many pencil beam surveys
has produced the Eso Slice Project (hereafter {\it ESP})
(Vettolani et al., 1994), that we have recently
analyzed. We find that, analyzing
the number-redshift relation, this sample
does not seem to show any
tendency towards homogenization up
to its depth ($\:600-800 Mpc$)
(Pietronero \& Sylos Labini, 1994;
Sylos Labini \& Pietronero, 1995a). The distribution of galaxies
is fractal with dimension $\:D \approx 2$
up to the sample limits. This value is larger
than the value previously found for {\it CfA1}, but it is
in agreement with various other deeper redshift surveys such as
{\it CfA2}
(Park et al., 1994), {\it Peruses-Pisces} (Guzzo et al., 1992)
 and {\it QDOT} (Moore et al., 1994).

\subsubsection{Multifractal properties of luminous matter distribution}

The fractal picture that we have discussed
in the previous sections has encountered an
unreasonable resistance probably because it puts into question
one of the fundamental hypotheses of cosmology like homogeneity.
On the contrary, the concept of
multifractal seems to be accepted in a smoother way and this is
quite curious (Jones et al., 1988; Martinez \& Jones, 1990).
Actually this probably happens for the wrong reasons.
In fact we have the impression that some authors may consider
multifractal as a sort of compromise in which scale invariant
structures at all scales may coexist with homogeneity
(Jones  et al., 1988).
This is actually not the case and multifractals
are in contrast with homogeneity exactly like fractals.
In fact the multifractal picture is a refinement and
generalization of the fractal properties
 (Paladin \& Vulpiani, 1987; Benzi et al., 1984).
 In the simple fractal case
one refers to the properties of  a set of points
and one needs only an exponent.
 In the more
complex case, when the scaling properties
can be different for different regions of the
system, one has to introduce a continuous
 set of fractal indexes to characterize the
 system  (the multifractal spectrum).
One refers to this case with the term "multifractality".
The discussion that we
have presented in the previous section,
was meant to distinguish between homogeneity and scale invariant
properties and, for this purpose, it is perfectly appropriate even if
the galaxy distribution would be multifractal.
In this case the correlation functions we have considered
would correspond to a single exponent of the multifractal spectrum,
but the issue of homogeneity versus scale invariance
(fractal or multifractal) remains exactly the same.

In the previous section  we have established  that
the basic characteristic of the observable galaxy distribution
is that
the two point $\:number-number$ correlation function
is a power law
up to the sample limit for galaxy and cluster distributions:
\be
\label{luc1}
G(r) = <n(r)n(0)> \sim r^{-(3-D)}
\ee
A second important observational feature is the
galaxy mass function:
this function determines the probability of having a mass in the
range between $\:M$ and $\:M+dM$ per unit volume,
and can be described by the Press-Schechter
function that shows a power law behaviour
followed by an exponential cut-off for large masses
(Press \& Schechter, 1974):
\be
\label{luc2}
n(M)dM \sim M^{\delta-2} exp(-(M/M^{*})^{2\delta})dM
\ee
with $\:\delta \sim 0.2$.
Hence the distribution of masses is also characterized
by a power law corresponding to self-similarity of
different nature. These two properties are naturally unified
by the concept of multifractality. This concept
naturally arises if the distribution of matter,
as given by both positions and masses, has self similar properties.
Indeed, masses of different galaxies
can differ by as much as a factor $\:10^{6}$
and it is important to include these mass values
in order to describe
the entire matter distribution
and not just the galaxy positions.

The distribution of visible matter is described, in a certain
sample of depth $\:R_{s}$, by the density function:
\be
 \label{luc3}
\rho(\vec{r}) = \sum_{i=1}^{N} m_{i} \delta(\vec{r}- \vec{r_{i}})
\ee
where $\:m_{i}$ is the mass of the $\:i$-th galaxy.
This distribution corresponds to a measure defined on the set of
points which have the correlation properties described by Eq.\ref{luc1}.
It is possible to define the normalized density function:
\be
\label{luc4}
\mu(\vec{r}) = \sum^{N}_{i=1} \mu_{i} \delta(\vec{r}-\vec{r}_{i})
\ee
with $\:\mu_{i} = m_{i}/M_{T}$ and $\:M_{T} = \sum^{N}_{i=1} m_{i}$.
The quantity $\:\mu(\vec{r})$ is dimensionless.
Suppose that the total volume of the sample
consists of a {\em 3-}dimensional cube of size $\:L$.
We divide this volume into boxes of linear size $\:l$. We label each
 box by the index $\:i$ and construct for each box the function:
\be
\label{luc5}
\mu_{i}(\epsilon)  = \int_{{\it i-th box}} \mu(r)dr
\ee
where $\:\epsilon = l/L$.
In the case of a MF distribution if in the {\em i-}th
box there is a singularity of type $\:\alpha$
then in the limit $\:\epsilon \rightarrow 0$,
the measure goes as:
\be
\label{luc6}
\mu_{i}(\epsilon) \sim \epsilon^{-\alpha (\vec{x})}
\ee
In Eq.\ref{luc6} the  exponent $\:\alpha (\vec{x})$
 (a sort of local fractal dimension) fluctuates widely
with the position $\:\vec{x}$.
For an homogeneous mass distribution, with a uniform
density, $\:\alpha = 3$, while for a simple
fractal with dimension $\:D$, $\:\alpha = D$.
 In general we will found
 several boxes with a measure that scales with the
same exponent $\:\alpha$. These boxes form a fractal subset
 with dimension  $\:f(\alpha)$.
Hence the number of boxes that have a measure that scale
with exponent in the range [$\:\alpha , \alpha + d\alpha$]
vary with $\:\epsilon$ as:
\be
\label{luc7}
N(\alpha, \epsilon)d\alpha \sim \epsilon^{- f(\alpha )} d\alpha
\ee
The spatial density of these singularities in the total volume
of the sample $\:L^{3}$ vary with $\:\epsilon$ as:
\be
\label{luc8}
n(\alpha, \epsilon) d\alpha \sim \epsilon^{3- f(\alpha )} d\alpha
\ee
The function $\:f(\alpha)$ is usually (Paladin \& Vulpiani, 1987)
a single humped
function with the maximum at:
\be
\label{luc9}
max_{\alpha} f(\alpha) = D(0)
\ee
where $\:D(0)$ is the dimension of the support. In the case
of a single fractal, the function $\:f(\alpha)$ is reduced to
a single point:$f(\alpha) = \alpha = D(0)$

A characteristic value of the spectrum is $\:\alpha_{min}$
and the corresponding value of $\:f(\alpha)$.
This formalism is suitable for the analysis of any distribution,
even for a regular (analytic) one, because it is completely
general and without any a priori assumption.
\begin{figure}
\vspace{8cm}
\caption{\footnotesize Example of a multifractal measure:
the (x,y) coordinates
are the spatial ones, while the z axis is the measure. The support of the
measure is itself a fractal.}
\label{mf1}
\end{figure}
The MF implies a strong correlation between spatial
and mass distribution (Fig.\ref{mf1}) so that the number of objects with
mass $\:M$ in the point $\:\vec{r}$
 per unit volume $\:\nu(M,\vec{r})$, is a function
of space and mass and is not separable
in a space density multiplied by a mass function (Binggeli et al., 1988).
This means that  we {\it cannot}
express the number of galaxies $\:\nu(M,x,y,z)$  lying in
volume $\:dV$ at {\it (x,y,z)} with mass
between {\it M} and {\it M + dM} as:
\be
\label{luc11}
\nu(M,x,y,z) dM dV = n(M)D(x,y,z)dM dV
\ee
where {\it D(x,y,z)} is the density of galaxies of any luminosity.
Moreover  we cannot
define a well defined average density, independent from
sample depth as for the simple fractal case.

It can be shown (Sylos Labini \& Pietronero, 1995b)
that the mass function of
a MF distribution, in a well defined volume,
has indeed a Press-Schechter behaviour whose
exponent $\:\delta$ (Eq.\ref{luc2}) can be related to the
properties of  $\:f(\alpha)$.
Moreover the fractal dimension of the support
is $\:D(0)=f(\alpha_{s})=3-\gamma$ (Eq.\ref{luc1}).
Hence with the knowledge of the
whole $\:f(\alpha)$ spectrum one obtains information on the
correlations in space as well as on the mass function.

If one wants to perform the analysis of the mass distribution of galaxies,
obviously one needs to know the density distribution
 $\:\rho(\vec{r})$. Usually an estimate of this quantity
is obtained
assigning to each galaxy a mass proportional to its
luminosity. Clearly this is a crude approximation,
 however a better relation between luminosity and
 mass should not change the MF nature of the mass
 distribution, if it is present in the sample, but only the
parameters of the spectrum. The analysis,
 carried out on $\:CfA 1$ redshift surveys
(Coleman \& Pietronero, 1992), provide unambiguous
evidence for a MF behavior. In Fig.\ref{mf2} is shown the
$\:f(\alpha)$-spectrum derived by Coleman \& Pietronero (1992).
\bef
\vspace{8cm}
\caption{\footnotesize The multifractal spectrum $\:f(\alpha)$ derived
from the $\:CfA1$ redshift survey (Coleman \& Pietronero, 1992).
The strongest singularity in the distribution of visible matter
is characterized by an exponent $\:\alpha_{min}=0.65$
and the corresponding fractal dimension is $\:f(\alpha_{min}) \approx 0$.}
\label{mf2}
\enf
This result has very deep physical implications and
can naturally resolve several puzzling problems, as we shall see
in the following.
\bigskip

{\it 3.4.4.1. Luminosity function}
\smallskip

The differential Luminosity Function (LF), $\:\phi(L) dL$, gives the
probability of finding a galaxy with luminosity
in the range $\:[L,L+dL]$ in the unit volume ($\:Mpc^{-3}$) and it
can be described by
the Schechter luminosity function (Binggeli et al., 1988):
\be
\label{luc12}
\phi (L) =\phi^{*}  (L/L^{*})^{- \alpha} exp(-L/L^{*})
\ee
 where  $\:L^{*}$ is
 the cut-off, $\:\phi^{*}$ is the normalization
constant and $\:\alpha$ is the exponent.
In the literature (see for a review Binggeli et al., 1988)
one finds several methods to
determine the LF for field galaxies and cluster
galaxies.
The standard analysis of the LF is based on the
assumptions that the distribution of galaxies is
homogeneous  and that galaxian
luminosities are not correlated with spatial locations.
These a priori assumptions lead to several problems in the
analysis of the LF.
In fact, due to the strong inhomogeneities present in all the
available redshift samples,
the average density, i.e. the amplitude of the LF,
is not well defined and hence the
classical methods
intend to separate the determination of the shape and the amplitude of the LF.
In particular, the so called {\it inhomogeneity-independent methods}
have been developed with the intent to determine only the shape
of the LF. These avoid the problems due to the presence of strong
inhomogeneities and the sample depth dependence of the amplitude
(for a more detailed discussion see Sylos Labini \& Pietronero 1995b).

If the distribution is a simple fractal, then the amplitude
$\:\phi^{*}$ of the LF
scales as $\:\sim r^{-(3-D)}$ as the average density (Eq.\ref{l5}).
The LF in this case can be
written as:
\be
\label{mz7}
\phi = A r^{-(3-D)} (L/L_{*})^{-\alpha} exp(-L/L^{*})dL/L_{*}
\ee
where $\:A$ is a constant,
and we assume the functional form in $\:L$
as an experimental fact.

If we want to consider the whole luminosity-space distribution
we have to analyze the MF case. In this case the amplitude scales
again as $\:r^{-(3-D)}$, but we can obtain
information also on the functional form
of the LF. In fact
if the whole mass
distribution is MF,
the exponent of the mass function is uniquely determined by
the shape  of
the $\:f(\alpha)$-spectrum.  Qualitatively we can say that
also the exponent of the LF is linked to the multifractal spectrum,
if one transforms masses into luminosities. However from the data
analysis one determines the $\:f(\alpha)$ spectrum of the luminosity
distribution rather than of the mass one, so that one can determine in a
 different way from the classical methods,
 the exponent of the luminosity function
(Sylos Labini \& Pietronero, 1995b)

Moreover if the distribution is MF it can be shown
(Coleman \& Pietronero, 1992) that the brightest luminosity $\:L_{max}$
in a sample is related to its depth $\:R_{s}$ by the relation:
\be
\label{luc13}
L_{max} \approx R_{s} ^ {3-\alpha_{min}}
\ee
Also this relation can be tested in real  redshift surveys,
even if it requires a high statistics because it deals
with strongly fluctuating quantities.
\bigskip

{\it 3.4.4.2. Number counts}
\smallskip

The count-magnitude relation (i.e. the number of galaxies
with apparent magnitude lower than $\:m$ versus $\:m$) is used
to test the uniformity of galaxy distribution
in space at small distances,
and the galaxy luminosity properties in deeper samples.
We show the behaviour of
count-magnitude relation for an homogenous, fractal
and multifractal distribution in the three dimensional Euclidean space.
Indeed at small distances it is possible to neglect the effect of space-time
evolution.

In the case of homogenous distribution the number of
galaxies with magnitude less than $\:m$, from Eq.3.71, goes as:
\be
\label{nbc1}
\log(N(<m)) \sim \alpha m + const
\ee
with $\:\alpha=0.6$. If the distribution is fractal with
dimension $\:D$, it is simple to show that (Peebles 1993):
\be
\label{nbc2}
\alpha = \frac{D}{5}
\ee
This relation generalizes the previous one and, if $\:D=3$,
we readily obtain $\:\alpha=0.6$.
It seems from Eq.\ref{nbc2} that knowing the luminosity properties of
galaxy distribution as expressed by the count-magnitude relation,
it is possible to reconstruct the fractal exponent of the
 space distribution. This is possible only
if there is not any correlation between space locations and luminosities
of galaxies. This is not the case for multifractal distributions.
Indeed in this case it is possible to show
(Sylos Labini \& Pietronero, 1995b, c)
that the exponent $\:\alpha$ is not simply
related to the fractal dimension of the support as in Eq.\ref{nbc2},
but it is a complex function of the whole multifractal spectrum.
In this case the exponent is greater than
the value given by Eq.\ref{nbc2}.

In any case to check the validity of Eq.\ref{nbc2} in real data, one
has to consider {\it volume limited subsamples}, where it is possible
to study the correlation properties without any bias, rather than
magnitude limited subsamples as usually done
(see {\it section 4.1.4.}).
\bigskip

{\it 3.4.4.3. Magnitude-redshift relation}
\smallskip

Taking the Schechter function
(Eq.\ref{luc12}) as the correct LF in
the case of homogenous distribution,
the joint distribution in galaxy redshifts $\:z$
and energy flux densities $\:f$ observed in region of sky with unit
solid angle is (Peebles, 1993)
\be
\label{mz3}
\frac{d^2 N}{dz df} = \int \phi_{*} (L/L_{*})^{-\alpha}
e^{-L/L_{*}} dL/L_{*} r^2 dr
\delta(z-H_{0}r) \delta(f - L/4 \pi r^2)
\ee
This ignores the effect of space curvature.
Integrating Eq.\ref{mz3} one obtains
\be
\label{mz4}
\frac{d^2 N}{dz df} = \frac{4\pi}{L_{*}} \left( \frac{c}{H_{0}} \right)^5
 z^4 \phi(kz^2) \equiv D(z,f)
\ee
where $\:k = 4\pi f c^2 / H_{0}^{2}L_{*}$.
The mean value of the
redshifts of galaxies with given flux $\:f$ is:
\be
\label{mz5}
<z> = \frac{\int z D(z,f) dz }{ \int D(z,f)  dz} = A_{*}f^{-1/2}
\ee
($\: A_{*}$ is constant) or in magnitude ($\: f \sim 10^{-0.4 m}$):
\be
\label{mz6}
<z> = 10^{0.2m + a_{*}}
\ee
where $\:a_{*}$ is a constant.
In the case of a fractal distribution we can use the
form of the LF given by Eq.\ref{mz7}.
Inserting this expression
in Eq.\ref{mz3} one obtains instead of Eq.\ref{mz4}:
\be
\label{mz8}
D(z,f) \equiv \frac{d^2 N}{dz df}  \sim  z^{1+D}
\ee
and instead of Eq.\ref{mz6}
\be
\label{mz9}
<z> = 10^{0.2 m + a_{f}}
\ee
for every value of $\:D$ ($\:a_{f}$ is constant for fixed depth,
but it  depends on the sample depth and it is
related to the fractal dimension - Sylos Labini \& Pietronero, 1995b).
Hence this is the theoretical
expectation in the case of a fractal distribution
in a three dimensional Euclidean space that has to be checked by the data (see
{\it section 4.1.2.}).
The behavior of the $\:<z>-m$ relation has the same exponent in an
homogenous and in a fractal distribution whit the constant
different in the two
cases: it is related to the behavior of the average density.
 \bigskip

{\it 3.4.4.4.  Luminosity segregation}
\smallskip

Dressler (1984), and Einasto \& Einasto (1987) found that
the brightest galaxies lie preferentially in dense
environments, in the core of groups and clusters of galaxies.
The fact that the giant galaxies are "more clustered" than the dwarf
ones has
given rise  to the proposition that larger
objects may correlate up to larger length scales
and that the amplitude of the correlation function is larger for
giants than for dwarfs. This effect actually is
a consequence of multifractality. For example in Fig.\ref{mf1}
one can see that the largest peaks of the distribution are located
in largest clusters. This arises naturally, due to the
self-similarity of matter distribution. Moreover the multifractality
implies that the largest  peak has  a lower fractal dimension than
the smaller ones: a trend is this direction has been
found (Giovanelli et al., 1986)
in the steeper  angular correlation
function for the early type galaxies (elliptical) than
for late type (spirals). We recall that the exponent of the
angular correlation function is not changed by projection
(for a more
detailed discussion
see Sylos Labini \& Pietronero, 1995c).
\bigskip

{\it 3.4.4.5. A simple stochastic model  for the
formation of a MF distribution}
\smallskip

{}From a theoretical point of view one would like to identify
the dynamical processes that lead to such a MF distribution.
In order
to gain some insight into this complex problem
 we have developed
a simple stochastic model (Sylos Labini \& Pietronero, 1994a; 1994b;
Sylos Labini et al.; 1995b)
that includes the basic properties of
the aggregation process and allows us to
pose a variety of interesting questions
concerning the possible dynamical origin of the MF distribution.
The dynamics is characterised by some parameters,
that have a direct physical meaning in terms of cosmological processes.
In this way we can relate the input parameters of the dynamics
to the properties of the final configuration and produce a sort of
phase diagram.

In  our model the formation of structures
proceeds by merging
of smaller objects.
When two particles collide, in order to form
a bound state, they have to dissipate a certain amount of energy.
The basic physical mechanism responsible for energy dissipation
in  collisionless and pressureless dustlike particles,
interacting only via gravitational force, is the
{\em dynamical friction}: a test particle moving
through a cloud of other background particles
undergoes a systematic deceleration effect due to the
gravitational scattering (Chandrasekhar, 1943).
Due to this effect of energy dissipation, the aggregation process
depends on the {\em environment}
in which it takes place, and it is more efficient in
denser region. Hence
when two particles collide, they have a probability $\:P_{a}$
of irreversible aggregation and probability $\:1-P_{a}$
to scatter (Fig.\ref{p1}).
\bef
\vspace{6cm}
\caption{\footnotesize The probability of making an irreversible
aggregation during a collision is greater in denser regions
(a) than in sparser ones (b)}
\label{p1}
\enf
 We find
that the environment dependence of the dynamical friction,
and then of the aggregation probability,
{\em breaks the spatial
symmetry} of the aggregation
process and it is one of the fundamental elements
that can give rise to a fractal (and multifractal) distribution.

This model shows an asymptotic fractal distribution
in a certain range of the dynamical parameters.
The non-linear dynamics leads spontaneously the
self-similar (multifractal) fluctuations
of the asymptotic state, so
that there is not any crucial dependence on initial conditions.
The fractal dimension of the asymptotic state depends only on
the parameters of the non-linear dynamics.
We find therefore that
the necessary ingredients for a dynamics in order to generate a fractal
(multifractal including masses) distribution
are the breaking of the spatial symmetry,
and the Self-Organised nature of the dynamical mechanism.
This is why the formation of a multifractal distribution is not due to
an amplification of the small amplitude initial fluctuations, but
the generation of such complex structures
is intrinsically generated by the
non linear dynamics, that has an
asymptotic critical state.
\bigskip

{\it 3.4.4.6.  Gravitational lensing in a fractal distribution}
\smallskip

The usual approach to the study of  the
gravitational lensing statistics
is based on the assumption that the matter distribution
 is homogenous
(Turner et al., 1984).  Baryshev et al., (1995b)
considered the gravitational lensing effect inside a
fractal distribution of matter showing that there are
large differences between this case and the case of
homogenous distribution of total (visible and dark) matter
(Fig.\ref{lens1f}.).

In the case of point mass lenses the
 differential optical depth is (Baryshev et al., 1995b):
\be
\label{lens1}
d\tau_{g} =2 \pi A  R_{H}^{(D-1)} R_{g}
\frac{\left( x_{S} x_{L}^{(D-2)}
-x_{L}^{(D-1)}\right)}
{x_{S}} dx_{L}
\ee
where the average density scales as (see {\it section 3.4}):
\be
\label{lens2}
<n> = \frac{3A}{4\pi} R^{D-3}
\ee
and $\:D$ is the fractal dimension,
$\:R_{g} = 2GM_{*}/c^{2}$ is the lens gravitational radius,
and $\:x_{S}$ and $\:x_{L}$ are source and lens
dimensionless distances
($\:R_{H}$ is the Hubble radius).
Integrating this equation along
the line of sight to the source
one obtains the total optical depth
of lensing between the observer in the
origin of the coordinate system and the
source at distance $\:R_{s}= x_{s}R_{H}$:
\be
\tau_{g} =2 \pi A R_{H}^{(D-1)} R_{g}
\frac{ x_{S}^{(D-1)}}{D(D-1)}
\ee
The differential optical depth, in the case of isothermal galaxies is :
\be
d\tau_{g} = A \sigma_{0} R_{H}^{D}
\frac{( x_{S}-x_{L})^{2}
x_{L}^{D}}{x_{S}^{2}} dx_{L}
\ee
where $\:\sigma_{0}$ depends on the  profile used (Turner et al., 1984)
for modeling the isothermal galaxy.
As in the previous case we calculate the
total optical depth of lensing:
\be
\tau_{g} = A R_{H}^{D} \sigma_{0}  x_{S}^{D}
\frac{2}{D(D+1)(D+2)}
\ee

Hence, in principle, it is possible to distinguish between the
homogeneous and the fractal case, even if the currently available data
are still scarce. Hopefully, in the next years
there will be new data on gravitational lenses so that, in principle,
it will be possible to obtain information, using quasars,
essentially over the
entire causally connected universe.
\bef
\vspace{11cm}
\caption{\footnotesize $\:(a)$ the differential optical depth, normalized
to the total one, for the homogenous
(dotted line) and fractal ($\:D=2$, solid line) case for point masses lens,
 versus the distance of the lens $\:x_{L}$: the source is at $\:z_{s}=0.3$.
$\:(b)$: the total optical depth for the homogenous
(dotted line) and fractal ($\:D=2$, solid line) case for point masses lens,
versus the normalized distance of the source.
$\:(c)$ The same of $\:(a)$ but  for the case of isothermal galaxies.
$\:(d)$ The same of $\:(b)$ but  for the case of isothermal galaxies.
The difference between the fractal and the homogeneous case is
appreciable (from Baryshev et al., 1995b).}
\label{lens1f}
\enf

\subsubsection{Newtonian fractal models}

The idea of modeling the Universe as a hierarchical
structure is an old concept, that now
can be discussed from the point of view of
Fractals. The concept of fractal or self-similarity
has behind it the same scaling idea of the old
hierarchical clustering.
The first person who introduced this idea
was Fournier D'Albe (1907). Charlier
(1908,  1922) applied this idea to a
Universe model in order to explain
Olber's paradox:
in fact (see {\em section 3.1.2.}), if the fractal dimension
is less than two, this paradox is
naturally avoided without introducing the redshift.
We have discussed in {\it section 3.1.2.} that
if the fractal dimension is one,
the gravitational paradox, in the Newtonian
theory of gravitation  also is resolved, as
the virial velocity inside the cluster
does not depend upon $\:r$.

G. de Vaucouleurs (1970) proposed a hierarchical cosmology
as a way of explaining his finding
that galaxies seem to follow an average density power
law with negative slope.  Haggerty \& Wertz (1971, 1972),
developed further the Newtonian hierarchical models
and proposed some possible observational tests. The hierarchical
model proposed by Wertz (1970) was conceived
at a time when fractal ideas had not yet appeared,
 but he assumed
the De Vaucouleurs' density
power law as the fundamental
empirical fact to be taken into account
in order to develop a hierarchical cosmology. He described a
{\it "polka dot model"},
that is a structure in discrete
hierarchy, where the elements of a clusters are all
of the same mass and are distributed regularly
in such a way that we can call now deterministic
fractal structure. He showed that regular hierarchic models
are fully defined by a single parameter - the fractal dimension.

\subsubsection{General Relativity  Fractal Models }

The standard approach to cosmology using General Relativity
(GR) assumes
that a well defined mean density exists in the observable Universe.
We have discussed in {\it section 3.4.3.} that it is
 clear that the large-scale structure
of the Universe does not show itself as a smooth and homogenous distribution
of luminous matter, but it has a well defined fractal nature.
In other words the mean
density scales according to de Vaucouleurs' formula (1970):
\be
\label{111}
<\rho> \sim r^{-(3-D)}
\ee
The exponent of this law has a fundamental physical meaning
and needs to be explained by the theory. The general solution
of Einstein's equation for spherically symmetric dust in
comoving coordinates is the Tolman solution (Tolman, 1934).
 This solution is
spherical symmetric about {\em one} point, so that
a cosmological model based
on this solution does  not consider the equivalence of
all points in the Universe.
For this reason Ribeiro (1992a, 1992b, 1993) constructs a model,
called the {\it Swiss cheese model},
with an
interior solution provided by the Tolman metric
surrounded by a Friedmann space-time:
with this scheme it is possible to
 save the equivalence of all points
in the Universe. Such a model needs to solve the junction condition
between the two metrics.

The Tolman metric with $\:\Lambda=0$ and $\:G=c=1$ may be written as:
\be
\label{112}
ds^{2} =  dt^{2} - \frac{R'^{2}}{f^{2}} dr^{2} -R^{2}d\Omega^{2}
\ee
with $\:r \ge 0$ and $\:R(r,t) \ge 0$, where $\:d\Omega^{2}$ is the
differential solid angle and $\:f(r)$ is an arbitrary function.
Einstein's field equations for this metric reduce to a single equation:
\be
\label{113}
2 R \frac{\partial R}{\partial t} + 2 R(1-f^{2}) = F
\ee
where $\:F(r)$ is another arbitrary function.
There are three classes of
Tolman solutions ($\:f^{2}=1$, parabolic ; $\:f^{2} >1$
hyperbolic and  $\:f^{2}<1$ elliptic)
that correspond respectively to flat, open
and closed Friedmann models. In order to obtain  fractal solutions
in the Tolman model, Ribeiro had performed some numerical calculations,
and the criteria for choosing and accepting the solutions are:
the linearity of the distance-redshift relation for $\:z<1$ with the
Hubble constant in the range $\:40-100 km sec^{-1}Mpc^{-1}$,
the constraints of the fractal dimension
to be in the range $\:[1,2]$
and the obedience to the de
Vaucouleurs' density power law $\:<\rho>_{v}
\sim d_{l}^{-\gamma}$, where $\:d_{l}$ is the
luminosity distance.
The reason for choosing this distance is that
this is an observable quantity.

Ribeiro found fractal behaviour in hyperbolic models
choosing appropriately the arbitrary
functions of the Tolman model. In this model
$\:H_{0} \sim 80 km sec^{-1}Mpc^{-1}$ and $\:D=1.4$ up to
$\:d_{l} \sim 270 Mpc$.  It seems that
this solution is highly dependent on the
parameters used, whose change can
 produce a qualitative change in its behaviour.
If  a Friedmann metric is joined to the Tolman-fractal solution  it is
found that $\:\Omega_{0} \sim 0.002$.
It is interesting that it is not a good
modeling of a possible crossover to
homogeneity in a fractal system, because Ribeiro
showed that the Friedmann metric looks
very inhomogeneous at larger scales
when measured along the past light cone:
this implies that this external solution is not really mandatory.
Clearly this is a simple model but it represents a first attempt
to model a fractal Universe within the GR framework.
In particular it is found that the idea of a vanishing global density
is compatible with the GR
solution. In fact the average density is made not
at a space-like hypersurface of constant time, but the average is
calculated along the backward null cone, and in this case
the observable density can go to zero.
This model provides a description
for the possible fractal structure, but it does not
provide an answer to the question of the origin of such structure.

\subsubsection{Field Fractal Models}

As a possible way to construct a new type of
cosmological model we consider here the so-called
Field Fractal model based on a relativistic
tensor field ($\:\psi^{ik}$) gravitation theory in flat
Minkowski space-time ($\:\eta^{ik}$).
The widespread opinion on the full coincidence
of Tensor Field Theory (TFT)
and General Relativity (GR)
(e.g. Misner, Thorn \& Wheeler,  1973)  is
based on the assumption of the uniqueness of the
energy-momentum tensor (EMT) of the gravitational
field, which was used for the iteration procedure.
However, in the framework of Lagrangian formalism
of relativistic field theory, the EMT of any field
is not defined uniquely, but can be transformed
(e.g. Landau \& Lifshiz, 1973; Bogolubov \& Shirkov, 1976):
\be
\label {ffm1}
T^{ik} \Rightarrow T^{ik}+ \Phi^{ikl}_{, l}
\ee
for $\:\Phi^{ikl}= - \Phi^{ilk}$, and one needs additional physical
restrictions to choose the final form of the EMT;
for example such conditions as  a positive energy,
a EMT with null trace, or some symmetries. It is apparent that different
EMTs would lead to different nonlinear
theories of gravitation.

To illustrate the above discussion let us consider the nonlinear
generalization of Poisson's equation for the case of
 distributed source, describing the negative
and the positive energy density of the gravitational field:
\be
\label{ffm2}
\Delta \varphi = - \frac{(\nabla \varphi)^{2}}{c^{2}}
\ee
and
\be
\label{ffm3}
\Delta \varphi = + \frac{( \nabla \varphi)^{2}}{c^{2}}
\ee
The solution of Eq.\ref{ffm2} is:
\be
\label{ffm4}
\varphi = c^{2} \ln \left(1 - \frac{GM}{rc^{2}}\right)
\ee
\be
\label{ffm5}
\frac{d\varphi}{dr} = \frac{GM}{r^{2}(1 - GM/rc^{2})}
\ee
whereas the solution of Eq.\ref{ffm3} is:
\be
\label{ffm6}
\varphi = - c^{2} \ln \left(1 + \frac{GM}{rc^{2}}\right)
\ee
\be
\label{ffm7}
\frac{d\varphi}{dr} = \frac{GM}{r^{2}(1 + GM/rc^{2})}
\ee
 From Eq.\ref{ffm4} - \ref{ffm7} we see two possible ways to
construct the nonlinear TFT. The first one, based
on the negative energy density of the gravitational field,
leads to the infinite gravity force at the finite distance
$\:R_{g} = GM/c^{2}$. The second way, based on positive energy,
gives TFT without singularity.

In GR  the energy density
of
the gravitational field is a poorly defined concept as a consequence
of the geometrical interpretation of gravity (pseudotensor
character of the gravitational EMT)
For example, Landau-Lifshiz (1973) pseudotensor gives
$\:T^{00} =  - 7(\nabla \varphi_{N})^{2}/8 \pi G$ but
the Grischuk-Petrov Popova (1984) gives
$\:T^{00} =  - 11(\nabla \varphi_{N})^{2}/8 \pi G$
for the spherically symmetric static (hereafter SSS)
weak field in harmonic
coordinates.

In the TFT approach there is a real tensor quantity for the energy
density of the gravitational field. For example the
canonical EMT gives
$\:T^{00} =  + (\nabla \varphi_{N})^{2}/8 \pi G$
for an SSS weak field and it corresponds to a
quantum description of the gravitational field as an
aggregate of gravitons in flat space-time.
Gravitons are massless particles, i.e. some kind of
matter in space-time, which carry positive
energy and momentum in the space-time. Besides the sum of
two tensor $\:\psi^{ik} + \eta^{ik} =  g^{ik}$ is not a metric
tensor because the covariant components of this tensor
$\:\psi_{ik} + \eta_{ik} =  g_{ik}$ provide the mixed
components $\:\psi^{i}_{k} + \eta^{i}_{k} =  g^{i}_{k} \ne \delta^{i}_{k}$
and the trace $\:g^{ik}g_{ik} = 4 +2 \psi + O(\psi^{2}_{ik}) \ne 4$.
It means that the TFT is a scalar-tensor theory,
and not a pure tensor one, and hence it includes spin 2
and spin 0 gravitons.

Hence the first hypothesis in this approach is
to choose the TFT as the gravitation theory.
Let us consider the relativistic symmetric
field $\:\psi^{ik}$ in Minkowski's space-time
$\:\eta^{ik}$. As the really observed gravitational fields
are the weak ones ($\:|\varphi| \ll c^{2}$), it
is natural to begin the construction of the TFT for the weak
field case. In this case we have a very close analogy with the
electromagnetic field and can use the standard Lagrangian formalism
of the relativistic field theory (below
we utilize the notation of the Landau-Lifshiz textbook).
We begin with the action integral in the form:
\be
\label{ffm8}
S = S_{(g)} +  S_{(int)} + S_{(p)} = \frac{1}{c}
\int \left(\Lambda_{(g)} + \Lambda_{(int)} +
 \Lambda_{(p)} \right) d\Omega
\ee
where{\it (g), (int)} and {\it (p)} indicate respectively the
gravitational field, the interaction and the particles parts
of the actions and Lagrangians. The Lagrangians
are given by the following expressions:
\be
\label{ffm9}
\Lambda_{(g)} =  - \frac{1}{16 \pi G} \left( 2 \psi_{nm}^{,n}
\psi^{lm}_{,l} - \psi_{lm,n} \psi^{lm,n} +
2 \psi_{ln}^{,l}\psi^{,n} + \psi_{,l} \psi^{,l}\right)
\ee

\be
\label{ffm10}
\Lambda_{(int)} =  - \frac{1}{c^{2}} \psi_{lm} T_{(p)}^{lm}
\ee

\be
\label{ffm11}
\Lambda_{(p)} =  -  \eta_{ik} T^{ik}_{(p)}
\ee
It has been shown by Kalman (1961) and Thirring (1961) that the total
EMT  of the system consists of three parts, which correspond
to those of the action integral in Eq.\ref{ffm8}:
\be
\label{ffm12}
T _{(\Sigma)}^{ik}= T_{(g)}^{ik} +  T_{(int)}^{ik} + T_{(p)}^{ik}
\ee
where the canonical EMT of the gravitational field for
Lagrangian Eq.\ref{ffm9} has the form:
\be
\label{ffm13}
T_{(g)}^{ik} =  \frac{1}{8 \pi G} \left( ( \psi^{lm,i} \psi_{lm}^{,k}
- \frac{1}{2} \eta^{ik} \psi_{lm,n} \psi^{lm,n}) - \frac{1}{2}
( \psi^{,i}\psi^{,k} -\frac{1}{2} \eta^{ik}\psi_{,l} \psi^{,l})  \right)
\ee
the interaction EMT is
\be
\label{ffm14}
T_{(int)}^{ik} =  \frac{2}{c^{2}}T_{(p) l}^{i}  \psi^{lk} -
\frac{1}{c^{2}} T^{ik}_{(p)} \psi_{lm} u^{l} u^{m}
\ee
the point particles EMT is
\be
\label{ffm15}
T_{(p)}^{ik} = \sum_{a} m_{a}c^{2}\delta(\vec{r} - \vec{r_{a}})
\left(1 - \frac{v_{a}^{2}}{c^{2}}\right)^{1/2} u^{i}_{a} u^{k}_{a}
\ee
Therefore the nonlinear EMT must include the interaction Lagrangian
in the form:
\be
\label{ffm16}
\Lambda_{(int)} = - \frac{1}{c^{2}} \psi_{lm}T^{lm}_{(\Sigma)}
\ee
The weak field condition allows us to use the linear approximation
as the first step and then to make nonlinear correction (Post-Newtonian
- hereafter PN- TFT).
The variation of the gravitational potentials in the action integral
Eq.\ref{ffm8}, where for fixed sources in the
PN approximation we can use the interaction Lagrangian in Eq.\ref{ffm16},
yield the PN field equation in the form:
\be
\label{ffm17}
- \psi^{ik,l}_{l} + \psi^{il,k}_{l} +\psi^{kl,i}_{l} - \psi^{,ik}
-\eta^{ik} \psi^{lm}_{,lm} + \eta^{ik}\psi^{,l}_{,l}  =
\frac{8 \pi G}{c^{2}} T_{(\Sigma)}^{ik}
\ee
Eq.\ref{ffm17} automatically requires the conservation of
the total EMT and leads to the equation of motion for particles in
the form $\:T^{ik}_{(\Sigma),k}=0$.
The field equations are invariant
(for fixed sources) under the gauge transformation
$\:\psi^{ik} \Rightarrow \psi^{ik} + \theta^{i,k}+ \theta^{k,i}$ and one
achieves the Hilbert gauge in the form
 $\:\psi^{ik}_{,k} = \frac{1}{2} \psi^{,i}$.
In this case the field equation (Eq.\ref{ffm17}) becomes:
\be
\label{ffm18}
\Box  \psi^{ik} =  \frac{8 \pi G}{c^{2}} \left(
T^{ik}_{(\Sigma)} - \frac{1}{2} \eta^{ik}T_{(\Sigma)}\right)
\ee

In the case of an SSS weak field the
first approximation EMT has the very simple form
$\:T^{ik}_{(\Sigma)} = diag (\rho_{0}c^{2},0,0,0)$
and the solution of  Eq.\ref{ffm18}
is the Birkhoff's potential:
\be
\label{ffm19}
\psi^{ik} = \varphi_{N} diag(1,1,1,1)
\ee
where $\:\varphi_{N}$ is the Newtonian potential, i.e.
$\:\varphi_{N} = - GM/r$
outside the gravitating body.

Using the SSS solution Eq.\ref{ffm19} and the corresponding EMT expression
Eq.\ref{ffm13}, \ref{ffm14}, \ref{ffm15} we find the total energy
 densities of the system from Eq.\ref{ffm12}:
\be
\label{ffm20}
T^{00}_{(\Sigma)} =  T^{00}_{(p)} + T^{00}_{(int)} + T^{00}_{(g)} =
\left( \rho_{0} c^{2} + e \right) + \rho_{0}\varphi_{N} +
\frac{1}{8 \pi G} (\nabla \varphi_{N})^{2}
\ee
where $\:(\rho_{0}c^{2} + e)$
is the rest mass and the kinetic energy density,
$\:\rho_{0}$  is the interaction energy density,
$\:\nabla \varphi_{N}^{2} /8 \pi G$
is the energy density of the gravitational field.
The total energy of the system will be:
\be
\label{ffm21}
E_{(\Sigma)} = \int T^{00}_{(\Sigma)} dV =  E_{(0)} + E_{(k)} + E_{(p)}
\ee
where $\:E_{(0)} = \int \rho_{0} c^{2} dV $ is the rest-mass energy,
$\:E_{k} = \int (e) dV$ is the kinetic energy, and $\:E_{p}$ is the
classical potential energy that equals the sum of the interaction
and the gravitational field energy:
\be
\label{ffm22}
E_{p} = E_{(int)} + E_{(g)} = \int
\left( \rho_{0}\varphi_{N} + \frac{1}{8 \pi G}
(\nabla \varphi_{N})^{2} \right) dV =
\frac{1}{2} \int \rho_{0} \varphi_{N} dV
\ee

Up to now TFT has been developed only for weak field approximation because
there is no exact expression for the total EMT (Eq. \ref{ffm12})
in the case of the  strong
field. But in TFT there is the  possibility  to  consider  the
case of weak force
 ($\: \nabla \varphi \rightarrow 0$) while $\:\varphi
\rightarrow c^{2}/2 $.
It is just  what  we
have for the
cosmological problem. In this  case  we  can  get  some  qualitative
results from post Newtonian TFT.
 Let us consider the case of a static homogeneous ($\:\rho=const$)
dust-like
cold matter ($\:p=0,e=0$) distribution within infinite space.
Inserting Eq.\ref{ffm20} in
the field equation Eq.\ref{ffm18} and taking into account the traceless
of  the  field
and interaction EMTs, we get the equation for the $\:\phi =\psi^{ 00}$
component  in the form
\be
\label{y26}
\Delta \varphi = 4\pi G\left(\rho + 2\rho \frac {\varphi}{c^{2}}+
\frac {2(\nabla \varphi)^{2}}{8 \pi G c^{2}}\right)
\ee
In our case the main terms in the  right-hand  side  of  Eq.\ref {y26}
are  the
positive rest mass density $\rho$ and the  negative  interaction
mass  density
$\:(2\rho\varphi)/ c^{2}$, because for $\:\varphi \rightarrow const$
its gradient
$\:\nabla \varphi \rightarrow 0$. Hence  we
can consider the following simple equation
\be
\label{y27}
\Delta \varphi - \frac {8 \pi G \rho \varphi}{c^{2}} = 4 \pi G \rho
\ee
Eq.\ref{y27} is equivalent to Einstein's equation Eq.3.7:
\be
\label{y28}
\Delta \varphi-\lambda \varphi = 4 \pi G \rho_{0}
\ee
Comparing Eq.\ref{y27} and Eq.\ref{y28} we conclude that $\:\Lambda-term$
in TFT is
\be
\label {y29}
\lambda= \frac {8 \pi G \rho}{c^{2}}
\ee
and it corresponds to the  interaction  EMT.
  According  to  Einstein (1917)  the
cosmological solution of Eq.\ref{y28} with Eq. \ref{y29} will be
\be
\label {y30}
\varphi=- \frac {4 \pi G \rho}{\lambda} = -\frac{c^{2}}{2}
\ee
 The cosmological solution Eq.\ref{y30} can be derived  also  as  a
limiting  case
($\:R \rightarrow \infty $) of the exact solution  of  Eq.\ref{y27}
for a finite  ball  radius  $\:R$  (see
Baryshev \& Kovalevskii, 1990).
So  within   TFT   there   is a natural static
cosmological  solution.  Note  that  this  solution  is  true  for  any  mass
distribution including a fractal
 one. For sufficiently small distances, we  have
a quasi Newtonian
 behavior of the gravitational potential, due to mass  density
fluctuations.

In the case of static Universe the cosmological redshift could be  due  to
the global gravitational effect (see {\it section 2.2.3}).
 For  having
a linear  redshift-distance relation the  fractal  matter  distribution needs
to have fractal
dimension $\:D=2$ (Baryshev, 1981, 1994). As a first qualitative step
to understand
a new possibility in
the cosmological solution, let us consider  a  generalization
of the Baryshev \& Kovalevskii (1990) expression for  the
total  gravitational  mass
inside the homogeneous ball of radius $\:R$ to the following form in
the case  of a
fractal distribution with $\:D=2$:
\be
\label {y31}
M(R) = M_{H} x \left(1-\frac {1-e^{-4x}}{4x}\right)
\ee
where $\:x = R / R_{H}$ is the metric distance $\:R$
in the units  of  the  Hubble  radius
$\:R_{H} = (c^{2} \pi G \rho_{0} R_{0})/2$, $\:\rho_{0}$ and $\:R_{0}$
are the  lower  cutoff  of  the
fractal  structure  (mass  density  and  radius  of  average   galaxy),   and
$\:M_{H} = (c^{2}R_{H})/2G$ is the Hubble mass.
   From    Eq.\ref{y31}    it    follows    for     small
distances     ($\:x \ll 1$)
$\:M(R) = 2 \pi\rho_{0} R_{0} R^{2}$ and for large ones ($\:x \gg 1$)
$\:M(R) = (c^{2}R)/(2G)$.
By combination of equation for mass (Eq.\ref{y31}) and the equation  for
gravitational redshift (Eq.2.8) one can find the
general redshift-metric
distance relation
\be
\label {y32}
{\cal L}(z) = \frac {l}{R_{H}}= \log(1+z)
\ee
where $\:l = R$.
So for small $\:z$ from Eq.\ref{y32}
we get the linear Hubble law
\be
\label {y34}
z = \frac {H_{g}}{c}l
\ee
where $\:H_{g}$ may be called the gravitational Hubble constant
\be
\label {y35}
H_{g} = \frac {2 \pi G \rho_{0} R_{0}}{c} = 69 km s^{-1} Mpc^{-1}
\ee
and the value corresponds to mass density
$\rho_{0}=5.2\cdot 10^{-24}g/cm^{3}$
and radius $R_{0}=10 kpc$.
   In the case of a
gravitational cosmological redshift  the  angular  and  the
bolometric distances are given by Eq.\ref{50}.
Hence the angular size-redshift
relation has the form
\be
\label {y36}
\Theta (z) = \left( \frac{d}{R_{H}}\right) \frac{(1+z)}{\log(1+z)}
\ee
and the magnitude-redshift relation is
\be
\label {y37}
m_{bol}(z) = 5 \log \left( \frac{\log(1+z)} {(1+z)}\right) + const
\ee
Hence we have shown  that within Field Fractal Models one can construct
definite theoretical predictions
but we emphasize  that  many  questions  are
still open and need further development.

\subsection{Other possibilities}

   Besides Steady State and Fractal models there are several
other cosmological theories proposed as alternatives to the
standard hot big bang model (see e.g. the review of Narlikar, 1987).
The motivation for exploring alternatives comes from the existence
of paradoxes in the standard model and from the necessity of analysis
of possible initial hypotheses (see Fig.1).
   The Tired Light (TL) cosmological model was first suggested
by Zwicky (1929) and it is based on the idea that the cosmological
redshift is caused by some unknown physical process in which
traveling photons continuously undergo an energy depletion or
aging effect. So the cosmological redshift is given by Eq.2.15.
Hubble and Tolman (1935) proposed several observational tests to
permit a decision between recessional and tired-light causes for
the cosmological redshift. These are the angular size-magnitude,
the surface brightness-redshift and the number counts-magnitude
relations. Later, many attempts have been made to
construct a physical mechanism for the photon energy loss and to
compare available observational data with the predictions of the
TL model (see e.g. Jaakkala et al.,1979; La Violette,1986; Vigier, 1988).
They
claimed that the TL model gives a good fit to
observations.  However, Sandage \& Perelmuter (1991) have
shown that the surface brightness of the giant elliptical
galaxies have $\:(1+z)^{-4}$ behavior while for the TL model
the $\:(1+z)^{-1}$ dependence was predicted. Nevertheless within
the scope of the TL model one can introduce an evolution of the
surface brightness to explain this observation. So the question
about validity of the TL cosmology is still open
 and  needs
further crucial tests.

The idea about perfect symmetry between matter and antimatter
in the Universe was proposed by Alfven and Klein
(Alfven, 1989). The baryon symmetrical plasma was supposed to have
been separated by hydromagnetic processes. For the modern
version of the model see Stecker (1982). Future tests of the
theory may come from observations of the cosmic neutrino
background by underwater detectors.
   Note also such ideas as variable physical constant
cosmologies (Dirac, 1937; Brans \& Dicke, 1961; Troitskii, 1987),
variable mass cosmology (Narlikar \& Arp, 1993),
chronometric cosmology (Segal, 1976).
   All these alternative cosmologies stimulate observational
checking of possible theories and need construction of the
crucial tests.

\section{CONFRONTATION OF WORLD MODELS WITH OBSERVATIONS}

In this section we consider
the data that are currently available
on classical and post-classical cosmological tests.
For a more detailed discussion we refer the reader
to more specific reviews such as Sandage (1987, 1988),
Yoshii \& Takahara (1988), Arp et al. (1990), Peebles et al. (1991),
Coleman \& Pietronero (1992), Coles \& Ellis (1994),
Efstathiou (1994).
According to our  classification (see {\em section 2})
we shall divide cosmological tests
into two groups: parametric and crucial.

\subsection{Parametric cosmological tests}
Parametric tests permit the determinations of the parameters of
each model. All these tests together can put strong constraints
on the various parameters of a model so that as a whole these can be
considered as a crucial test.
We consider in detail classical parametric tests
discussing the most recent observational data. Moreover we consider the
post-classical parametric tests that, in the last few years, have been
strongly developed from observations.

\subsubsection{Redshift-distance relation and the Hubble constant}

The first main observational result is the
linearity of the Hubble law at small distances (Fig.\ref{dz2}),
 deep inside
the inhomogeneity cell, in the range between
$\:2 h_{50}^{-1}Mpc$ up to $\:80 h_{50}^{-1}Mpc$
(Sandage, 1986, 1991; Peebles, 1993).  The uncertainty in the
value of Hubble's constant does not affect the test of
the linearity of the
redshift-distance relation. In Fig.\ref{dz1}
Hubble's law is shown
for clusters (Mould et al., 1991) and in Fig.\ref{dz2}
for individual galaxies (Peebles, 1988b).
There are some evidences
of large fluctuation in the linear $\:z-d$ law in the
Virgo and Great Attractor direction  (Lynden Bell et al., 1998;
Teerikorpi et al., 1992) so that
further and more detailed studies are needed: the filled circles
in Fig.\ref{dz1} represent clusters in the neighborhood
of the Great Attractor.

\bef
\vspace{8cm}
\caption{\footnotesize Hubble's law for nearby galaxies
(from Peebles 1988b). The
distance are determined using the infrared Tully-Fisher method:
$\:d= cz/H_{0}$.}
\label{dz2}
\enf

\bef
\vspace{8cm}
\caption{\footnotesize Hubble's law for clusters. The distance is determined
using the Tully-Fisher relation (from Mould et al. 1991).}
\label{dz1}
\enf

The second important
observational result is the clear tendency
to the higher value of the Hubble constant. Recent observations
converge to the value of
$\:H_{0} \approx 80 km sec^{-1} Mpc^{-1}$.
Pierce et al. (1994) found a value of  $\:87 \pm 7 km sec^{-1} Mpc^{-1}$
with the observations of three Cepheid variables in a galaxy of the Virgo
Cluster;  Schmidt et al. (1994)
found $\:\sim 75 km sec^{-1} Mpc^{-1}$ with the type II supernovae.
Moreover Freedman et al. (1994) found
$\:H_{0} = 80 \pm 17 km sec^{-1}Mpc^{-1}$
with the Cepheids  method.
However, according to Sandage (1994) the value
of $\:H_{0}$ is about $\:40-50 km sec^{-1}Mpc^{-1}$.

Hence all cosmological models
need to incorporate the linear
$\:z-d$ relation for small distance with,
 probably, a high value of $\:H_{0}$.
The observed linear
$\:z-d$ relation is well fitted by the standard
Friedmann, Steady State, Tired light models.
But problems arise for  SM , SSM and TL models
when one attempts to take into account the strong
fractal-like inhomogeneities at small distances
where the $\:z-d$ relation is linear.

Also there is  a conflict
between the large
ages of globular clusters and distant radio
galaxies in comparison with the age
of the Universe that comes out in the Friedmann
models (Dunlop et al., 1988; Chambers et al., 1990).

\subsubsection{Angular size versus redshift relation}
After Hoyle's (1959) suggestion
of testing cosmological models by the angular size-redshift
relation there were a lot of papers devoted to this topic.
The main problem of such type of  test is the
identification of the {\it "standard stick"}
for which one knows reliably its
non-variation with redshift, or its variation with
a well known evolutionary scheme.

Up to now in optical waveband there are
data only for relative  small redshifts ($\:z \le  1$).
In Fig.\ref{tz1} (from Djorgovski \& Spinrad, 1981)
one sees how the observed optical
sizes of giant elliptical galaxies are fitted
by Standard Friedmann, Steady State and non standard model
with $\:\Lambda$-term models.
\bef
\vspace{8cm}
\caption{\footnotesize Optical angular sizes
 of giant galaxies versus redshift
(from Djorgovski \& Spinrad 1981).}
\label{tz1}
\enf
In radio wavelengths the situation
is in more doubt, because there is not
 a good standard stick such as the optical galaxy size
(which is stable for a sufficiently long time). If one
considers classical double radio sources are obtained
the results shown in Fig.\ref{tz2} (from Kapahi, 1987).
The recent discussion on double sources in the
angular size test has been given by Nilsson et al. (1993)
\bef
\vspace{8cm}
\caption{\footnotesize Double radio sources angular size  versus redshift
(from Kapahi, 1987). }
\label{tz2}
\enf
However for compact VLBI structures
Kellerman (1993) found a deviation
from the Euclidean behaviour
of the angular size-redshift relation (Fig.\ref{tz3}).
The compact VLBI radio source is
produced by relativistic
jets, ejected from the central energy
machine of  the quasars, and it is unlikely that such
jets have a well defined linear size for
all sources.
\bef
\vspace{8cm}
\caption{\footnotesize $\Theta-z$ relation for
VLBI structures (from Kellerman, 1993).}
\label{tz3}
\enf
This is why we need more observational information about the
$\:\Theta-z$ relation. The most promising
could be joint radio and optical
observations of a well defined sample of
radiogalaxies.

\subsubsection{Apparent magnitude-redshift relation }

The slope of the magnitude-redshift ($\:<m>-z$)
 relation was found to be $\:5$ within limits of the error
(Sandage, 1988 for a review) (see Fig.\ref{mz1}a,b). However this behaviour
does not permit to discriminate between
alternative models and in particular between
fractal
and homogeneous distributions, as they have the same slope
(Eq.\ref{mz6} and Eq.\ref{mz9}). A possible test in this direction
can be the $\:<m>-z$ diagram at largest redshifts.
The SM in fact predicts a correction to the $\:5\log(cz)$
law of first order in $\:z$ (Eq.\ref{86}). Such a
measurement in principle permits the determination of $\:q_{0}$.
\bef
\vspace{9cm}
\caption{\footnotesize $\:<m>-z$ relation
in (a) the B band, and in (b) the K band.
Thick and thin lines represent the
models for evolving and nonevolving
galaxies. Number beside each thin line is the
value of $\:q_{0}$ and numbers in parentheses
are the values of $\:q_{0}$ and $\:z_{F}$ (from Yoshii \&
Takahara, 1988).}
\label{mz1}
\enf
 In this case the problem is in estimating
the evolutionary corrections for stellar evolution theory
but no definitive answers have been obtained in the visible
band (Sandage, 1988). In the near-IR there are
several observations at large redshift: in this band the evolutionary
corrections are predicted to be smaller than in the visible band.
A summary of the K-data is given by Yoshii \& Takahara (1988).
Even in this case, as in the case of number counts,
 the situation is problematic
due to the unknown evolutionary model that has to be taken into account.
Clearly the argument is circular
if we use the standard model predictions to prove that evolution
has occurred without having a priori the proof
that the standard model is correct.

\subsubsection{Galaxy counts}

The usual analysis of the galaxy counts
is performed in magnitude limited subsamples. In Fig.\ref{nc2a}
the whole behaviour of $\:\log(N(<m)) \approx \alpha \cdot m$
in the B-band is shown .
\bef
\vspace{9.5cm}
\caption{ \footnotesize Differential
galaxy number counts in the B-band.
The symbols represent the count data
 taken from various surveys.
 For $\:12 \ltapprox m \ltapprox 18$
the exponent is $\:\alpha \approx 0.6$, while in the range
$\:m > 19$ $\:\alpha \approx 0.45$. This exponent has been
established in magnitude limited samples
(from Yoshii, 1993).}
\label{nc2a}
\enf
It has been found (Shanks et al., 1984; Tyson, 1988;
Broadhurst et al., 1990a,1990b;
Colles et al., 1990; Lilly et al., 1991; Cowie et al., 1993)
that  $\:\alpha \approx 0.6$ in the range $\:15 < m < 18$
and $\:\alpha \approx 0.45$ for $\:m \sim 19$
up to $\:m \sim 27$. The fact that $\:\alpha \approx 0.6$
at intermediate magnitudes has been invoked as a proof of homogeneity
of the
matter distribution (Peebles, 1993), while the subsequent change of slope
has been interpreted as an effect of galaxy evolution,
or a combined effect of galaxy and space-time evolution
(Broadhurst et al., 1990a, 1990b; Yoshii, 1993; Lilly et al., 1991;
 Metcalfe et al., 1991; Lilly, 1993; Babul \& Rees, 1992; Efstathiou
et al., 1991).

In the infrared K-Band $\:\alpha \sim 0.67$ for $\:12 < K < 16$
and the slope changes at $\:K \sim 18$ to $\:\alpha \approx 0.26$ as shown in
Fig.\ref{nc2b}. Hence this is in contrast with the B-band counts
 (Gardner et al. 1993,
Mobasher et al., 1986; Jenkins \& Reid, 1991).
\bef
\vspace{8cm}
\caption{\footnotesize Galaxy number counts in the K-band.
The counts
are fitted with two separate
power laws, with an exponent of $\:0.67$ in the
range $\: 10 < K < 16$ and $\:0.26$ in the range $\: 18 < K < 23$
(from Gardner et al. 1993).}
\label{nc2b}
\enf
There are several
biases and selection effects that should be taken into account
in the study of the galaxy luminosity properties especially for the
observations of faint galaxies
(see {\it section 2.4}) and we refer
to Yoshii (1993) and McGaugh (1994) for a detailed discussion.

If one considers the
galaxy counts at small and intermediate redshifts ($\: z< 0.2$)
the galaxy and space-time evolution shall be neglected.
In principle it should be possible to verify experimentally the relation
$\:\alpha = D/5$, that has never been tested.
To do this one needs to analyze the luminosity properties
together with the correlation properties in volume limited subsamples,
so that one does not introduce any bias due to observational selection effects
(for a detailed discussion see Sylos Labini \& Pietronero 1995c).
For example, analyzing the $\:CfA1$ redshift survey we have found that
$\:\alpha \sim 0.4$ for the whole magnitude limited catalog, while
it is $\:\alpha \sim 0.25 \approx D/5$ ($\: D \approx 1.4-1.5 $)
for a volume limited subsample.
Clearly in this case we have only a small range of magnitudes
available.

In order to identify the effect of galactic and space-time evolution
one has to clarify the space and
luminosity properties and the possible correlation
between them. In other words to
isolate the evolutionary effect on the galaxy counts,
certain properties of the spatial
 distribution of nearby galaxies need to be well known.
If one defines evolution as the deviation from the
Euclidean behavior ($\:\alpha=0.6$) there is the a priori
assumption that the galaxies distribution becomes homogenous
and that the space locations are not correlated with the luminosities
of galaxies, while this seems not to be the case in the available samples.

\subsubsection{Other parametric tests}
\bigskip

{\em 4.1.5.1. Ages of objects}
\bigskip

There are two methods to measure the age of the Universe $\:T_{0}$:
(i) determination of the
age of the oldest stars in a galaxy and (ii) determination
of the age of the chemical elements.
 Regarding the first, stellar evolution provides
a stringent limit to the age of the oldest stars. These stars are
observed in Globular Clusters
 and the age of these objects can be estimated
by fitting isochrones to the Hertzprung-Russell diagram.
This method is fairly accurate but it is based
on the theory of stellar evolution. It yields
 a typical age in the range $\:(14-16) Gyr$ but there are some
controversial estimates of $\:20 Gyr$.  Significantly lower ages
are in conflict with the generally accepted theory of stellar evolution
(Van der Berg, 1983, 1990; Sandage, 1988).

Radioactive decay ages are based on the
radioactive isotopes of $\:^{235}U$ in lunar and meteoric rocks.
Typical estimates give $\:T_{0} \sim (9-16) Gyr$ but there is
a big uncertainty depending on
the adopted models of galaxy evolution and various
other assumptions (Symbalisky \& Schramm, 1981).
\bigskip

{\it 4.1.5.2. Galaxy bulk flows}
\smallskip

To study the distribution of mass and not
only of the visible matter, one has to look not just
to the spatial position of galaxies, but also at their peculiar velocities.
In fact if peculiar velocities are generated
by inhomogeneities in the distribution of matter, they can
be used to estimate the total amount of mass.
The total redshift of a galaxy will be in general
\be
\label{vel1}
(1+z) \approx  (1+ z_{H}) (1 + z_{p}) \approx   z_{H}  + z_{p} + 1
\ee
where $\:z_{H}$ is the Hubble-redshift and
$\:z_{p}$ is the redshift due to the peculiar velocity
along the line of sight and the last expression is true for small $\:z$.
The problem is that one must measure the galaxy distance using
some indicator other than the redshift. Knowing the redshift and the distance
one readily obtains $\:z_{H}$ and
then $\:v_{p}=c\cdot (z-z_{H})$ from Eq.\ref{vel1}.  Actually there
 are two main methods
of measuring distances that have been applied in the study of
the peculiar velocities.
The first is based on the Tully-Fisher
 relation (Tully \& Fisher, 1977) for measuring the
distances of spiral galaxies, while the second
utilizes the $\:D_{n}-\sigma$ relation
(Lynden-Bell et al., 1988) for elliptical galaxies.
In the  last few years
the peculiar velocities here have been measured
for $\:\sim 4000$ galaxies (Efstathiou, 1994 for  a review).
Of course it is a very difficult task to produce an "homogenized"
catalogue of the peculiar velocity field because the
sampling of the peculiar velocity field is very patchy.

{}From the first controversial
finding of Rubin et al. (1976) of a large-scale flow,
called "Rubin-Ford"-effect, of  $\:600 km sec^{-1}$ in a direction
nearly orthogonal to the CMBR dipole, a lot of
evidences of very large scale coherent motions have been collected.
 Davis \& Peebles (1983) suggested the existence of
a systematic
inflow of galaxies towards the center of the Virgo cluster, but
Dressler et al. (1987) showed that the
phenomena are
more complicated than suggested
by simple Virgo infall model.
In fact there  are important evidences of a systematic motion of
$\:\sim 600 km sec^{-1}$ with respect to the CMBR
that suggests the existence for a large mass concentration,
called the "Great Attractor" (GA),
 in the direction of the Hydra-Centaurus supercluster
(Lynden-Bell et al., 1988).

An important kind of large-scale measurement of
motions is the so-called "bulk" flow that is the estimate
of the net velocity $\:<V>$ of galaxies within a large volume.
There are evidences
(Courteau et al., 1993) of large-scale coherent flows, and
find a coherent motion on very large
volume ($\:\sim 150 h^{-1}Mpc$) that implies an increase of $\:<V> $
on scales much larger than that of the GA.
If the CMBR dipole is due to Doppler effect then the
Local Group (LG) has a velocity of $\:620 km sec^{-1}$
with respect to the
rest frame represented by the CMBR. The origin of this motion can be
due to some anisotropic mass fluctuation whose amplitude is expected
(in an homogenous picture) to decrease with increasing scale. Hence the bulk
flows of galaxies contained in very large volume should be at rest with
respect to the CMBR. The results of Lauer \& Postman
show on the contrary that the LG motion relative to
 an Abell clusters sample
is inconsistent with the velocity of the LG inferred form the CMBR dipole,
and imply that the CMBR dipole anisotropy
 is generated by a very large
scale mass concentration
beyond $\:100 h^{-1} Mpc$.
Moreover Mathewson et al.(1992a) found that there is not backside infall
into the GA and that there is evidence of a bulk flow of $\:600 km sec^{-1}$
in the direction of the GA on  a
scale at least $\:60 h^{-1} Mpc$. Willick (1990)
found a bulk flow of $\:450 km sec^{-1}$ in the opposite part of the sky
and both these results suggested a bulk flow
in the supergalactic plane over very large
scale greater than $\:130 h^{-1}Mpc$
(Mathewson et al., 1992b). Very recently Mathewson \& Ford (1994)
found that the flow is not uniform over the GA region and that it seems
to be associated with a denser region that participates in the flow too.

{}From the data now available it
emerges that the {\it full extent of the galaxies flows
is still uncertain} and not detected and the origin of
these large amplitude and coherent length peculiar motions
is very unclear in the standard scenario. In fact
it is very hard to reconcile these results
with an homogenous picture,
in which the bulk flows have to be small on large scales
as the mass fluctuations have to be of small amplitude on the
large scale.
On the contrary in a fractal distribution,
that is intrinsically inhomogeneous,  there are fluctuations
in the distribution of mass at all scales; in this case
large-scale coherent flows are limited only by the
the property  of  local isotropy that
characterized a fractal structure and implies that the dipole
(the net gravitational force)
saturates beyond a certain scale (Sylos Labini, 1994).
\bigskip

{\em 4.1.5.3. Gravitational lensing}
\smallskip

The gravitational lensing effect is a new tool to test
the very large-scale distribution of both visible and dark matter.
One can use the lensing effect to map the total
amount of mass of the lens-object, and use
statistical arguments to constrain the large-scale
distribution of matter.
The cosmological applications of gravitational
lensing and modern observational data have been
considered by Blandford \& Narayan (1992) and
Refsdal \& Surdej (1994).
It has been shown by Baryshev et al. (1995b)
that, in principle, it is possible to distinguish between the
homogeneous and the fractal distribution of lenses in the Universe
(see {\it section 3.4.4.6}).
Unfortunately the currently available data
are still scarce, but  hopefully in the next few years
there will be new data on the gravitational lenses so that, in principle,
it will be possible to obtain information about large-scale
structure of the matter distribution in the Universe
essentially over the
entire causally connected universe.
\bigskip

{\em 4.1.5.4. Light element abundances}
\smallskip

The determination of pregalactic
(assumed primordial)
abundances of light elements provides
a probe to test the assumptions of
cosmological models.
At this aim, one needs to
infer the cosmic abundances
from the observed ones.
Measurements of these abundances
are carried out in our galaxy
($\:^{3}He$ and $\:^{7}Li$),
and also in other galaxies
($\:^{4}He$ and $\:H$).
For the $\:D$, the usual local ($\:< 1kpc$)
measurements have been very recently,
complemented by the first
probable detection of $\:D$
at a large redshift, i.e. $\:z\sim 3.3$, (Songaila et al., 1994).
In any case, unless for this
last measurement, the abundances of
the elements that
we observe today, are affected by chemical
evolution of the galaxies. This is a complex
process to model, as it involves the cycling
of interstellar gas through generations of stars.
Consequently, the inferred primordial
abundances of the light elements
are strongly model dependent
results.
The method for the detection
of $\:D$ is the same for either
the measurements in the local interstellar
medium or for the recent
one at high redshift.
In both cases, we study the
absorption spectrum of an astrophysical
light source (stars or quasars).
The presence of $\:D$ is
detected through measurements of the
isotopic shift of atomic HI lines
(Boesgaard \& Steigman, 1985).
If we study the absorption
spectra of high-redshift
quasars, we can analyse
the $\:D$ abundance in a variety
of very distant and fairly
chemically unevolved
environments.
In this case one can bypass
the need for the correction
of the chemical evolution of the Galaxy and
have a direct measurement of the ratio
$\:D/H$.
This last measurement has given
a value of  $\:D/H$ as high as $\:2.5 \cdot 10^{-4}$,
roughly $\: \sim10$ times the value observed
in the interstellar medium.
(Songaila et al., 1994a).

The abundance of $\:^{4}He$
has been determined in a number of
astrophysical sources: the atmospheres
of young stars, the atmospheres of old
stars, planetary nebulae, HII
regions in our galaxy and in
other galaxies, and in isolated
extragalactic HII regions.
Usually  to deduce
the primordial abundance
from the observed one,
models of stellar and
galactic chemical evolution are needed.
Low-metallicity, extragalactic
HII regions reasonably to provide
a "nearly" primordial sample of
the helium abundance. This is
determined from the analysis of the
HII region emission spectra,
as the intercept
(at zero metallicity) of the regression line
of the relation between helium
and oxygen abundances.
The primordial helium mass fraction $\:Y_{p}$
is determined to be, with a $\:95 \% $ confidence
limit, $\: 0.228 \pm 0.005$
(Pagel et al., 1992).

Lithium has been observed
in hundreds of PopI stars
of various ages, in  a less number of
PopII stars, in chondritic
meteorites and in the
interstellar gas.
All of the abundances are
based on measurements of the
equivalent width of the
Li I resonance doublet at
{\it 6707.761 A} and {\it
6707.912 A}.
As a star evolves, its
surface Lithium is subject to
destruction and dilution.
However, the observed
abundance in sufficiently
metal-poor halo stars, seems,
presumably, provide a nearly
primordial sample.
The inferred primordial
abundance is bounded by
$\:10^{-10} <(^{7}Li/H)_{p}
 < 8 \cdot 10^{-10}$
(Sandage, 1988).

The measurements
 of $\:^{3}He$ abundance
are restricted, substantially, to the solar
system only. Measurements in the
galactic
HII regions are very few and
are very hard to
perform.
Additionally, the situation
is more complex because during the
galactic evolution there is a competition
between destruction, survival and production
of $\:^{3}He$.
Therefore, the
value of $\:^{3}He$ as a probe
of primordial nucleosynthesis
is unclear, although some models
of galactic evolution seem
to confirm that $\:^{3}He$ and
$\:D + ^{3}He$  remain close to
the primordial values.
(Steigman \& Tosi, 1992).
 The observed data seem
to imply a value of $\:(D + ^{3}He)/H
\sim 4 \cdot 10^{-5}$.
This value is hard to reconcile
with the high value of $\:D/H$, found
at high redshift.
This contradiction can show how
the value of primordial abundances
(as in the case of $\:^{3}He$), inferred
from a galaxy evolution model, can be inadequate.
Using these bounds, it is possible to
investigate the consistency
of the big bang nucleosynthesis model
and derive constraints on the
nucleon-to-baryon ratio and so on the
baryon density $\:\Omega_{b}$.
At the present, a clear result is
that the inferred  $\:\Omega_{b}$
is much lower than the critical one
and the recent data on the
high redshift deuterium implied
a lower density, barely more
than the baryons
density we know to be present
in luminous stars or high-redshift
quasar absorber.

\subsection{Crucial cosmological tests}

The main task of observational cosmology is to select
a cosmological theory which is consistent with observational
data.  From the preceding consideration of the parametric tests
it is clear that the possibility
exists always to add some new
parameters in any cosmological theory and hence to explain
a special behavior of the observational data.  That is why we have
to pay special attention on the crucial cosmological tests which deal with
the initial hypotheses of the competing cosmological theories.
   According to our classification of cosmological theories (see
{\it Fig.\ref{fg1}}) we will consider as crucial the following subjects:
1) experimental basis of the gravity theory;
2) cosmological redshift nature; 3) large-scale matter
distribution; 4) CMBR nature.

\subsubsection{Experimental testing relativistic gravity}

Experiments in our Solar System can test relativistic gravity
only in the weak-field limit, i.e. the post-Newtonian (PN)
relativistic effects. So any given relativistic theory of
gravity which predicts the correct values of the experimentally
measured PN effects could be considered as a possible future
gravitation theory. Such a theory can deviate from General
Relativity (GR) in the strong gravitational field and lead to
an alternative cosmological model.
   GR is a geometrical theory of gravitation
which describes gravity as a property of space-time. GR
predictions such as the perihelion advances of a
planetary orbit, the
bending and delay of light rays passing near the Sun, and  the
gravitational redshift of spectral lines had been experimentally
checked in the Solar System with an accuracy of
about one percent and
it was a great success of that theory.
   The recently discovered pulsars in gravitationally bound binary
orbits, provide new astrophysical laboratories for testing
gravity at much more stronger fields.
In the case of the binary pulsar PSR1913+16
the GR prediction for the combined effect of the gravitational
redshift and the special relativistic time dilation has been
measured with an accuracy better than 0.07\% and the advance of
the periastron  with an accuracy of 0.0004\% (Taylor et al., 1992).
Moreover the new relativistic effect-energy loss via
gravitational radiation was discovered. According to Taylor et
al. (1992) the observed energy loss exceeds the theoretical prediction
for the quadrupole gravitational radiation of 0.96\% with an
accuracy 0.4\%. It has been shown by Damour \& Taylor (1991) that
one must take into account the effect of the galactic rotation
and the proper motion of the pulsar. The distance to PSR1913+16
is a very sensitive parameter in the calculation of the galactic
effect. If the distance is in the interval $\:3-8 kpc$ the
galactic effect is 0.11\% - 0.69\% respectively, so that
the problem of
the excess of the orbital energy loss needs further study.

Tensor field gravitation theory (TFT) describes gravity as a
physical interaction caused by exchange of gravitons in flat
space-time. All the post-Newtonian effects in the Solar System
and in the binary pulsars are the same as in GR, but the
interpretation of the effects differs. For example the 16.7\% of the
periastron advance is due to the positive energy density of the
gravitational field distributed around the gravitating neutron
stars. Hence the energy density of the static gravitational field
has been measured with an accuracy better then 0.002\%.
Within the TFT there are tensor quadrupole and scalar monopole
gravitational radiations. In the case of the  binary pulsar
PSR1913+16 the monopole radiation provides 0.735\% energy loss
excess (Baryshev,1994b).
Within the scope of the TFT there is a new
interpretation of the Planck mass and a possibility
exists for
quantum gravity effects in
a weak gravitational field  (Baryshev \& Raikov, 1994)
and hence for testing the quantum nature of the gravitational
interaction.

\subsubsection{Nature of redshift}

In {\it section 2.2.3} we discussed possible mechanisms
of the cosmological redshift.
About sixty years ago
Hubble \& Tolman (1935) suggested several tests
to discriminate between different redshift mechanisms,
but only now the observational technique
have reached an adequate level.
One of such tests is the surface brightness versus
redshift test. In the cases of space expansion, Doppler
and gravitational redshift the surface brightness
(SB) of a standard source decreases with increasing
redshift as $(1+z)^{4}$. In the case of tired light
the SB should vary with redshift only as $(1+z)$.

Recently Sandage \& Perelmuter (1991) considered observational
data for the first ranked galaxies in 56 nearby
clusters and groups. It was demonstrated that after
removing all biases and reducing the SB values
to a standard condition,  there was a $(1+z)^{4}$
dependence of the SB on redshift. This is
a first piece of evidence that the cosmological redshift
is caused by space expansion, Doppler, or
gravitational effects and that it is not the tired
light effect. However if one introduces an evolution
of the SB  within the scope of Tired Light cosmology
it is possible to incorporate the result in the scheme.

Another test of the redshift nature was proposed by
Sandage (1962) and is related to observations of
the redshift of the same object at different epoch,
i.e. $z(t)$ dependence. If the Universe expands
with a certain deceleration parameter $\:q_{0}$
the redshift of a fixed object
will vary with time so that
$(dz/dt)_{0} \approx -H_{0} q_{0}z$. In terms of frequency
variation it will be $1.7 \cdot 10^{-14}$ $(day)^{-1}$.
This test can distinguish between Doppler and
gravitational effects.

In {\it  section 3.2.5} we have considered the inhomogeneity
paradox, i.e. the linearity of the redshift-distance
relation deeply inside the fractal-like inhomogeneity
cell . The problem is that in the SM the linear
Hubble law is a consequence of the homogeneity.
So the measurements of distance independent from
redshifts and the measurements of the
large-scale structure at the same distances
could be considered as a test of the redshift nature.

\subsubsection{Large-scale matter distribution}

The existence of very large-scale structures
has been raised form many redshift surveys
(Tully, 1986, 1987; Paturel et al., 1988, 1994).
These structures are limited only by the extent of the survey
in which they are detected. The new correlation analysis
that we have discussed in {\it section 3.4} reconciles
the statistical studies with the observed LSS.
We have extensively discussed in {\it section 3.4.3}
and {\it section 3.4.4.} the main properties of
visible matter distribution and we refer to these sections
for an detailed analysis of the experimental data.
Here we summarize the main results.
\begin{itemize}
\item $\:D \approx 2$ . The fractal dimension of galaxy distribution
approaches to 2 in many independent redshift surveys such as
CfA2 (Park et al., 1994), QDOT (Moore et al., 1994), Perseus-Pisces
(Guzzo et al., 1992) and ESP (Sylos Labini \& Pietronero, 1995a).

\item $\:\lambda_{0} \gtapprox R_{s}$ where $\:\lambda_{0}$ is the
scale of homogeneity and $\:R_{s}$ is the depth of the various currently
available
redshift surveys. No clear cut-off towards homogenization has yet been
identified in the available samples
($R_{s}\gtapprox 200-400 Mpc$).
In the next few years there will be available many
new
redshift surveys (see for a review Efstathiou, 1994)
that will cover a fraction of the entire
Hubble radius. From these data a
definitive measurement of the eventual scale of homogeneity
will be soon available.

\item $\:D \approx 1.5$ for the clusters distribution, even
though there are a lot of uncertainties in the exact
value of the fractal dimension due to the poor statistics
of cluster catalogues. Recently Borgani et al. (1994)
found that the fractal dimension turns
out to be $\:D \sim 2.2.$ for both Abell and ACO clusters.

\item The galaxy
luminosity function is well described by a Schechter function.

\item Low-surface-brightness and dwarf galaxies seem to
fall into the structures delineated by the luminous ones
and there is no evidence that
these galaxies fill voids (see {\it section 2.4}).

\item The whole mass distribution can be investigated by its
gravitational effects studying the large-scale bulk flows
({\it section 4.1.5.2.}) and
the gravitational lensing ({\it section 4.1.5.3.}).
The full extent of these flows
is still unknown and it is not clear if the dark matter
is traced by the luminous one. The gravitational lensing
 is a new tool that is at the beginning
of its development.
\end{itemize}

\subsubsection{Cosmic Microwave Background Radiation}
There are several kind of observations
that have been done on the CMBR from its discovery in 1968 until now
(Penzias \& Wilson, 1965; see for a review
Melchiorri \& Melchiorri, 1994).
 The absolute radiometry intends
to determine the temperature and the spectrum of the CMBR. The most
accurate spectrum is obtained with the observation of a region
of low dust density content and it was found by the COBE team
(Mather et al., 1990, 1994) and independently by Gush (1990)
that it is a perfect black body spectrum with
a temperature:
\be
\label{cmbr1}
T = 2.726 \pm 0.010 K
\ee
This is the most accurate determination of the CMBR in the
millimetric and submillimetric region ($\:1-20 cm^{-1}$).
Deviations from this blackbody are less than 1 \% of
the peak brightness.
 The situation is more
complex at higher wavelengths (Melchiorri \& Melchiorri, 1994).
The CMBR is remarkably close to isotropic, but
there is a clear indication of an existence of a dipole anisotropy.
 COBE
 (Fixsen et al., 1994) has provided both the spectrum and the direction of
dipole anisotropy with high accuracy. The observed
temperature difference between two regions in the sky, which lie in
the
opposite directions,
is $\:\Delta T =  3.343 \pm 0.016 mK$.
If this anisotropy is
due to the motion of our galaxy with respect to the CMBR
this is an important cosmological tool when compared with the observed
peculiar velocities of other galaxies (see {\it section 4.1.5.2.}).

 The detection of small as well as large-scale anisotropies
is a very difficult task (Melchiorri \& Melchiorri, 1994).
 The main problem is to identify the possible
source of spurious
signals that can be instrumental and local environmental effects,
atmospheric, Solar System, galactic and extragalactic disturbances.
 Few groups have detected CMBR anisotropies at various angular scales
while many other have obtained upper limits only. The amplitude
of these fluctuations, after having taken into account the
various spurious contributions,
is of the order of some $\:\Delta T/T \sim 10^{-5}$. Hence it
is very hard to decide if the signals observed by Relict 1
 (Strukov et al., 1992), by
COBE (Smoot et al., 1992) and other groups
are CMBR anisotropies or a mixture of CMBR and spurious signals.
These results pose some trouble for the  SM . In fact
in the standard scenario of galaxy formation, small amplitude primordial
perturbations to the energy density are amplified by gravitational instability
as the Universe expands and they are predicted to
leave a detectable imprint on the CMBR. Such low
amplitude measured fluctuations are not
compatible with the standard baryonic matter
and hence in  many theories
of galaxy formation one must introduce some
exotic kind of non baryonic matter that
has a weaker interaction with photons
than the baryonic one. Moreover on large angular scale (few degrees)
the dominant mechanism by which density fluctuations induce anisotropy in the
CMBR is the Sachs-Wolf effect
(Sachs \& Wolf, 1967). This is a gravitational effect
due to the presence of matter between the CMBR and the observer.
For this reason the existence
of large-scale structures can represent  a
serious problem due to their
incompatibility with such low amplitude
temperature fluctuations.

According to the Hot Big Bang (HBB) model
the temperature of the CMBR must scale in proportion to $\:(1+z)$,
where $\:z$ is the redshift at which the radiation is measured.
Clearly such a measurement is a crucial one because it is
a direct proof in favor (or not) of the HBB scenario. This
 effect can be tested by measuring the populations
of excited fine-structure lines
in the absorption spectra of distant quasars.
Observations of this kind  is strongly limited
by the signal-to-noise ratio of the available telescopes
(Meyer et al., 1986).
Recently (Songaila et al., 1994a, 1994b) found a meaningful
$\: 2 \sigma $
upper limit of $\: T=13.5 K$ for the temperature of
the CMBR in a cloud
that contains singly ionized carbon at $\:z = 2.9$, and
have measured the excitation temperature
of $\:7.4 \pm 0.8$ for the first
fine-structure level of neutral carbonic
atoms in a cloud at redshift $\:1.776$.
 These upper
limit are very close to the temperature expected
on the basis of the HBB scenario that is $\:T=10.7 K$
and $\:T = 7.58 K$, of course assuming that
no other significant sources of excitation (collisions, radiation) are
are present.
It is to be hoped that  more definitive measurements
will be soon available because a firm measurement
of the CMBR temperature at high redshift
would permit to
rule out or not the HBB theory.

\section{DISCUSSION AND CONCLUSIONS}

Cosmology as an experimental science must
be based on experimentally checked hypotheses, and
 the main task of this paper is to
analyze what are facts and what are
only ideas in contemporary
 cosmology. To do this, one requires a
comparison of cosmological models with
observational data and we have divided
({\it sec. 2.1.,2.2})
the possible experimental tests into two
different kinds.
The first ones are the crucial tests and deal with the fundamental
basis of any cosmological theory. They check the validity of
the initial assumptions and hypotheses of  various theories.
The second kind of tests are the parametric ones and give estimates
of the parameters of different models. The whole set of parametric tests
can play the role of a crucial one if no other free parameters of the
model  are available.
We have classified different cosmological theories (Fig.1.)
according to their answer to the following questions: what is gravity,
how matter is distributed in space, what is the nature of the redshift,
what is the nature of the CMBR and what is evolution and the
arrow of time. Therefore according to this classification,
 the crucial tests
deal with these fundamental matters: gravitation,
matter distribution, redshift nature, and Cosmic
 Microwave Background Radiation
(CMBR) nature.

Experiments on gravitation are performed only in
weak field approximation.
In this limit General Relativity (GR)
gives predictions that are in good agreement with the results
of various experiments. We discuss ({\it sec. 2.2.1.})
different approaches to the gravitation
theory that give basically the same results of GR in a weak field,
but that can be different in stronger fields. Among these approaches
we have discussed in particular the tensor field theory in flat space-time
that can represent an alternative direction with
respect to the geometric theories.

The matter distribution ({\it  sec. 2.2.2.})
in space represents the
most powerful test of the basic initial
hypothesis
of the main cosmological theories: the Standard Friedman
Model (SM) and the Steady-State Model (SSM). In fact, both theories
assume, beyond a certain scale $\:\lambda_{0}$,
the homogeneity of matter distribution. The initial hypothesis of homogeneity
has been justified as required by the {\it Cosmological Principle}
 (CP)
in order to avoid any privileged observer in the Universe.
In our opinion, nowadays,  the main experimental problem is
the value of $\:\lambda_{0}$. We have discussed
in detail this point stressing that from many redshift
surveys there is no clear evidence towards homogenization,
but on the contrary  a well defined fractal behavior is found.
Moreover the condition of local isotropy, that is satisfied in a
fractal structure, is the necessary condition in order to ensure the
statistical
equivalence of all the observers. Hence the CP can be saved
in a fractal distribution in a more weaker version, i.e.
without the strong request of the complete
transitional and rotational invariance (homogeneity).

Another initial hypothesis of the SM and SSM is the {\it Expanding Space
Paradigm}. According to this
assumption the redshift and the linearity of the Hubble law
are due to space expansion. Actually ({\it sec. 2.2.3.})
there are
four different kinds of physical mechanisms that
can produce a redshift: the Doppler, the gravitational,
 the space expansion and
the tired light effects. Of these, only the first two are
tested in laboratory experiments, while
the space-expansion and the tired light redshifts have never been
experimentally proved.
We have discussed in detail the
gravitational redshift in a homogeneous
and in a fractal structure starting from
the spherically symmetrical Bondi-Tolman model.
We have stressed that for a fractal with dimension
$D=2$, it is possible to obtain
a linear {\it z-d} relation,
whose amplitude is related to the
lower cut-offs of the fractal distribution.
We have particularly
emphasized that the observational quantity is the redshift and
not the expansion velocity.

The CMBR ({\it sec. 2.2.4.})
together with the three dimensional space distribution
of galaxies is the most important fact of modern cosmology.
The main characteristics are its present temperature, its nearly perfect
blackbody spectrum, its extraordinary isotropy and its energy content that is
of the same order of magnitude as local sources.
The extragalactic nature of the CMBR is without doubts
for the observed transparency of the Universe up
to the distance of  Quasars,  but there
is not clear experimental evidence, but only upper limits,
 that its temperature scales
linearly with redshift,
as predicted by the Hot Big Bang (HBB) scenario.

The discussion on modern cosmological ideas begins
({\it sec. 3.1.})
from the well known paradoxes of  Newtonian cosmology such as the
gravitational, the Olber's and the thermodynamics paradoxes. These
paradoxes pose fundamental problems. To resolve a
paradox
 one needs to consider the initial postulates of the model used.

The SM ({\it sec. 3.2.})
is based on the  following hypothesis: the GR is the correct
gravity theory, the density of matter is constant (at fixed time)
so that the Universe is
homogenous beyond a certain scale $\:\lambda_{0}$,
the redshift and the linearity of the Hubble law ($\:z-d$
relation) are due
to space expansion and, finally the laws of thermodynamics
hold in the expanding space.
The main success ({\it sec. 3.2.2.})
of the SM lies in the prediction of the
blackbody spectrum of the CMBR, while it does not determine its
temperature and does not explain its isotropy. Another great success
of the SM is the predictions of light elements abundances that
are in good accordance with observational constraints.
The crucial tests ({\it sec. 3.2.3.})
of the SM concern the GR in a
strong field, the detection of the homogeneity scale
$\:\lambda_{0}$ of matter
distribution, the reality of space expansion
and then that of the expansion-redshift, the experimental
verification of the relation $\:T(z) = T_{0} (1+z)$,
the comparison of the age of the Universe with the age of the oldest
objects and finally the comparison of theoretical
predictions with more
stringent data on chemical compositions.
The classical parametric tests
that we have considered ({\it sec. 3.2.4.})
for a comparison with experimental data
 are the redshift-distance,
redshift-angular size, redshift-magnitude,
number-magnitude, number-redshift and time-redshift relations.
There are 5 paradoxes of the SM ({\it sec. 3.2.5.})
of which 4 are
well-known and the last
one has only recently been discussed. In fact the flatness, the isotropy,
the superluminal velocity and the global energy paradoxes are
widely discussed in the literature and there
are attempts to resolve some of these paradoxes in the framework of
Big Bang models. The so-called {\it inhomogeneity paradox}
 is the following:
if the Hubble law is a consequence of homogeneity, then there is a
contradiction of the
nearly linear $\:z-d$ relation at the same scales (at least $\:2-20 Mpc$)
where one observes well defined fractal behavior and large-scale
departures from homogeneity.
It is possible to characterize quantitatively
the expected behavior of the $\:z-d$ relation
in the case of fractal distributions of matter: the result
is that inside  the "inhomogeneity-cell", i.e. for
distances $\:\ltapprox \lambda_{0}$  a strong non-linearity
is expected.
To resolve this paradox one should
hypothesize that the dark matter is homogenous at very small scales
(of order of some Mpc) and hence one
introduces a very large amount of dark matter that is
 in conflict with
the dynamical estimates in the galactic halos and galaxy  clusters.

To resolve some of  the problems of the SM
non
standard models have been introduced ({\it sec. 3.2.6.})
that are the $\:\Lambda-term$ model ({\it sec. 3.2.6.1.})
and the inflationary
Universe. In both these models one
avoids some particular problems of the SM
but introduces some new hypothesis that has
to be checked experimentally.
In the case of the $\:\Lambda-term$ model one can make the age
of the Universe larger than in the SM (for a fixed density)
and modify the geometry
of the Universe, so that a flat space is allowed also with a density parameter
lower than 1 but with $\:\Omega+\Lambda =1$.
The problem, here, is to explain the particular value of
the cosmological constant used, i.e. the
famous {\it cosmological constant problem}:
The theoretical expectation value for the cosmological constant exceeds
the observational limits by some 120 orders of magnitude.
In the inflationary scenario ({\it sec. 3.2.6.2.})
one introduces a phase of exponential expansion
during the evolution of the Universe that can explain the flatness
and the isotropy paradoxes. The scalar field, that is central in this scenario,
comes from the Grand Unified Theories (GUT ) of weak and strong interactions.
The problems in this approach are
the strong initial assumptions and the poor experimental
verifications of GUT theories.
In any case this model does not resolve the other paradoxes of
the SM.

 In the Steady-State Model (SSM) ({\it sec. 3.3. and 3.3.1.})
it is assumed that the density of matter is
constant in space and in time (Perfect Cosmological Principle),
 that a modified
version of GR is the gravity theory and that the redshift is due
to the expansion of  space. There are predictions ({\it sec. 3.3.2.})
for the classical
parametric tests that, at the first order in $\:z$,
do not differ
from the  SM 's ones. The flatness and the
 isotropy paradoxes are avoided,
but there are still the superluminal, the global energy
and the inhomogeneity paradoxes.
In the new version of SSM, the "Quasi- Steady State Model"
({\it sec. 3.3.3.})
there
is attempt to explain the CMBR as
a post-stellar scattered radiation thermalized by
a special kind of intergalactic medium. The thermalising process
can be one of the crucial aspects of this model, together with
the crucial tests that we have discussed for the  SM .

A fractal model ({\it sec. 3.4.}
has not been developed yet. Hence, first of all we
have focused ({\it sec. 3.4.1.})
on the main properties of  non-analytic and
self-similar distributions stressing which are the new
concepts that can have an important impact not only in theoretical models
but also in data analysis. In fact fractals are intrinsically
non-analytical distributions
characterized by having self-similar structures and voids
at all scales
({\it sec. 3.4.2.}).
For this kind of distributions, the correlations properties
are described by power law functions. The main point is that
one cannot discuss self-similar structures in terms of
amplitude of correlations, neither in a theoretical model
nor in the data analysis.
The only physically meaningful property is the exponent
of the correlation function, and this is
the crucial quantity that a theoretical
model needs to explain. The amplitude of the
correlation function, as well as other related
quantities such as
the normalized density fluctuations ($\:\delta N / N$), are spurious.
We have stressed therefore that for a self-similar structure it
has no physical meaning to discuss the dynamics in terms
of "linear" and "non-linear" regimes, that are identified
by the scale at which the fluctuations become negligible with respect to the
average (i.e. $\: \delta N \sim N$) because the average itself
is not well defined, because
it depends from the sample size.
Hence the concept of  "big" and "small"   amplitudes are misleading
for a self-similar distribution, because
 is not defined a constant  average density.
We have discussed in detail ({\it sec. 3.4.2.})
the analysis
{\it without assumptions} that we have performed
on real redshift survey, such as {\it CfA1,
Abell, CfA2, ESP}.
{}From this discussion it is clear that
the new redshift surveys data imply a deep change of
the theories of galaxy formation, such as the
biased galaxy formation: the clusters correlations
are just the continuation of the galaxy correlations
at larger scales and the galaxy-cluster mismatch,
i.e. the different amplitude of correlations function for
galaxies and clusters,
is simply solved from this point of view.

If one considers the whole mass distribution,
 the concept of multifractal (MF)  naturally arises
({\it sec. 3.4.4.})
that is just a refinement of the concept of fractal
and, indeed, it is not in contrast with it.
Hence if the whole mass distribution is self-similar
the space locations of galaxies are correlated
with their luminosities, and this correlation can be
studied and quantitatively characterized with  the
MF  formalism.
Analyzing the properties of  fractal (and MF) distributions
 in the three dimensional Euclidean space,
one can make predictions for  some of the
classical parametric tests, such as the
number-counts and the redshift-magnitude relations. We find that
in this case, at small redshifts, there is a good agreement with the
available data. Actually the situation at intermediate and large
redshifts is not clear because of  selection effects in the
data, in the data analysis  and  the possible evolutionary effects
that can be relevant at these distances. The problem here lies in the fact
that if one defines evolution as
the difference between theoretical predictions and
experimental data, one should be sure that the model has
been widely verified.
The correlation between
space locations and masses of galaxies is the important element
that should  be taken into account not only
in theoretical models, but also for the data analysis.
 The luminosity function (LF) of galaxies
can be naturally
related to the MF properties and in
particular  its
exponent can be related to the MF spectrum.
The amplitude of the LF  is related
to the average density and hence it is not
constant but depends upon the sample depth.
These considerations have an important impact also in the
experimental
methods for determining the luminosity function.
Moreover we have described  a simple stochastic model based
on the aggregation of particles. The main aim of such a model  is
to study which are the characteristics of the dynamical process that
can give rise to a fractal, and a MF if one
includes masses distribution. We have identified in
the breaking of the spatial
symmetry of the aggregation process and the Self-Organized
nature of the dynamical mechanism, the
key elements in order to generate such structures.

There have been some interesting
 attempts towards the so-called "Newtonian fractal models"
({\it sec. 3.4.5.})
in particular because they permit us to
solve two famous paradoxes, the gravitational
and the Olbers ones, without
need to modify  the Newtonian gravity theory.
In the framework of General Relativity (GR) there have been some attempts
to describe fractal distributions ({\it sec. 3.4.6.}).
In particular using the
inhomogeneous spherical
symmetric
metric (the Bondi-Tolman metric)
joined with a Friedmann metric (i.e. the so-called {\it Swiss cheese model})
one avoids any privileged observer and saves the equivalence of all points.
In this model it is possible to find solutions
that agree with two main experimental facts: the linear redshift-distance
relation at small redshift and  the fractal behavior along the backward
 null cone
with the right experimental values of the Hubble
constant and the fractal dimension. Therefore it is possible
to describe such complex structures in GR even if
this model does not provide an answer to the origin problem.
But this is the first methodological example
of including fractal models in the framework of GR.
Moreover it clarifies the point that in GR
it is possible to describe systems with a vanishing
average density: the key-point is that
the average density is made not
at a space-like hypersurface of constant
 time, but such averaging is carried
out along the backward null cone.

We considered also a possibility of the construction of
a cosmological model based on the tensor field gravitation theory (TFT)
 and fractal distribution. In particular, we note
that TFT, with the additional condition of the
positiveness of the energy density of the
gravitational field, substantially differs from GR in the case of
strong fields and infinite matter distribution.
In weak fields approximations, it gives the same predictions
of GR.
As an example, we construct a static
 cosmological model, with cosmological redshift
due to the gravitation of the fractal matter distribution
with fractal dimension $D=2$ {\it sec. 3.4.7.}.

There are several other alternative
cosmologies, which have been discussed
 in the literature. Usually
there is no a definitive formulation
 of the initial hypotheses of the models.
Further development of the models
 to make it possible a comparison of
its predictions with observations
is needed ({\it sec. 3.5}).

According to our classification of observational data, parametric
tests permit the determination of the various parameters
of a model. The whole set of parametric tests can be considered
as a crucial one if no other free parameter can be added to the model
({\it sec. 4.1.}).

Up to now the SM only has definite predictions
for classical parametric tests.
The  most important parametric test is the redshift-distance relation
({\it sec. 4.1.1.}).
The main results are
the linearity of the $\:z-d$ Hubble-law in the range
$\:\sim 2-100 Mpc$ and the
value of the Hubble constant that, from various independent
measurements, seems to converge now to the value of
$\:\sim 80 km sec^{-1}Mpc^{-1}$.
This high value may represent a problem for the SM unless
one does allow the density parameter less than one.
Moreover the linearity of this relation
is difficult to reconcile, if due to space expansion,
with a highly inhomogeneous distribution of matter at the
same scales. This is the  {\it inhomogeneity-paradox} that we
have previously discussed.
On the contrary if we consider
the cosmological redshift
as due to the gravitational redshift
effect, without space expansion,
one obtains a linear $z-d$ relation
in highly inhomogeneous
matter distribution.
On the angular size-distance relation there are different results
according to the different "standard-sticks" used
({\it sec. 4.1.2.}). The experimental
data are therefore problematic and no
definitive answer on the geometry of the Universe
is accepted. The slope of the magnitude-redshift ($\: <m> \sim \log(z)$)
relation has been
found to be 5 with fair accuracy
({\it sec. 4.1.3.}). This result does not permit
us to discriminate
between a fractal and an homogenous distribution.
In fact, we have shown that
both distributions at low redshift (Euclidean space)
have the same behavior. Moreover
there is a discrepancy between the results in the B-band
and  the K-band.
At larger
redshifts the experimental situation is highly uncertain and
an evolutionary model has probably to be taken into account.
 The problem
of which particular evolutionary model one chooses,
 as discussed before, is very
difficult. The exponent of the galaxy counts ($\:log(N<m) \sim \alpha m$)
is connected
with the fractal dimension of matter distribution only
if it comes from the
analysis of the number-counts in
volume limited samples and under the
assumption that there is no
correlation between space locations and luminosities of
galaxies. We have discussed in detail
these effects that have never been taken into account
either in the analysis, nor in the theoretical models
({\it sec. 4.1.4.}).
Even in this case no definitive answers on the
value of the fractal dimension and on the geometry of the space
are available.
Moreover the evolutionary,
the bias and selection effects can play an important role,
and can explain the different slopes measured in different
frequency-bands.

The age of the oldest objects is a very stringent test
 for the validity of
 SM  models
({\it sec. 4.1.5.1.}).
In fact typical estimates of globular cluster ages and
radioactive decay ages give respectively $\:T_{0} \sim 14-16 Gyr$
and $\:T_{0} \sim 9-16 Gyr$. In the  SM
these ages are in conflict with a high value
of the Hubble constant ($\:H_{0} \gtapprox 75 km sec^{-1}Mpc^{-1}$)
unless one introduces a cosmological
constant different from zero.

To study the whole mass distribution,
including the dark matter, it is
important to consider  the bulk flows, i.e. the net peculiar velocity
of galaxies averaged over big volumes ({\it sec. 4.1.5.2.}).
 The main results
that in these last years have been obtained
from many independent surveys, is that the
full extent of these flows is  still uncertain and it is limited only
by the sample size. In fact
very large-scale coherent flows  have been detected,
 over scales of $\:\sim 150 Mpc$.
It is very hard to reconcile these results with an homogeneous
picture, where the
mass fluctuations have to be small on large scales.
In a fractal distribution the fluctuations are intrinsic
and the extent of these flows can be limited only by the scale
of isotropy $\:\lambda_{i}$
that has not been clearly detected in the available samples.

A new tool to study the whole mass distribution is
represented by the gravitational lensing
({\it sec. 4.1.5.3.}). We have shown
the difference between an homogenous and a fractal distributions
of  lenses. The experimental data are still scarce, but
in the near future this may represent a powerful test to study
matter distribution on very large scales.

The light elements abundances estimations are affected by
the chemical evolution of the galaxy
({\it sec. 4.1.5.4.}). Hence the inferred
primordial abundances are strongly model dependent
results. Only recently the estimates of extragalactic
abundances such as deuterium are available. In any case
the predicted abundance of helium and deuterium are well in
agreement with the predictions of the standard Hot Big-Bang
nucleosynthesis. Also the revised version of the SSM
is claimed to  reproduce the observed abundances.

As we stressed previously experiments on
gravitation were performed only in
weak field approximation.
The recently discovered pulsars
in gravitationally bound
binary orbits provide a new
laboratory
for testing predictions  of GR as of TFT
({\it sec. 4.2.1.}).
In fact, in the case of the
binary pulsars PSR 1913+16, it is
predicted, within the TFT, the
existence of 0.735\% excess of
gravitational radiation, due to the
scalar gravitation waves.
It is very important for
cosmology that in the near future, observational
technology will reach the necessary level for testing
of the redshift nature. Recent observations of
the surface brightness of giant elliptical
galaxies (Sandage \& Perelmuter, 1991)
showed a first experimental indication
against tired light (TL) mechanism
of cosmological redshift.
However, these observations can not
discriminate between space expansion, Doppler and
gravitational effects
({\it sec. 4.2.2.}).

Among the crucial tests, the investigations  of the
large-scale matter distribution has been strongly developed in the last
decade ({\it sec. 4.2.3.}). Now
 there are now available three dimensional redshift surveys that
cover a large fraction of the Hubble radius. Moreover in the next
few years new and larger redshift surveys will be
 completed. From these catalogues emerges the new picture of the Universe
that is dominated by large-scale structures and voids of all scales.
 In fact, the extent of
the largest structures and voids recently discovered is limited only by the
boundaries of the surveys. The homogeneity scale $\:\lambda_{0}$
has not been identified and seems to shift to a very large value,
at least 10 times greater than the previously believed scale
inferred from the angular catalogues ($\: \sim 5 Mpc$).
This new situation clearly poses new crucial
 problems
to the classical paradigm.
The fractal dimension
of the visible matter distribution converges to
the value $\:D \approx 2$ from many independent redshift surveys.
The galaxy cluster distribution shows power law correlations
that are the continuation of the galaxy correlations
at more deeper scales. Hence clusters and galaxies are
part of the same
self-similar and scale-invariant distribution.
A very important observational cosmology, in our opinion, is
the determination of the eventual homogeneity scale
$\:\lambda_{0}$ and this is a crucial point because the
assumption of homogeneity is the cornerstone of the SM and SSM.

The Cosmic Microwave Background Radiation represents
the other well studied observational fact in modern cosmology
 ({\it sec. 4.2.4.}).
Its temperature and spectrum has been determined very accurately.
The isotropy of  the CMBR blackbody spectrum has been
intensively studied. The  result is  that the anisotropies
are probably detected at $\:\Delta T/T \sim 10^{-5}$.
The extragalactic origin of this radiation is without doubt. The
crucial test for the Hot Big-Bang scenario is the
measurements of the CMBR temperature at large redshift, and
in particular the experimental proof that the temperature scales
as $\:(1+z)$. Actually  only upper limits are available, but
with the use of new large telescopes in the following
years it is hoped to obtain more stringent
results.
\bigskip

Our conclusion is that some of the basic assumptions of the
 SM  and SSM are in conflict with the new observations
and tests. It is important therefore to focus the attention of the
experimental crucial tests that can give
a definitive answer to the validity of the classical paradigm.
Moreover it should be stressed that the fractal behavior
of matter has changed our view of galaxy correlations
that are now compatible with the observed large-scale
structures. This
is the new and fundamental element that
has to be taken into account in any cosmological model.
The significant growth of the observational data in cosmology
leads to a new situation, because the Standard Model has to use
many ad hoc parameters to explain observations. But such
procedures can be made within other alternative models
and the exceptional role of the Standard Model disappear.
That is why now it is a good time for more careful analysis of
initial hypotheses of different cosmological theories.
To distinguish between alternative possibilities
one must develop the crucial observational tests
which allow us to check the initial postulates of the models.
We emphasized that such tests will be fulfilled in the near future.
\bigskip

{\it Acknowledgments}
\smallskip

We are particularly grateful to Dr. P. Vettolani,
for giving us the possibility of analyzing
the ESP survey and to Dr. P. Coleman for
continuous discussions and collaborations.
We thank for useful discussions, criticisms and
suggestions L. Amendola, P. Bak, R. Cafiero,
R. Capuzzo-Dolcetta, M.Costantini,
P. de Bernardis, P. Grassberger,
M. Geller, R. Kolb,
A. Linde, V. Loreto, B. Mandelbrot,
F.Melchiorri, S.A. Oschepkov, G.Paladin,
A.A. Raikov, R. Scaramella, D. Schramm, M. Ribeiro,
A.G. Sergeev, G. Setti, P. Teerikorpi,
D. Trevese, S.A. Tron',
N. Vittorio, G. Zamorani.

\newpage

\section*{FACTS AND IDEAS IN MODERN COSMOLOGY}
\vspace{0.9cm}

\hspace{0.6cm}{1. \bf INTRODUCTION}
\vspace{0.9cm}

{2. \bf GENERAL BASIS OF COSMOLOGY}
\vspace{0.9cm}

{\bf 2.1 Empirical basis for cosmological theories}
\bigskip

{\bf 2.2. A classification of cosmological theories}
\smallskip

{\rm 2.2.1. Gravitation theories}
\smallskip

{\rm 2.2.2. Matter distribution}
\smallskip

{\rm 2.2.3. The nature of the redshift}
\smallskip

{\rm 2.2.4. CMBR nature}
\smallskip

{\rm 2.2.5. Evolution and the arrow of time}
\smallskip

{\rm 2.2.6. Main rival cosmological theories}
\bigskip

{\bf 2.3. Classification of cosmological tests}
\bigskip

{\bf 2.4. Biases and selection effects in astronomical data}
\vspace{0.9cm}

{\bf 3. THEORETICAL PREDICTIONS OF COSMOLOGICAL MODELS}
\bigskip

{\bf 3.1. Paradoxes of Newtonian cosmology and origin of modern
cosmological ideas}
\bigskip

{\bf 3.2. Big Bang models}
\smallskip

{\rm 3.2.1. Initial hypotheses of the Standard Friedmann Model}
\smallskip

{\rm 3.2.2. Successes of the Standard Model}
\smallskip

{\rm 3.2.3. Crucial tests}
\smallskip

{\rm 3.2.4. Parametric tests}

\hspace{3cm} {\it 3.2.4.1. The proper distance-redshift relation}

\hspace{3cm} {\it 3.2.4.2. The angular size-redshift relation}

\hspace{3cm} {\it 3.2.4.3. The magnitude-redshift relation }

\hspace{3cm} {\it 3.2.4.4. The time-redshift relation }

\hspace{3cm} {\it 3.2.4.5. Count-redshift and count-magnitude relations}
\smallskip

{\rm 3.2.5. Paradoxes of the standard model}
\smallskip

{\rm 3.2.6. Non-standard models}

\hspace{3cm} {\it 3.2.6.1. $\:\Lambda$-term}

\hspace{3cm} {\it 3.2.6.2. Inflationary Universe}
\bigskip

{\bf 3.3. Steady-State Models}
\smallskip

{\rm 3.3.1. Basic hypothesis of the SSM}
\smallskip

{\rm 3.3.2. Predictions for testing}
\smallskip

{\rm 3.3.3. Modern developments of the SSM}
\bigskip

{\bf 3.4. Fractal Models}
\smallskip

{\rm 3.4.1. Fractal geometry}
\smallskip

{\rm 3.4.2. Fractal structures}

\hspace{3cm}{\it 3.4.2.1. Mathematical Self-similarity}

\hspace{3cm}{\it 3.4.2.2. "Linear and non-linear dynamics"}

\hspace{3cm}{\it 3.4.2.3. Fair sample}

\hspace{3cm}{\it 3.4.2.4. Isotropy, homogeneity and Cosmological Principle}

\hspace{3cm}{\it 3.4.2.5. Power Spectrum}
\smallskip

{\rm 3.4.3. Fractal Properties of visible matter}

\hspace{3cm}{\it 3.4.3.1. Galaxies and clusters}

\hspace{3cm}{\it 3.4.3.2. The CfA2 redshift survey}

\hspace{3cm}{\it 3.4.3.3. Pencil beams and very deep surveys}
\smallskip

{\rm 3.4.4. Multifractal Properties of luminous matter distribution}

\hspace{3cm}{\it 3.4.4.1. Luminosity function}

\hspace{3cm}{\it 3.4.4.2. Number counts}

\hspace{3cm}{\it 3.4.4.3. Magnitude-redshift relation}

\hspace{3cm}{\it 3.4.4.4. Luminosity segregation }

\hspace{3cm}{\it 3.4.4.5. A simple stochastic
model for the formation of a MF distribution}
\smallskip

\hspace{3cm}{\it 3.4.4.6. Gravitational lensing in a fractal distribution}

{\rm 3.4.5. Newtonian fractal models }
\smallskip

{\rm 3.4.6. General relativity Fractal models }
\smallskip

{\rm 3.4.7. Field Fractal models}
\bigskip

{\bf 3.5. Other possibilities}
\vspace{0.9cm}

{\bf 4. CONFRONTATION OF WORLD MODELS WITH OBSERVATIONS}
\bigskip

{\bf 4.1 Parametric cosmological tests}
\smallskip

{\rm 4.1.1. Redshift-distance relation and the Hubble constant}
\smallskip

{\rm 4.1.2. Angular size-redshift relation}
\smallskip

{\rm 4.1.3. Apparent magnitude-redshift relation}
\smallskip

{\rm 4.1.4. Galaxy counts}
\smallskip

{\rm 4.1.5. Other parametric tests}

\hspace{3cm}{\it 4.1.5.1. Ages of objects}

\hspace{3cm}{\it 4.1.5.2. Galaxy bulk flows}

\hspace{3cm}{\it 4.1.5.3. Gravitational lensing}

\hspace{3cm}{\it 4.1.5.4. Light element abundances}
\smallskip

{\bf 4.2 Crucial cosmological tests}
\smallskip

{\rm 4.2.1. Experimental testing of the relativistic gravity}
\smallskip

{\rm 4.2.2. Redshift nature}
\smallskip

{\rm 4.2.3. Large-scale matter distribution}
\smallskip

{\rm 4.2.4. Cosmic Microwave Background Radiation}
\bigskip

{\bf 5. DISCUSSION AND CONCLUSIONS}
\bigskip

{\bf 6. REFERENCES}
\bigskip

\end{document}